\documentclass[12pt]{article}
\usepackage[colorlinks,linkcolor=Cobalt,citecolor=Cobalt,urlcolor=Cobalt,bookmarks,bookmarksnumbered]{hyperref}
\usepackage[scaled=0.85]{helvet}
\usepackage{XoohmE,accents}
\usepackage{graphicx,color}
\usepackage{booktabs,multirow,placeins,array}

\let\4=\boxplus
\def\1{\kern.5pt}
\def\brk#1{\texttt{[\!\![}#1\texttt{]\!\!]}}
\def\cA{\mathcal{A}}
\def\BB{{\boldsymbol B}}
\def\bB{{\bf B}}
\def\sC{\mathscr{C}}
\def\ssC{{\mathscr{C}\mkern-11mu|\mkern5mu}}
\def\BD{{\boldsymbol D}}
\def\BDb{\7{\>\rule[-1pt]{7pt}{.67pt}}{\BD}}
\def\rD{{\rm D}}

\def\ddt{\partial_\tau}
\def\cF{{\cal F}}

\def\BF{{\boldsymbol\F}}

\def\cI{\mathcal{I}}
\def\bJ{{\boldsymbol\J}}

\def\cM{{\cal M}}
\def\sN{\mathscr{N}}
\def\ssH{{\mathscr{H}\mkern-14mu|\mkern8mu}}

\def\cV{{\cal V}}
\def\bX{{\boldsymbol\X}}
\def\BX{{\boldsymbol{\mit\X}}}
\def\fR{\mathfrak{R}}

\def\Spin{\mathop{\textsl{Spin}}}
\def\SU{\mathop{\textsl{SU}}}
\def\spin{\mathop{\textrm{spin}}}

\def\SS#1#2{\mathfrak{Sp}^{\!^{#1|#2}\!}}
\def\htS#1#2#3{\S^{{\sss#1}#2}_{#3}}
\def\pp{{\vphantom{+}\smash{\mathchar'75\mkern-9mu|\mkern5mu}}}
\def\mm{{=\,}}

\def\Dt#1{\accentset{\hbox{\LARGE.}}{#1}}	
\def\ad{{\accentset{\hbox{\large.}}{\a}}}
\def\bd{{\accentset{\hbox{\large.}}{\b}}}
\def\DELB{\textsl{\small DE}}
\def\sDE{\textsl{es\small DE}}
\def\fc#1#2{\relax\ifmmode{\scriptstyle\frac{#1}{#2}} 
                    \else$\scriptstyle\frac{#1}{#2}$\fi}    
\def\fRc#1#2{{}^{\sss#1}\mkern-5mu/\mkern-4mu_{\sss#2}}
\def\vC#1{\vcenter{\hbox{\hss#1\hss}}}
\def\sm#1{\begin{smallmatrix}#1\end{smallmatrix}}
\def\bm#1{\left[\begin{smallmatrix}#1\end{smallmatrix}\right]}
\def\pxC#1{\arraycolsep=1pt\left[\begin{array}{@{\,}c|c@{\,}}#1\end{array}\right]}
\def\pC#1{{\protect\scriptsize\pxC{#1}}}
\def\Bm#1{\begin{bmatrix}#1\end{bmatrix}}

\def\Lx#1{\makebox[0pt][l]{#1}}
\def\Cx#1{\makebox[0pt][c]{#1}}
\def\Rx#1{\makebox[0pt][r]{#1}}
\def\Pgf#1{\paragraph{{\slshape p}\,+\,{\slshape q}\,=\,#1:}}
\let\MC=\multicolumn
\let\MR=\multirow

\definecolor{Red}    {rgb}{1.00,0.00,0.00} 
\definecolor{Green}  {rgb}{0.00,0.75,0.00} 
\definecolor{Blue}   {rgb}{0.00,0.00,1.00} 
\definecolor{Orange} {rgb}{1.00,0.72,0.00} 
\definecolor{Purple} {rgb}{0.50,0.00,0.50} 
\definecolor{Gold}   {rgb}{1.00,0.85,0.00} 
\definecolor{Magenta}{rgb}{1.00,0.00,1.00} 
\definecolor{Turque} {rgb}{0.00,0.88,0.88} 
\definecolor{Seaweed}{rgb}{0.00,0.25,0.00} 
\definecolor{Brown}  {rgb}{0.50,0.13,0.00} 
\definecolor{Cobalt} {rgb}{0.00,0.00,0.50} 
\definecolor{Sage}   {rgb}{0.00,0.50,0.38} 
\definecolor{grey1}  {rgb}{0.20,0.20,0.20} 
\definecolor{grey2}  {rgb}{0.40,0.40,0.40} 
\definecolor{grey3}  {rgb}{0.60,0.60,0.60} 
\definecolor{grey4}  {rgb}{0.80,0.80,0.80} 
\definecolor{grey5}  {rgb}{0.90,0.90,0.90} 
\def\C#1#2{{\ifcase#1\or
             \color{Red}\or\color{Green}\or\color{Blue}\or
              \color{Orange}\or\color{Purple}\or\color{Gold}\or
             \color{Magenta}\or\color{Turque}\or\color{Seaweed}\or
               \color{Brown}\or\color{Cobalt}\or\color{Sage}\or
                 \color{grey1}\or\color{grey2}\or\color{grey3}\or
                 \color{grey4}\else\color{grey5}\fi#2}}
\definecolor{gray}{rgb}{.7,.7,.7}
\def\XXX{\colorbox{yellow}{\color{red}\bf X\kern-4pt{\Large$\bs*$}\kern-4.125ptX}}
\def\cb#1#2{\setlength\fboxsep{.3333pt}\colorbox{#1}{\color{#1}\fbox{\color{black}#2}}}
\def\cB#1{\hbox to0pt{\setlength\fboxsep{0pt}\hss\color{gray}\fbox{\cb{white}{#1}}\hss}}
 \SfTitles 
 \allowdisplaybreaks
\begin{document}
\thispagestyle{empty}
 \begin{center}
{\LARGE\bsf\boldmath
  Weaving Worldsheet Supermultiplets\\[1mm]
  from the Worldlines Within
 }\\*[2mm]
{\bsf T.\,H\"{u}bsch
}\\*
\parbox[t]{80mm}{\small\centering\it
      Department of Physics \&\ Astronomy,\\[-3pt]
      Howard University, Washington, DC 20059\\[-3pt]
     {\small\tt  thubsch@howard.edu}}
\\[5mm]
{\sf\bfseries ABSTRACT}\\[3mm]
\parbox{145mm}{\addtolength{\baselineskip}{-2pt}\parindent=2pc\noindent
Using the fact that every worldsheet is ruled by two (light-cone) copies of worldlines, the recent classification of off-shell supermultiplets of $N$-extended worldline supersymmetry is extended to construct standard off-shell and also unidextrous (on the half-shell) supermultiplets of worldsheet $(p,q)$-supersymmetry with no central extension. In the process, a new class of error-correcting ({\em\/even-split\/} doubly-even linear block) codes is introduced and classified for $p{+}q\leqslant8$, providing a graphical method for classification of such codes and supermultiplets. This also classifies quotients by such codes, of which many are not tensor products of worldline factors. Also, supermultiplets that admit a complex structure are found to be depictable by graphs that have a hallmark twisted reflection symmetry.
}
\end{center}
\vspace{3mm}
\noindent
\parbox[t]{60mm}{PACS: 11.30.Pb, 12.60.Jv}\hfill
\parbox[t]{100mm}{\raggedleft\small\baselineskip=12pt\sl
            And there in the warp and the woof is the proof of it.\\[-0pt]
            |\,Elwyn Brooks White (``Charlotte's Web'')\hphantom{.}}
\ToC{3}{12.6pt}
\section{Introduction, Rationale and Summary}
Supersymmetry has been utilized in physics for about four decades\cite{r1001,rPW,rWB,rBK}, as it stabilizes the vacuum and simplifies renormalization or even eliminates the need for it. Nevertheless, and although the use of off-shell fields is paramount in quantum theories, off-shell formulations of supersymmetric models are still known only for relatively low total number of supercharges, $N$ (counting each real spinor component separately). This situation has remained largely unchanged in the past two decades (as reviewed, e.g., in Ref.\cite{rES-SuGra}), and has been recognized as a major remaining challenge\cite{rGLPR}.

To this end, Refs.\cite{rGR-1,rGR0} proposed:
 ({\small\bf1})~classifying the off-shell worldline supermultiplets (which are technically simpler owing to the simplicity of the Lorentz group $\Spin(1,0)\simeq\ZZ_2$), and then
 ({\small\bf2})~determining which of these extends to higher-dimensional spacetimes and reconstructing the models built from them.
It turns out that worldlines admit myriads\ft{For $N\leqslant32$, there are ${>}\,10^{12}$ inequivalent families of supermultiplets, within each of which the supermultiplets have the same {\em\/chromotopology\/} (see below) but where the number and variety of the component (super)field relative engineering (mass-)dimensions grows hyper-exponentially with $N$\cite{r6-1,r6-3,r6-3.2,r6-1.2,r6-3.1}. Recently provided numerical\cite{rFIL,rFL} and graphical\cite{rGH-obs} criteria demonstrate that a rather small fraction of off-shell worldline supermultiplets extends to $(3{+}1)$- and higher-dimensional spacetime.} of inequivalent {\bsf adinkraic} supermultiplets\cite{rA,r6-1,r6--1,r6-3c,r6-3,r6-3.2,r6-1.2} of $(N\,{\leqslant}\,32)$-extended worldline supersymmetry; see also Refs.\cite{rPT,rT01,rT01a,rCRT,rKRT,rKT07,rGKT10}. In such supermultiplets, each component field is mapped by each supercharge to precisely one other component field or a derivative thereof; graphical depictions of such supermultiplets are Adinkras. The familiar supermultiplets from the standard literature on models in 3+1-dimensional spacetime are either themselves adinkraic or may be built from adinkraic ones\cite{rUMD09-1,rUMD09-2,rGHHS-CLS,rUMD12-3}.

Worldsheet supersymmetry is essential in string theory\cite{rGSW1,rGSW2,rJPS,rBBS}, and is very rich in structure\cite{rHSS,rGSS}. Worldsheet theories include worldlines and worldline-restricted ({\bsf unidextrous}) fields in several inequivalent ways\cite{rUDSS01,rUDSS04,rUDSS06,rUDSS08,rUDSS09,rAD-Uni,rHP1,rHT-UCS}, which provides for exceptional constructions on the worldsheet not possible in spacetimes of any other dimension and which provides for much of the richness and complexity of string theory and its $M$- and $F$-theoretic extensions.
 In addition, extending worldline supersymmetry to a worldsheet is a stepping stone in the realization of the original proposal\cite{rGR-1,rGR0} of studying higher-dimensional supersymmetry via dimensional extension of worldline results.
 Ref.\cite{rGH-obs} provides a simple criterion for extending worldline off-shell supermultiplets to worldsheet supersymmetry, which then suffices for many string theory applications.
 This also significantly enhances the efficiency of the numerical criteria of Ref.\cite{rFIL,rFL} for extending to supersymmetry in higher-dimensional spacetimes.

Complementary to the filtering approach of Ref.\cite{rGH-obs}, the constructive approach presented herein produces for all $p,q\geqslant0$:
\begin{enumerate}\itemsep=-3pt\vspace*{-3mm}\addtolength{\leftskip}{1pc}
 \item off-shell (ambidextrous) supermultiplets of worldsheet $(p,q)$-supersymmetry, and
 \item on the half-shell (unidextrous) supermultiplets of ambidextrous $(p,q)$-super\-symmetry,
\end{enumerate}\vspace{-2mm}
by tensoring a left- and a right-handed copy of worldline supermultiplets,
 and projecting to their quotients by certain discrete symmetries.
 It is gratifying to note that the lists obtained in such complementary ways in fact coincide, at least for the low enough values of $p{+}q$, where comparisons could be made by inspection. A computer-aided mechanized computation is clearly desirable, generalizing the one performed for {\em\/worldline\/} supermultiplets\cite{r6-3,r6-3.2,rRLM-codes}.
 
In many cases, the resulting tensor-product Adinkras exhibit one or more $\ZZ_2$ symmetries, which are encoded by \sDE-codes (see Eqs.\eq{e:DE;E,E} for the definition). These are the {\em\/even-split\/} refinements of the error-correcting (binary) doubly-even linear block (\DELB-)codes of Refs.\cite{r6-3,r6-3.2,rRLM-codes}, being an encryption theory consequence of the filtering condition of Ref.\cite{rGH-obs}. Very much like in the case of worldline supermultiplets\cite{r6-3,r6-3.2}, passing to the quotient of such a $\ZZ_2$ symmetry provides a new, half-sized supermultiplet, and one may do so repeatedly using mutually commuting such $\ZZ_2$ symmetries. A list of these symmetries for $p{+}q\leqslant8$ is depicted in Figure~\ref{f:1} using a graphical method\cite{r6-3c} that may also be used for $p{+}q>8$.

The main results presented herein are:
\begin{enumerate}\itemsep=-3pt\vspace{-2mm}
 \item Constructions~\ref{C:pxq>pq} and~\ref{C:N>N0} for off-shell representations, and Construction~\ref{C:N>Nq} for unidextrous (on the half-shell) representations of worldsheet $(p,q)$-supersymmetry, and their listing for $p{+}q\leqslant8$;
 \item the definition (and a $p{+}q\leqslant8$ listing) of {\em\/even-split (binary) doubly even linear block\/} (\sDE) codes (see Section~\ref{s:esDE}) that encode possible $\ZZ_2$ quotients of tensor product supermultiplets, many of which not themselves tensor products (see Section~\ref{s:FnF});
 \item the definition of a {\em\/twisted\/} $\ZZ_2$ symmetry in Adinkras, which implies a complex structure;
 \item a demonstration that some worldsheet supermultiplets depicted by topologically inequivalent Adinkras are nevertheless equivalent, and by (super)field redefinition only;
 \item a demonstration that the same Adinkra may depict distinct supermultiplets of the same $(p,q)$-supersymmetry, though at least some of them can be shown to be equivalent, and by (super)field redefinition only;
 \item an independent confirmation of the conclusion of Ref.\cite{rGH-obs}, that ambidextrous off-shell supermultiplets of ambidextrous supersymmetry must have at least three levels\cite{rCRT,r6-1}, \ie, their component (super)fields must have at least three distinct, adjacent engineering dimensions.
\end{enumerate}

The paper is organized as follows:
 The remainder of this introduction presents the requisite definitions,
 and Section~\ref{s:WSSm} then presents the three constructions of off-shell and on the half-shell representations of worldsheet $(p,q)$-supersymmetry.
 Section~\ref{s:WSCodes}  discuses the role of \sDE\ error-correcting codes in the proposed framework for classifying off-shell representations of worldsheet supersymmetry; in particular, Section~\ref{s:LowSpC} catalogs the maximal such codes|and thus the minimal such supermultiplets|for $p{+}q\leqslant8$.
 Our conclusions are summed up in Section~\ref{e:coda}, and technically more involved details are deferred to the appendices.

\paragraph{Definitions and Notation:}
We will consider only supersymmetry algebras without central extension, and will construct linear and finite-dimensional off-shell or on the half-shell~(see below) representations of $(1,1|p,q)$-super\-sym\-metry. This is worldsheet $(p,q)$-extended super-Poincar\'e symmetry, generated by $p$ real, left-handed superderivatives\ft{While not strictly necessary to use superdifferential operators to study supersymmetry, we find it simpler to do so, and there is no loss of generality: supersymmetry implies that superspace exists\cite{rHTSSp08}.} $\rD_{\a+}$, $q$ real right-handed ones, $\rD_{\bd-}$, and the light-cone worldsheet derivatives $\vd_\pp$ and $\vd_\mm$. On the worldsheet, the indices $\a$ and $\Dt{\a}$ count ``internal'' (not spacetime) degrees of freedom, which may well stem from a dimensional reduction of a higher-dimensional spacetime symmetry. The defining supercommutators of these algebras are:
\begin{equation}
 \SS{1,1}{p,q}:\quad
 \big\{\,\rD_{\a+}\,,\,\rD_{\b+}\,\big\}=2i\,\d_{\a\b}\,\vd_\pp,\qquad
 \big\{\,\rD_{\ad-}\,,\,\rD_{\bd-}\,\big\}=2i\,\d_{\ad\bd}\,\vd_\mm,
 \label{e:pqSuSy}
\end{equation}
and all other supercommutators vanish. These generators act as first order differential operators on functions (superfields) $\BF,\bJ$, \etc, over $(1,1|p,q)$-superspace.
 The component fields
\begin{equation}
  \f\Defl\BF|,\quad
  \j_{\a+}\Defl i\rD_{\a+}\BF|,\quad
  \j_{\ad-}\Defl i\rD_{\ad-}\BF|,\quad\cdots\quad
  F_{\a\b\pp}\Defl \frc i2[\rD_{\a+},\rD_{\b+}]\BF|,\quad\etc,
 \label{e:CompF}
\end{equation}
are|up to numerical factors chosen for convenience|defined by projecting to the purely bosonic and commutative $(1,1|0,0)$-dimen\-si\-o\-nal worldsheet the
\begin{equation}
  \rD^{\bf a|b}\Defl \rD_{1+}^{~a_1}\wedge\cdots\wedge\rD_{p+}^{~a_p}
                      \rD_{1-}^{~b_1}\wedge\cdots\wedge\rD_{q-}^{~b_q},\qquad
  a_\a,b_\ad\in\{0,1\},
 \label{e:MonD}
\end{equation}
superderivatives of superfields.
In the definitions\eq{e:CompF}, the factor $i^{\brk{\bf a|b}}$ is included to insure that the component fields\eq{e:CompF} projected with the operators\eq{e:MonD} are real. We have
\begin{equation}
  \brk{\bf{a|b}}\Defl{\ttt\binom{{\bf|a|+|b|}+1}2},\qquad
  |{\bf a}|\Defl\sum_{\a=1}^p a_\a,\quad
  |{\bf b}|\Defl\sum_{\ad=1}^q b_\ad,
 \label{e:brk}
\end{equation}
where $|{\bf a}|{+}|{\bf b}|$ is the {\em\/Hamming weight\/}\cite{rCHVP} of the split binary number ${\bf a|b}$ with binary digits $a_1,\dots,a_p|b_1,\dots,b_q$.

 Being abelian, the worldsheet Lorentz symmetry $\Spin(1,1)\simeq\IR^\times$ (the multiplicative group of nonzero real numbers, \ie, the non-compact cousin of $U(1)$) has only 1-dimensional irreducible representations, upon which it acts by a multiplicative number\cite{rWyb,rFRH}. Eigenvalues of the only Lorentz generator will be called {\bsf spin} for simplicity\ft{The only generator of $\Spin(1,1)$ actually being a {\em\/boost\/} operation, this is a misnomer. However, this can cause no ambiguity since $\Spin(1,1)$ has no rotations with which to possibly confuse it.}. For example,
\begin{equation}
 \spin(\rD_{\a-})=+\inv2=-\spin(\rD_{\ad-}),\qquad
 \spin(\vd_\pp)=+1=-\spin(\vd_\mm),
 \label{e:spin}
\end{equation}
and we use the ``$\pm$'' subscripts to count this quantity in units of $\pm\frac12\hbar$; superscripts count oppositely.
 In addition to spin, all objects also have an {\bsf engineering dimension}, such as
\begin{equation}
  [\rD_{\a+}] = \inv2 = [\rD_{\ad-}],\qquad
  [\vd_\pp] = 1 = [\vd_\mm].
 \label{e:mass}
\end{equation}
These two functions, (\ref{e:spin}) and\eq{e:mass}, make the supersymmetry algebra\eq{e:pqSuSy} doubly $\ZZ$-graded, and all supermultiplets of interest are then finite-dimensional unitary representations of this bi-graded superalgebra.

 A superfield is {\bsf off-shell} if it is subject to no worldsheet differential equation (one involving $\vd_\pp$ and/or $\vd_\mm$, but neither $\rD_{\a+}$ nor $\rD_{\ad-}$).
 If it is subject to only {\bsf unidextrous} worldsheet differential equations\cite{rUDSS01,rHSS} (involving {\em\/either\/} $\vd_\pp$ {\em\/or\/} $\vd_\mm$ but not both),
 it is said to be {\bsf on the half-shell}\cite{rHP1}; such superfields are not off-shell on the worldsheet in the standard field-theoretic sense, but {\em\/are\/} off-shell on a unidextrously embedded worldline and provide for features not describable otherwise\cite{rHT-UCS}. A superfield, operator, expression, equation or other construct thereof will be called {\bsf ambidextrous} to emphasize that is not unidextrous. Following Ref.\cite{rUDSS01}, the $(p,0$)- and $(0,q)$-supersymmetries will continue to be called {\bsf unidextrous}. However, one must keep in mind that the absence of $\rD_{\ad-}$-superderivatives in $(p,0)$-supersymmetry results in the absence of $\vd_\mm$-generated unidextrous worldsheet constraints; the parity-mirror analogue holds for $(0,q)$-supersymmetry.

\section{Worldsheet Supermultiplets}
\label{s:WSSm}
To highlight the complexity of the classification of off-shell supermultiplets, we recall the comparatively much simpler study of {\em\/multiplets\/} of (global and local) symmetries in particle physics.

For any Lie group $G$, a $G$-multiplet is a collection of component fields which span a representation of $G$, \ie, within which the $G$-action closes. That is, each component field within the multiplet is transformed by any element of $G$ into a linear combination of componenet fields within the multiplet. For example, a general element of the color $\SU(3)_c$ symmetry group transforms any particular quark of any particular color into a linear combination of all three colors of the same quark. At any point in spacetime, the component fields in a multiplet thus span a vector space, which is a representation of the structure group: the red, blue and yellow version of a given quark form a basis for the 3-dimensional vector space of the $\SU(3)_c$ representation that particle physicists denote as ``{\bf3}''. This vector space then varies over spacetime, forming a vector bundle.

All Lie groups are products of factors that are either {\em\/simple\/} or are copies of the abelian group $U(1)$. All simple Lie groups have an infinite sequence of irreducible unitary finite-dimensional representations, but all of which can be constructed from only one or maybe two ``fundamental'' representations by means of the so-called {\em\/Weyl construction\/}\cite{rWyb,rJSH}, by:
 ({\small\bf1})~(internal) tensor product,
 ({\small\bf2})~``symmetrization'' in various ways\ft{More properly, this refers to projection on irreducible representations of the permutation group acting on the factors in the tensor product $V\otimes\cdots\otimes V$ of the fundamental representation $V$ with itself.}, and
 ({\small\bf3})~subtraction of ``traces'', \ie, contraction with invariant tensors specific to the given simple Lie group.
 These classification theorems rely on:
 ({\small\bf1})~the existence of a ``Cartan+ladder generator'' basis where the Cartan generators $H_i$ unambiguously identify the ladder generators $E_\a$ through the non-degenerate action $[H_i,E_\a]=\a_i E_\a$, and
 ({\small\bf2})~the existence of the positive-definite Killing metric of the given simple Lie algebra, $g_{ab}\Defl-f_{ac}{}^df_{bd}{}^c$.

However, the Killing metric defined from the structure constants of any supersymmetry algebra (without and also with central and other extensions, in any spacetime dimension and signature) tends to be degenerate and in fact vanishes completely for\eq{e:pqSuSy}: The action of the Cartan generators, $\vd_\pp,\vd_\mm$, on all supercharges and superderivatives is maximally degenerate|all commutators vanish.
 For the study of off-shell representations of supersymmetry, this obstructs both the standard Lie-algebraic methods and its ensuing standard and familiar classification theorems.

 In turn, we shall see that
 a fundamental result in Lie group representation theory---that a representation of a tensor product of two Lie groups is always a tensor product of representations of the respective factor groups\cite{rWyb,rFRH,rJSH}---does not hold for worldsheet supersymmetry\eq{e:pqSuSy}.

\subsection{Weaving Worldsheets from Worldlines Within}
\label{s:WWW}
The defining relations\eq{e:pqSuSy}|with all other (anti)commutators understood to vanish|clearly indicate that the worldsheet $(p,q)$-supersymmetry algebra is actually a direct sum of the left- and the right-handed parts
\begin{equation}
  \SS{1,1}{p,q} = \SS1p_+\oplus\SS1q_-,\qquad
   \bigg\{\begin{array}{rll}
           \SS1p_+&\Defl\SS{1,0}{p,0}&\ni\{\rD_{\a+},\vd_\pp\},\\[1mm]
           \SS1q_-&\Defl\SS{0,1}{0,q}&\ni\{\rD_{\ad-},\vd_\mm\},
          \end{array}
 \label{e:Sp11pq}
\end{equation}
where both $\SS1p_+$ and $\SS1q_-$ are isomorphic, respectively, to a {\em\/worldline\/} $p$- and $q$-extended supersymmetry algebra without central charges:
\begin{equation}
  \SS1N\ni\{\rD_I,\ddt\}:\quad
 \big\{\,\rD_I\,,\,\rD_J\,\big\}=2i\,\d_{IJ}\,\ddt.
 \label{e:SuSyN}
\end{equation}
Therefore, {\em\/all\/} representations of $\SS1N$ are also representations of $\SS1p_+$ and of $\SS1q_-$, and their (external) tensor product is a representation of $\SS{1,1}{p,q}=\SS1p_+\oplus\SS1q_-$; this reflects the ``bi-filtration'' of Ref.\cite{r6--1} and is expected from Lie group representation theory\cite{rWyb,rFRH,rJSH}. Akin to the situation with worldline supermultiplets,\cite{r6-3,r6-3.2}, such a representation may well have a symmetry that commutes with supersymmetry, allowing to construct the quotient supermultiplet:
\begin{cons}[off-shell]\label{C:pxq>pq}
Let\/ $\fR_+$ and $\,\fR_-$ denote off-shell representations of two copies of the (centrally unextended) worldline supersymmetry algebras, $\SS1p_+$ and\/ $\SS1q_-$ respectively, and let $Z$ be a symmetry of $\fR_+\,{\otimes}\,\fR_-$, covariant with supersymmetry\eq{e:pqSuSy} and including the trivial case, $Z=\Ione$.
 The $Z$-quotient of the tensor product\ft{Elements of $\fR_+\,{\otimes}\,\fR_-$ are worldsheet supermultiplets that transform as the respective factors under the separate action of the two summands in $\SS{1,1}{p,q}=\SS1p_+\oplus\SS1q_-$, but need not themselves factorize.} $(\fR_+\,{\otimes}\,\fR_-)/Z$ is then an off-shell representation of\/ $\SS{1,1}{p,q}=\SS1p_+\oplus\SS1q_-$, but when $Z\neq\Ione$ need not itself be a tensor product.
\end{cons}
When the $Z$-action involves both the $\rD_{\a+}$ and the $\rD_{\ad-}$, the $Z=(\ZZ_2)^k$ actions of Refs.\cite{r6-3.1} are specified by a refinement on the encryption codes of Ref.\cite{r6-3.1}, and the quotient $(\fR_+\,{\otimes}\,\fR_-)/Z$ is not a real tensor product. Unless otherwise stated, all representations and operations considered herein are real. At times---but not always---$(\fR_+\,{\otimes}\,\fR_-)/Z$ does turn out to be a (hyper-)complex tensor product of (hyper-)complex representations; see Section~\ref{s:WSCodes} for the details.

Construction~\ref{C:pxq>pq} is somewhat analogous to the familiar Weyl construction of Lie algebra representations\cite{rWyb,rFRH,rJSH} but exhibits important differences:
\begin{equation}
  \begin{tabular}{@{} p{80mm}|p{80mm} @{}}
   \bsf Weyl's construction\cite{rWyb,rFRH,rJSH}
 & \bsf Construction~\ref{C:pxq>pq}, as given here\\ 
    \toprule
   (internal) tensor product of representations\newline
   of the same Lie algebra
 & (external) tensor product of representations\newline
   of the left- and right-handed parts of\eq{e:pqSuSy}\\ 
    \midrule
   Young symmetrization: projection to\newline variously symmetrized and traceless parts
 & projection to quotients by \sDE-encoded\newline discrete symmetries\\
    \bottomrule
  \end{tabular}\vspace{1mm}
 \label{e:WvsI}
\end{equation}
 A physicist familiar with the Standard Model will find the results of Construction~\ref{C:pxq>pq} akin to, say, the quark doublet $(u_{\sss L},d_{\sss L})$, which represents the tensor product of the irreducible representations: the {\bf3} of color $SU(3)_c$ and the {\bf2} of weak $SU(2)_{\sss L}$, and the $(\fRc12,0)$ representation of the Lorentz group, $\Spin(1,3)\simeq\textsl{SL}(2,\IC)$.

\paragraph{Unidextrous Special Cases:}
By construction, $\fR_+$ is $\SS1p_-$-invariant: $(\rD_{\ad-}\fR_+)=0=(\vd_\mm\fR_+)$, and $\fR_-$ is $\SS1p_+$-invariant: $(\rD_{\a+}\fR_-)=0=(\vd_\pp\fR_-)$. There are then two interesting special cases of Construction~\ref{C:pxq>pq}:
\begin{cons}[unidextrous representations, on the half-shell]\label{C:N>Nq}
Selecting\/ $\fR_-\mapsto\Ione_-$ ($\SS1p_-$-constant), the tensor product representation of Construction~\ref{C:pxq>pq} becomes a {\bsf unidextrous} representations (on the half-shell) of the (centrally unextended) worldsheet ambidextrous supersymmetry $\SS{1,1}{N,q}$ for arbitrary $q>0$.
{\em Mutatis mutandis} for the parity mirror-image, $\SS1N\mapsto\SS{1,1}{p,N}$.
\end{cons}
In other words, by identifying\/ $(\rD_I,\ddt)\mapsto(\rD_{\a+}\vd_\pp)$,
 all off-shell representations of (centrally unextended) worldline $N$-extended supersymmetry $\SS1N$ automatically extend to (centrally unextended, left-moving) unidextrous representations of  (centrally unextended) ambidextrous worldsheet $(N,q)$-supersymmetry $\SS{1,1}{N,q}$ for {\em\/arbitrary\/} $q>0$:
 $\rD_{\ad-}(\fR_+\otimes\Ione_-)=0=\vd_\mm(\fR_+\otimes\Ione_-)$: such representations are constant in the right-moving $(\t{-}\s)$ light-cone direction on the worldsheet.

Such representations are not off-shell on the worldsheet in the standard field-theoretic sense, but are off-shell on a continuum of worldlines within the worldsheet: they are {\bsf on the half-shell}\cite{rHP1}.
\begin{corl}\label{c:NqUDmodels}
Every off-shell model with (centrally unextended) $N$-extended worldline supersymmetry automatically defines an $(N,q)$-supersymmetric worldsheet model on the half-shell, for arbitrary $q>0$.
 {\em Mutatis mutandis\/} for the parity mirror-image.
\end{corl}
In the special case of Construction~\ref{C:N>Nq} when $q=0$ ($p=0$), there are no $\rD_{\ad-}$'s (no $\rD_{\a+}$'s), and unidextrous annihilation by $\vd_\mm$ (by $\vd_\pp$) is not implied:
\begin{cons}[unidextrous supersymmetry]\label{C:N>N0}
By identifying\/ $(\rD_I,\ddt)\mapsto(\rD_{\a+},\vd_\pp)$,
all off-shell representations of (centrally unextended) worldline $N$-extended supersymmetry $\SS1N$ automatically extend to fully off-shell representations of (centrally unextended) unidextrous worldsheet $(N,0)$-super\-sym\-metry $\SS{1,1}{N,0}$. 
{\em Mutatis mutandis\/} for $\SS1N\mapsto\SS{1,1}{0,N}$.
\end{cons}
\begin{corl}\label{c:N0models}
Every off-shell model with (centrally unextended) $N$-extended worldline supersymmetry automatically defines an off-shell, (centrally unextended) unidextrous $(N,0)$-super\-sym\-met\-ric worldsheet model, as well as its unidextrous $(0,N)$-supersymmetric parity mirror-image.
\end{corl}

In turn, Construction~\ref{C:N>N0} may also be regarded as a prerequisite to Construction~\ref{C:N>Nq}:
\begin{corl}\label{c:N0>Nq}
Every off-shell supermultiplet of (centrally unextended) unidextrous $(N,0)$-super\-sym\-metry given by Construction~\ref{C:N>N0} extends to a worldsheet {\em\/unidextrous\/} supermultiplet on the half-shell of the (centrally unextended) {\em\/ambidextrous\/} $(N,q)$-supersymmetry, and for arbitrary $q>0$.
 {\em Mutatis mutandis\/} for the parity mirror-image.
\end{corl}

Note the key difference:
\begin{itemize}\itemsep=-3pt\vspace{-2mm}
 \item Construction~\ref{C:N>Nq} produces unidextrous representations $\bs\L\sim(\fR_+\,{\otimes}\,\Ione_-)$ of the ambidextrous worldsheet $(N,q)$-supersymmetry;
 such $\bs\L$ are necessarily on the half-shell, $\vd_\mm\bs\L=0$.
 \item Construction~\ref{C:N>N0} produces off-shell representations $\bs{A}\sim\fR_+$ of the unidextrous worldsheet $(N,0)$-supersymmetry: $\bs{A}$ need satisfy no particular worldsheet differential equation for the supersymmetry algebra\eq{e:pqSuSy} to close on it.
\end{itemize}

The products of Construction~\ref{C:N>N0} are representations only of unidextrous $(p,0)$- and $(0,q)$-supersymmetry, and so cannot be mixed with the products of Constructions~\ref{C:pxq>pq} and~\ref{C:N>Nq} that are designed for ambidextrous $(p,q)$-supersymmetry. In turn, worldsheet models with ambidextrous supersymmetry, constructed with a mix of results from Constructions~\ref{C:pxq>pq} and~\ref{C:N>Nq}, indeed exist: Refs.\cite{rHSS,rGSS,rHP1,rHT-UCS} discuss $(2,2)$-supersymmetric models that involve both off-shell ambidextrous representations (the familiar chiral, twisted-chiral superfields and their conjugates) and unidextrous representations (leftons and rightons) on the half-shell, and produce unique resulting target spaces.

Foreshadowing subsequent results, a few such supermultiplets are presented in Table~\ref{t:1}.
\begin{table}[ht]
$$
  \begin{array}{@{} rcc|ccl @{}}
 \text{\bsf name}
 & (\fR_+\otimes\fR_-)/\ssC & \MC2c{\text{\bsf Adinkra}} & (\fR_+\otimes\fR_-)/\ssC
 &\text{\bsf name} \\ 
    \toprule
 &&\MC2c{\text{\bsf off-shell}}\\
 \parbox{25mm}{\raggedleft\baselineskip=12pt\bsf chiral\\ \& herm.\ conj.}
 & \big(\vC{\begin{picture}(23,10)
             \put(1,.5){\includegraphics[trim=2mm 0mm 520mm 20mm,clip,width=24mm]{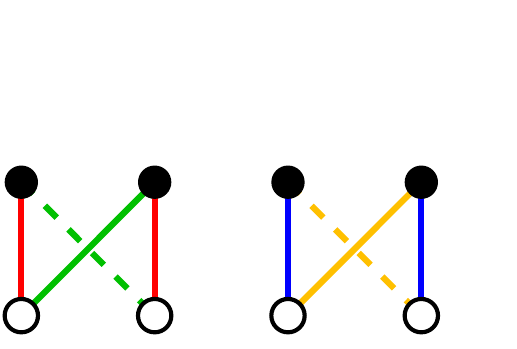}}
             \put(10,4){\small$\otimes$}
             \put(9,0){$\SSS+$}
             \put(21,0){$\SSS-$}
            \end{picture}}\big)\big/d^+_{2,2}
 & \vC{\includegraphics[height=12mm]{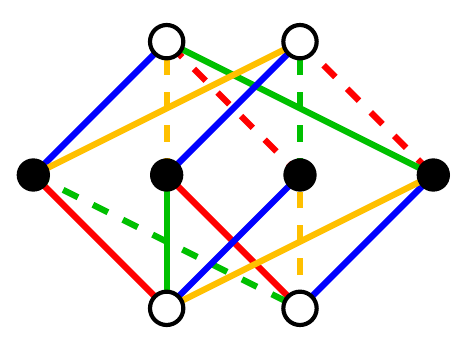}}
 & \vC{\includegraphics[height=12mm]{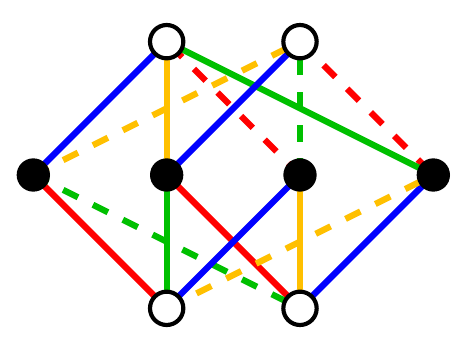}}
 & \big(\vC{\begin{picture}(23,10)
             \put(1,.5){\includegraphics[trim=2mm 0mm 520mm 20mm,clip,width=24mm]{22x22.pdf}}
             \put(10,4){\small$\otimes$}
             \put(9,0){$\SSS+$}
             \put(21,0){$\SSS-$}
            \end{picture}}\big)\big/d^-_{2,2}
 & \parbox{25mm}{\raggedright\baselineskip=12pt\bsf twisted chiral\\ \& herm.\ conj.} \\ 
    \midrule
 &&\MC2c{\text{\bsf on the half-shell}}\\
 \parbox{25mm}{\raggedleft\baselineskip=12pt{\bsf lefton}\\ $\in\ker[\vd_\mm]$}
 & \big(\vC{\includegraphics[height=7mm]{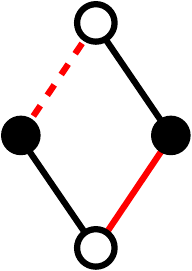}}_+\otimes\Ione_-\big)
 & \vC{\includegraphics[height=10mm]{N2T.pdf}} 
 & \vC{\includegraphics[height=10mm]{N2T.pdf}}
 & \big(\Ione_+\otimes\vC{\includegraphics[height=7mm]{N2T.pdf}}_-\big)
 & \parbox{25mm}{\raggedright\baselineskip=12pt{\bsf righton}\\
                  $\in\ker[\vd_\pp]$}\\ 
    \bottomrule
  \end{array}
$$
\caption{Some off-shell and on the half-shell supermultiplets of worldsheet $(2,2)$-supersymmetry. The chiral and twisted chiral supermultiplets admit a complex structure as they are; the lefton and righton supermultiplets may be complexified, thus doubling the number of their degrees of freedom.}
\label{t:1}
\end{table}

\subsection{Some Learning Examples}
\label{s:XMples}
For illustrative purposes, we restrict herein the otherwise general Constructions~\ref{C:pxq>pq}--\ref{C:N>N0} to using only adinkraic representations of (centrally undextended) worldline $N$-extended supersymmetry defined and explored in Refs.\cite{rA,r6-1,r6-3,r6-3.2,r6-1.2,r6-2,rUMD09-1}, which are easily depicted by Adinkras.

\paragraph{Adinkras and Worldline Supermultiplets:}
{\em Adinkraic} supermultiplets admit a basis of supersymmetry generators and component (super)fields $(\f_A|\j_B)$, such that each supersymmetry generator maps each component (super)field to precisely one other component (super)field or its (space)time derivative. With $n$ bosons $\f_A$ and $n$ fermions $\j_B$, this is a $(n|n)$-dimensional representation.

By contrast, in {\em\/non-adinkraic\/} supermultiplets the action of at least one supercharge on at least one component (super)field is bound to produce a linear combination of other component (super)fields and their derivatives|and there exists no (super)field redefinition that would turn the supermultiplet adinkraic. Examples of non-adinkraic {\em\/worldline\/} supermultiplets have been discussed in Ref.\cite{r6-3,rGHHS-CLS,rTHGK12,rGIKT12,rTHGK13}.
 In spacetime supersymmetry, nontrivial Lorentz covariance prevents many of the linear combinations of component (super)fields within a supermultiplet. While this tends to obstruct the non-adinkraic constructions {\em\/\`a la\/} Ref.\cite{rTHGK12,rGIKT12}, it also tends to obstruct compensating (super)field redefinitions. This leaves open the logical possibility that adinkraic supermultiplets do not exhaust the space of finite-dimensional unitary representations of spacetime supersymmetry.
 
 It is thus noteworthy that Constructions~\ref{C:pxq>pq}--\ref{C:N>N0} and Corollaries~\ref{c:NqUDmodels}--\ref{c:N0>Nq} apply to {\em\/all\/} representations, adinkraic or not. For now however, we focus on adinkraic supermultiplets.
 
\begin{table}[ht]
  \centering
  \begin{tabular}{@{} cc|cc @{}}
    \makebox[15mm]{\bsf Adinkra} & \makebox[40mm]{\bsf Supersymmetry Action} 
  & \makebox[15mm]{\bsf Adinkra} & \makebox[40mm]{\bsf Supersymmetry Action} \\ 
    \hline
    \begin{picture}(5,9)(0,5)
     \put(0,0){\includegraphics[height=11mm]{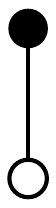}}
     \put(3,0){\scriptsize$A$}
     \put(3,9){\scriptsize$B$}
     \put(-1,4){\scriptsize$I$}
    \end{picture}\vrule depth4mm width0mm
     & $\rD_I\begin{bmatrix}\bJ_B\\\BF_A\end{bmatrix}
           =\begin{bmatrix}\Dt\BF_A\\i\bJ_B\end{bmatrix}$
  & \begin{picture}(5,9)(0,5)
     \put(0,0){\includegraphics[height=11mm]{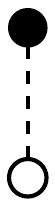}}
     \put(3,0){\scriptsize$A$}
     \put(3,9){\scriptsize$B$}
     \put(-1,4){\scriptsize$I$}
    \end{picture}\vrule depth4mm width0mm
     & $\rD_I\begin{bmatrix}\bJ_B\\\BF_A\end{bmatrix}
           =\begin{bmatrix}-\Dt\BF_A\\-i\bJ_B\end{bmatrix}$ \\[5mm]
    \hline
    \begin{picture}(5,9)(0,5)
     \put(0,0){\includegraphics[height=11mm]{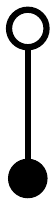}}
     \put(3,0){\scriptsize$B$}
     \put(3,9){\scriptsize$A$}
     \put(-1,4){\scriptsize$I$}
    \end{picture}\vrule depth4mm width0mm
     &  $\rD_I\begin{bmatrix}\BF_A\\\bJ_B\end{bmatrix}
           =\begin{bmatrix}i\dot\bJ_B\\\BF_A\end{bmatrix}$
  & \begin{picture}(5,9)(0,5)
     \put(0,0){\includegraphics[height=11mm]{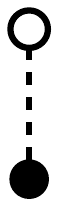}}
     \put(3,0){\scriptsize$B$}
     \put(3,9){\scriptsize$A$}
     \put(-1,4){\scriptsize$I$}
    \end{picture}\vrule depth4mm width0mm
     &  $\rD_I\begin{bmatrix}\BF_A\\\bJ_B\end{bmatrix}
           =\begin{bmatrix}-i\Dt\bJ_B\\-\BF_A\end{bmatrix}$ \\[5mm]
    \hline
  \multicolumn{4}{l}{\vrule height3.0ex width0pt\parbox{120mm}{\small\baselineskip12pt
   The edges are here labeled by the variable index $I$; for fixed $I$,
   they are drawn in the $I^{\text{th}}$ color.}}
  \end{tabular}
  \caption{Adinkras depict supermultiplets\eq{e:SM=SF} by assigning:
    (white/black) vertices\,$\iff$\,(boson/fermion) component (super)fields;
    edge color/index\,$\iff$\,$\rD_I$;
    solid/dashed edge\,$\iff$\,$c=\pm1$;
    nodes are placed at heights equal to the engineering dimension of the depicted component (super)field, determining $\l$ in Eqs.\eq{e:SM=SF}.}
  \label{t:A}
\end{table}
As done in\cite{rFGH}, we introduce a collection of otherwise {\bsf intact} (that is, unconstrained, ungauged, unprojected\dots) component superfields {\em\/\`a la\/} Salam and Strathdee\cite{rSSSS4}, and pair the supersymmetry transformations with superderivative constraint equations\ft{\label{fn:DQ}The pairing\eq{e:SM=SF} derives from the superspace relation $Q_I=i\rD_I+2\d_{IJ}\q^J\ddt$ between supercharges $Q_I$ and superderivatives, and the fact that if the $\rD_I$ act from the left then the $Q_I$ act from the right. It the follows that $\{Q_I,\rD_J\}=0$, so that mappings defined by means of $\rD$'s are manifestly supersymmetric.}
\begin{equation}
 \left. \begin{array}{r@{\>}l}
          \rD_I \,\BF_A &= ic\,(\IL_I)_A{}^B\,(\ddt^{1-\l}\bJ_B)\\
          \rD_I \,\bJ_B &= c\,(\IL^{-1}_I)_B{}^A\,(\ddt^\l\BF_A)
        \end{array}\right\}
 \quad\Iff\quad
 \left\{\begin{array}{r@{\>}lr@{\>}l}
          Q_I\,\f_A &=-c\,(\IL_I)_A{}^B\,(\ddt^{1-\l}\j_B),\quad
           & \f_A   &\Defl\BF_A|,\\
          Q_I\,\j_B &=-ic\,(\IL^{-1}_I)_B{}^A\,(\ddt^\l\f_A),\quad
           & \j_B &\Defl\bJ_B|,
        \end{array}\right.
 \label{e:SM=SF}
\end{equation}
where the exponent $\l=0,1$ depends on $I,A,B$, and the matrices $\IL_I$ have exactly one entry, $\pm1$, in every row and in every column. This type of (adinkraic) supersymmetry action is then depicted using the ``dictionary'' provided in Table~\ref{t:A}.
 For example,
\begin{subequations}\label{e:121}
 \begin{align}
 \rD_1\,\BF     &=i\,\bs\J_1,    & \C1{\rD_2}\,\BF   &=i\,\bJ_2,\label{Q.f1}\\*
 \rD_1\,\bJ_1   &=\Dt\BF,        & \C1{\rD_2}\,\bJ_1 &=-{\bf F},\label{Q.j1}\\*
 \rD_1\,\bJ_2   &={\bf F},       & \C1{\rD_2}\,\bJ_2 &=\Dt\BF,\label{Q.j2}\\*
 \rD_1\,{\bf F} &=i\Dt\bJ_2,
  \begin{picture}(30,0)(-12,0)
  \put(10,-2){\includegraphics[height=25mm]{N2T.pdf}}
  \put(22,-1){\small$\bs\F$}
  \put(4,10){\small$\bs\J_1$}
  \put(29,10){\small$\bs\J_2$}
  \put(22,20){\small$\bf F$}
 \end{picture}\quad
                                 & \C1{\rD_2}\,{\bf F} &=-i\,\Dt\bJ_1,\label{Q.f2}
 \end{align}
\end{subequations}
and
\begin{subequations}\label{e:22}
 \begin{align}
 \rD_1\,\bB_1 &=i\,\bX_1,     & \C1{\rD_2}\,\bB_1 &=i\,\bX_2,\label{Qf1}\\*
 \rD_1\,\bB_2 &=i\,\bX_2,     & \C1{\rD_2}\,\bB_2 &=-i\,\bX_1,\label{Qf2}\\*
 \rD_1\,\bX_1 &=\Dt\bB_1,     & \C1{\rD_2}\,\BX_1 &=-\Dt\bB_2,\label{Qj1}\\*
 \rD_1\,\bX_2 &=\Dt\bB_2,
  \begin{picture}(30,0)(-12,0)
  \put(10,0){\includegraphics[height=20mm]{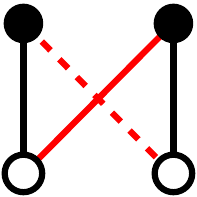}}
  \put(5,0){\small$\bB_1$}
  \put(5,18){\small$\bX_1$}
  \put(31,0){\small$\bB_2$}
  \put(31,18){\small$\bX_2$}
 \end{picture}\quad
                              & \C1{\rD_2}\,\bX_2&=\Dt\bB_1,\label{Qj2}
 \end{align}
\end{subequations}
define two clearly distinct {\em\/worldline\/} $N\,{=}\,2$ supermultiplets.

Given the comparative brevity and ease of comprehension, supersymmetry transformation rules such as\eqs{e:121}{e:22} will subsequently be depicted by Adinkras rather than written out explicitly, except for occasional examples to reinforce this relationship. This also permits identifying Adinkras with the supermultiplets that they depict, which is a faithful 1--1 correspondence except for a well-defined subclass where multiple Adinkras depict isomorphic supermultiplets: see the sections~\ref{s:1}--\ref{s:2} for the worldsheet extension of the worldline characterization of Refs.\cite{r6-3,r6-3.2}.

\paragraph{Adinkras and Worldsheet Supermultiplets:}
Adinkras such as\eq{e:121} and\eq{e:22} may well also depict worldsheet supermultiplets. To this end, the edge-colors must now be partitioned into those that depict the action of $p$ $\rD_{\a+}$'s (which square to $i\vd_\pp$) and those that depict the action of $q$ $\rD_{\ad-}$'s (which square to $i\vd_\mm$).
 As shown in\eq{e:CompF}, component fields themselves acquire spin, and the necessary and sufficient condition for an Adinkra to depict a worldsheet supermultiplet\cite{rGH-obs} insures that all component fields can be assigned a spin consistently with the $\rD_{\a+}$- and $\rD_{\ad-}$-action throughout the supermultiplet/Adinkra. 
\begin{defn}
An Adinkra together with the additional choices (partitioning of edge-colors into $p$ left- and $q$ right-moving and consistent assignment of spin) that make it depict a supermultiplet of $(p,q)$-supersymmetry is thus called a {\bsf{\upshape(}\1\text{\bsf p}\1,\1\text{\bsf q}\1{\upshape)}-Adinkra}.
\end{defn}
The Adinkras presented herein will not be further complicated by annotating the edges to signify their left/right-handed partitioning, nor will nodes be annotated to signify spin; this permits multiple duty for most of the illustrations herein.

\paragraph{Complex Structures:}
In the superdifferential systems\eqs{e:121}{e:22}, all superfields $\BF,\bJ_i,{\bf F},\bB_i,\bX_i$ may be chosen real, as seen by writing the superderivative action in terms of supercommutators:
\begin{subequations}
 \label{e:Cpx1}
\begin{align}
  (\rD_j\BF)\Defl[\rD_j,\BF],\quad&\To\quad
  (i\bJ_j)^\dag=[\rD_j,\BF]^\dag=[\BF^\dag,\rD_j^\dag]=-[\rD_j,\BF]
               =-i\bJ_j,\\
  (\rD_1\bJ_2)\Defl\{\rD_1,\bJ_2\},\quad&\To\quad
  ({\bf F})^\dag=\{\rD_1,\bJ_2\}^\dag=\{\bJ_2^\dag,\rD_1^\dag\}
   =+\{\rD_1,\bJ_2\}={\bf F},\quad\etc
\end{align}
\end{subequations}
However, note that the Adinkra\eq{e:22} exhibits a {\em\/twisted\/} horizontal $\ZZ_2$ symmetry: by simultaneously swapping $\bB_1\iff\bB_2$ and $\bX_1\iff\bX_2$, the $\rD_1$ (black edges) action is preserved, but the \C1{$\rD_2$}-action (\C1{red} edges) flips the overall sign, depicted by swapping of the solid/dashed parity of the corresponding edges. This may be seen to depict a pair of complex structures by defining
\begin{subequations}
 \label{e:Cpx}
\begin{equation}
  \BD\Defl\fc1{\sqrt2}[\rD_1+\cI\,\C1{\rD_2}],\qquad
  \BB\Defl\fc1{\sqrt2}(\bB_1+\cI\,\bB_2),\quad \text{and} \quad
  \BX\Defl\fc1{\sqrt2}(\bX_1+\cI\,\bX_2),
\end{equation}
with\ft{The two choices of the complex structure $\cI=\pm i$ only have a relative distinction.} $\cI=\pm i$, so that the left-hand half of the Adinkra\eq{e:22} plays the role of the real part, the right-hand side the imaginary part of the new, complex component (super)fields; {\em\/also\/}, the edges entirely within the left- or right-hand side play the role of the real part, and the edges criss-crossing from one to the other side play the role of the imaginary part of the complex supersymmetry transformation:
\begin{equation}
 \vC{\begin{picture}(75,18)(0,-2)
   \put(.5,0){\includegraphics[height=15mm]{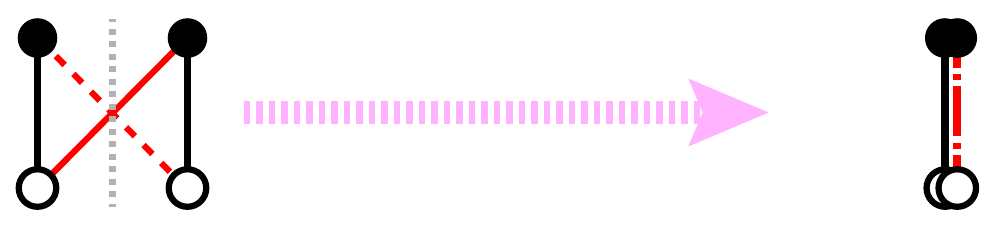}}
    \put(-3,0){\footnotesize$\bB_1$}
    \put(-3,13){\footnotesize$\bX_1$}
    \put(14.5,0){\footnotesize$\bB_2$}
    \put(14.5,13){\footnotesize$\bX_2$}
    \put(-10,7){\Rx{\parbox[c]{25mm}{\footnotesize\raggedleft\baselineskip=9pt
                   a real Adinkra with a twisted left-right $\ZZ_2$ symmetry}}}
    \put(56,-2){\footnotesize$(\bB_1+\cI\,\bB_2)$}
    \put(56,15){\footnotesize$(\bX_1+\cI\,\bX_2)$}
    \put(80,7){\Lx{\parbox[c]{17mm}{\footnotesize\raggedright\baselineskip=9pt
                   a complex rendition of the same Adinkra}}}
    \put(-2,6.5){\footnotesize$\rD_1$}
    \put(6,11){\footnotesize\C1{$\rD_2$}}
    \put(54.5,6.5){\footnotesize$(\rD_1+$}
    \put(66,6.5){\footnotesize$\C1{\cI\,\rD_2})$}
 \end{picture}}
 \label{e:CpxA}
\end{equation}
 With this, we compute
\begin{equation}
  \BD\BB=i\,\BX,\quad
  \BDb\BX=\skew3\Dt{\BB},\qquad\text{and}\qquad
  \BD\BX=0,\quad
  \BDb\BB=0.
\end{equation}
In fact, owing to the very last of these results, the supermultiplet $(\BB;\BX)=\big(\BB;({-}i\BD\BB)\big)$ may be considered the worldline $N=2$ antichiral supermultiplet. Combining these, the $N=2$ supersymmetry algebra\eq{e:SuSyN}
\begin{equation}
  \{\BD,\BD\}=0=\{\BDb,\BDb\},\quad\text{and}\quad\{\BD,\BDb\}=2i\ddt
 \label{e:CpxSuSy}
\end{equation}
\end{subequations}
is satisfied on $(\BB;\BX)$.
\begin{corl}[complex structure]\label{C:Cpx}
An Adinkra admits a conjugate pair of complex structures if it has a rendition that exhibits a {\bsf\slshape twisted horizontal \,$\PMB{\ZZ_2}$ symmetry}, where:
\begin{enumerate}\itemsep=-3pt\vspace{-3mm}\addtolength{\leftskip}{1pc}
 \item the intended `real (imaginary) part' nodes are in the left-hand (right-hand) half,
 \item the left-hand side half is identical to the right-hand half,
 \item edges criss-crossing between the halves come in solid/dashed parity-reversed pairs.
\end{enumerate}
\end{corl}
See also Section~\ref{s:Cpx}; also, the Appendix~A of Ref.\cite{rGH-obs} details a rather more involved example.

\paragraph{Tensor Product Adinkras:}
The tensor product of Adinkras refines the standard tensor product of graphs\cite{rR+T-Combi} by accounting for the fact that Adinkra nodes are bi-partitioned into bosons and fermions, drawn at a height determined by the engineering dimension, and that edges are either solid or dashed in such a way that every 2-colored quadrangle has an odd number of dashed edges. {\bsf\boldmath Tensor product ({\itshape p,q}\,)-Adinkras} (Adinkras with $p$ of the edge-colors depicting $\rD_{\a+}$-action and $q$ of them $\rD_{\ad-}$-action) are constructed as follows:
\goodbreak
\begin{cons}[tensor product (p,q)-Adinkras]\label{C:RxR}$~$
\begin{enumerate}\itemsep=-3pt\vspace{-3mm}\setcounter{enumi}{-1}
 \addtolength{\leftskip}{1pc}\addtolength{\rightskip}{1pc plus2pc}

 \item Given two Adinkras $\cA_+$ and $\cA_-$ depicting two adinkraic worldline supermultiplets, $\cA_+$ will depict the $\rD_{\a+}$-action, and $\cA_-$ the $\rD_{\ad-}$-action in $\cA_+\,{\otimes}\,\cA_-$. Each vertex in $\cA_\pm$ is drawn at the height proportional to the engineering dimension of the corresponding component field; each component field also has a definite spin.

 \item Draw a copy of $\cA_+$ in the place of every node of $\cA_-$, but flip the boson/fermion (node) and solid/dashed (edge) parity in the copies of $\cA_+$ that replace  fermionic nodes of $\cA_-$; as convenient, exaggerate the size of $\cA_-$.

 \item For every edge $E$ in $\cA_-$, redraw a copy of $E$ to connect like nodes in the  copies of $\cA_+$ that replaced the $E$-connected $\cA_-$-nodes.

 \item Revert any temporary size exaggeration from step~1 by repositioning the resulting nodes to their proper height, so all edges extend precisely one level up/down. In particular, the function of spin\eq{e:spin} is additive: the spin of a product is the sum of spins of the factors, and so is the function of engineering dimension\eq{e:mass}.
\end{enumerate}
\end{cons}
The reason for flipping the boson/fermion and solid/dashed parity as described in Step~1 is simple: Bosons correspond to the identity element of the $\ZZ_2\subset\Spin(1,1)$ Lorentz group, whereas fermions correspond to its nontrivial $(-1)$ element. Since edges represent the action of supersymmetry, between bosons and fermions, they also correspond to the nontrivial $(-1)$ element of $\ZZ_2\subset\Spin(1,1)$. The tensor product of a black (fermionic) node with an entire Adinkra thus necessarily flips the association with the $+1/-1\in\ZZ_2\subset\Spin(1,1)$ elements in that Adinkra.

To illustrate this, we now turn to construct the Adinkras depicting ambidextrous off-shell and unidextrous (on the half-shell) $(4{-}q,q)$-supermultiplets in this manner, for $q=0,1,2$.

\subsubsection{The Building Blocks}
\label{s:BB}
Tables~6 and~7 of Ref.\cite{r6-3.2} list 28 $N=4$ worldline Adinkras, without distinguishing dashed edges for simplicity and to save space.

Of these 28 Adinkras, 24 have 8 white and 8 black nodes, depicting supermultiplets with corresponding 8 bosonic and 8 fermionic component (super)fields. The edges in all of these 24 Adinkras form a tesseract (4-cube) and have only one equivalence class of edge-dashing: any choice of solid/dashed edge parity may be obtained from any other one by judicious sign-changes in component (super)fields and horizontal rearrangement of nodes that is inconsequential to the supermultiplets depicted. These Adinkras differ from each other solely by various height-positioning of the nodes, \ie, engineering dimensions of the various component (super)fields.
 To save space, these 24 Adinkras are not listed herein, and the Reader is referred to tables~6 and~7 of Ref.\cite{r6-3.2}.

The remaining four $N=4$ Adinkras are ``half-sized'' and each admits a twisted version:
\begin{equation}
 \vC{\begin{picture}(150,48)
   \put(0,0){\includegraphics[width=150mm]{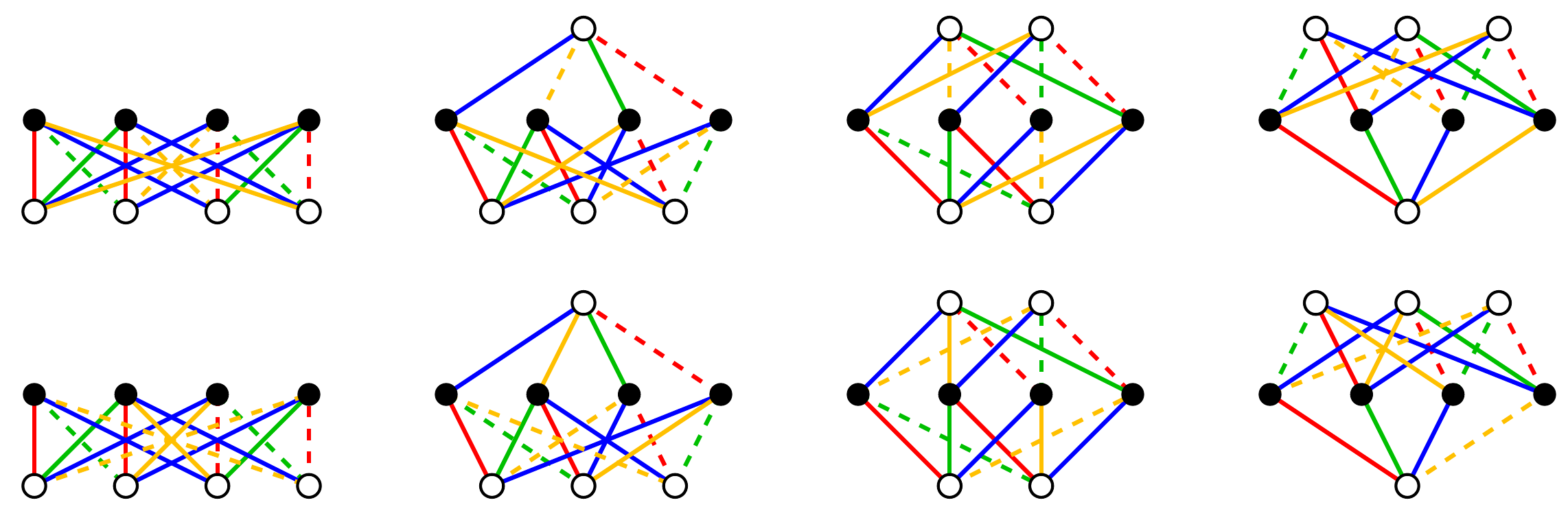}}
    \put(0,24){\color{Seaweed}\rule{150mm}{.4pt}}
    \put(2,20){\small twisted versions:}
 \end{picture}}\quad
 \label{e:N4half}
\end{equation}
where an Adinkra differs from its twisted variant in the solid/dashed parity in edges of an odd number of colors|here the orange-colored ones.
 Together with the 24 described in the previous paragraph, these eight inequivalent $N=4$ Adinkras add up to 32; together with their boson/fermion flips, the 28 $N=4$ Adinkras in tables~6 and~7 of Ref.\cite{r6-3.2} therefore represent 64 inequivalent $N=4$ Adinkras, all of which depict inequivalent off-shell supermultiplets of $N=4$ worldline supersymmetry.

In addition, we may make use of the $N=3$ Adinkras:
\begin{equation}
 \vC{\begin{picture}(140,22)(0,2)
   \put(0,0){\includegraphics[width=140mm]{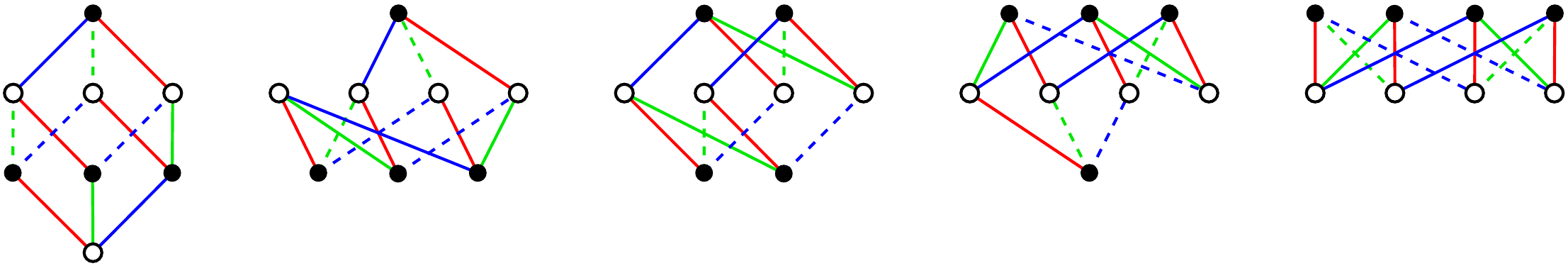}}
 \end{picture}}
 \label{e:N3list}
\end{equation}
their boson/fermion flips, as well as the $N=2$ Adinkras\eqs{e:121}{e:22}, their boson/fermion flips and the $N=1$ Adinkras in Table~\ref{t:A}.

\subsubsection{Tensor Product (4,0)-Adinkras}
\label{s:(4,0)}
By Construction~\ref{C:N>N0}, the 64 inequivalent $N=4$ worldline off-shell supermultiplets are interpretable as 64 inequivalent $(4,0)$- and $(0,4)$-Adinkras, depicting 64 inequivalent off-shell supermultiplets of unidextrous worldsheet $(4,0)$- and $(0,4)$-supersymmetry, respectively.

The size of these Adinkras|$8+8$ nodes in 24 of the $N=4$ Adinkras \textit{vs}.\ $4+4$ nodes in the remaining 8, lined up in\eq{e:N4half}|correlates with the following quality:
 The 8 Adinkras\eq{e:N4half} and their boson/fermion flips are all ``2-color decomposable'' in that it takes deleting all edges of any {\em\/two\/} colors for the Adinkra to decompose into disjoint Adinkras of lower supersymmetry. By contrast, the other 48 $N=4$ Adinkras are all ``1-color decomposable'': they decompose into two disjoint Adinkras of lower supersymmetry upon deleting the edges of any one color.

Below, we will see example Adinkras in which this $n$-color decomposability is not as uniform over the edge-colors. However, this quality is correlated with the fact that Adinkras that exhibit a higher $n$-color decomposability (corresponding to supermultiplets of smaller size) may be obtained from Adinkras of lesser $n$-color decomposability (corresponding to supermultiplets of larger size) by projection with respect to certain $\ZZ_2$ symmetries; these will be explored in Section~\ref{s:WSCodes}.

\subsubsection{Tensor Product (3,1)-Adinkras}
\label{s:(3,1)}
\begin{figure}[htp]
 \begin{center}
  \unitlength=1.067mm
     \begin{picture}(140,47)(1,7)
      \put(-5,.5){\includegraphics[width=160mm]{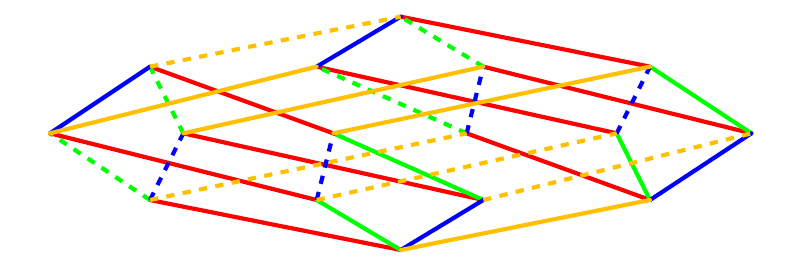}}
      \put(70,47.5){\cB{$\inv4\{[\rD_{1+},\rD_{2+}],\rD_{3+}\}\rD_-$}}
      \put(22,37.5){\cB{$\inv4\{[\rD_{1+},\rD_{2+}],\rD_{3+}\}$}}
      \put(54,37.5){\cB{$\inv2[\rD_{1+},\rD_{2+}]\rD_-$}}
      \put(86.5,37.5){\cB{$\inv2[\rD_{1+},\rD_{3+}]\rD_-$}}
      \put(118,37.5){\cB{$\inv2[\rD_{2+},\rD_{3+}]\rD_-$}}
      \put(4.75,25){\cB{$\frc i2[\rD_{1+},\rD_{2+}]$}}
      \put(31,24.5){\cB{$\frc{i}2[\rD_{1+},\rD_{3+}]$}}
      \put(57.5,24.5){\cB{$\frc{i}2[\rD_{2+},\rD_{3+}]$}}
      \put(83.5,24.5){\cB{$i\rD_{1+}\rD_-$}}
      \put(110,24.5){\cB{$i\rD_{2+}\rD_-$}}
      \put(135.75,25){\cB{$i\rD_{3+}\rD_-$}}
      \put(22,12){\cB{$i\rD_{1+}$}}
      \put(53.5,12){\cB{$i\rD_{2+}$}}
      \put(86.5,12){\cB{$i\rD_{3+}$}}
      \put(118,12){\cB{$i\rD_-$}}
      \put(70,3){\cB{$\Ione$}}
     \end{picture}
 \end{center}
 \caption{The tesseract of superderivative operators used in projecting component fields of worldsheet $(3,1)$-superfields. Edges are associated with the superderivatives: $\C1{\rD_{1+}\iff\text{red}}$, $\C2{\rD_{2+}\iff\text{green}}$, $\C3{\rD_{3+}\iff\text{blue}}$, $\C4{\rD_{-}\iff\text{orange}}$; see Table~\ref{t:A} for more details.}
 \label{f:31Ds}
\end{figure}
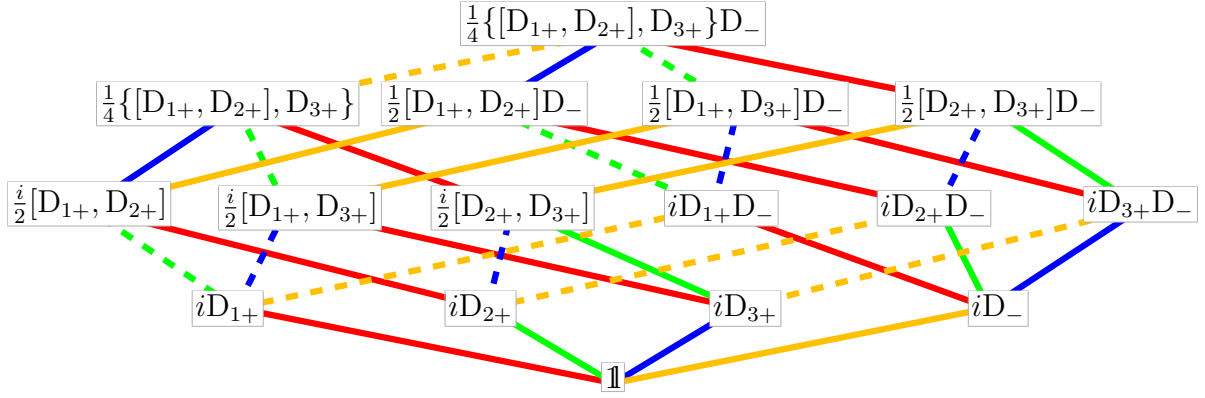

The non-trivial aspects of Construction~\ref{C:RxR} are illustrated by constructing $(3,1)$-Adinkras. We begin with
\begin{equation}
 \vC{\begin{picture}(150,37)(0,-2)
      \put(0,5){\includegraphics[width=150mm]{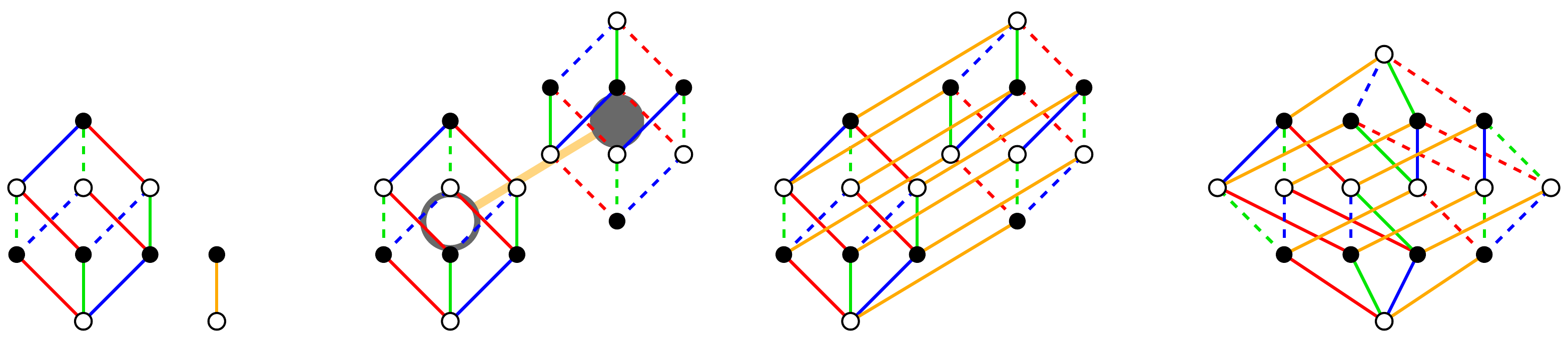}}
       \put(6,28){\small$\cA_+$}
       \put(19,16){\small$\cA_-$}
       \put(10,0){\small Step~0}
       \put(48,0){\small Step~1}
       \put(88,0){\small Step~2}
       \put(127,0){\small Step~3}
     \end{picture}}
 \label{e:11x1331}
\end{equation}
where we have temporarily exaggerated the size of $\cA_-$ in Step~1, retained the relative alignment of the nodes between Steps~1 and~2, arranging them finally at their proper heights in Step~3.
 Up to flipping the sign of the three right-hand side component (super)fields in the middle row and the top-most one\ft{Flipping the sign of a component (super)field depicted by the node $n$ also flips the solid/dashed parity assignment of each edge incident to $n$; edges connecting two sign-flipped nodes remain unchanged.}, the nodes in the Adinkra\eq{e:11x1331} depict the tesseract of superderivatives used to project component fields\cite{rHSS,r6-1}, shown in Figure~\ref{f:31Ds}.

The topology of the resulting Adinkra\eq{e:11x1331} and the one in Figure~\ref{f:31Ds} is by construction a tesseract, \ie, a 4-cube, being the tensor product of a usual cube and an interval\eq{e:11x1331}. The topology of an Adinkra together with a fixed edge-color assignments is called a chromotopology\cite{r6-3}; an Adinkra additionally exhibits the solid/dashed parity of the edges and the height arrangement of the nodes. In addition, to represent worldsheet $(p,q)$-supermultiplets, the collection of edges in a $(p,q)$-Adinkra is also split into those corresponding to the $\rD_{\a+}$ {\em\/vs.\/} those corresponding to the $\rD_{\ad-}$.

By virtue of the evident isomorphism between the Adinkra\eq{e:11x1331} and the one in Figure~\ref{f:31Ds}, the resulting $(3,1)$-Adinkra\eq{e:11x1331} is easily seen to depict the supermultiplet also represented by the intact $(3,1)$-superfield with component fields computed in the manner of\eq{e:CompF}. This same Adinkra also turns up in the list of Section~\ref{s:(4,0)}, the difference being that there all edges correspond to either $\rD_{\a+}$-action for $(4,0)$-supersymmetry or to $\rD_{\ad-}$-action in $(0,4)$-supersymmetry; here, the edges of all but one (orange) color correspond to $\rD_{\a+}$-action and edges of the fourth (orange) color correspond to $\rD_{\ad-}$-action.

The remaining $(3,1)$-Adinkras obtained as tensor products of $N{=}3$ and $N{=}1$ Adinkras are:\begin{equation}
 \vC{\begin{picture}(160,27)
      \put(0,5){\includegraphics[width=150mm]{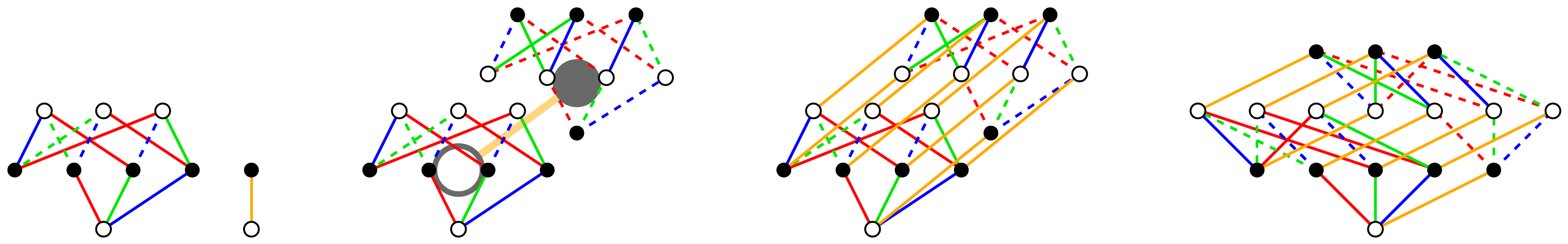}}
       \put(8,20){\small$\cA_+$}
       \put(21,14){\small$\cA_-$}
       \put(9,0){\small Step~0}
       \put(46,0){\small Step~1}
       \put(84,0){\small Step~2}
       \put(127,0){\small Step~3}
     \end{picture}}
 \label{e:11x143}
\end{equation}
\begin{equation}
 \vC{\begin{picture}(160,27)
      \put(0,5){\includegraphics[width=150mm]{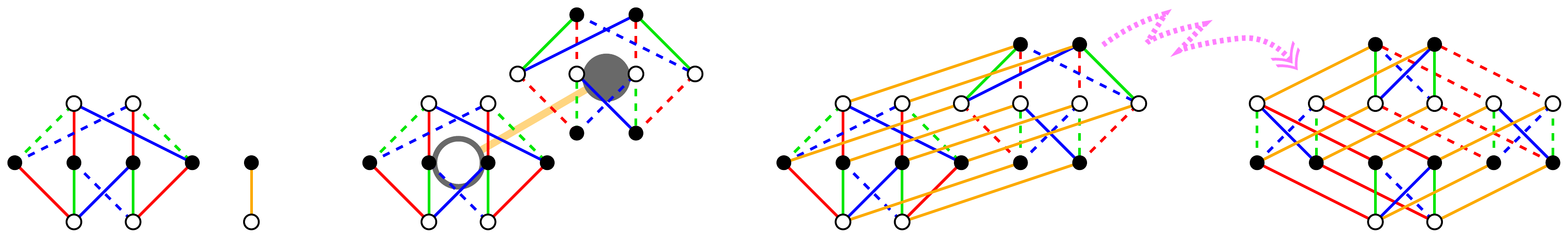}}
       \put(8,20){\small$\cA_+$}
       \put(21,14){\small$\cA_-$}
       \put(9,0){\small Step~0}
       \put(46,0){\small Step~1}
       \put(86,0){\small Step~2}
       \put(130,0){\small Step~3}
     \end{picture}}
 \label{e:11x242}
\end{equation}
where the zig-zagging arrow denotes some horizontal node rearrangements (see Section~\ref{s:Cpx}),
\begin{equation}
 \vC{\begin{picture}(160,30)(0,-1)
      \put(0,5){\includegraphics[width=150mm]{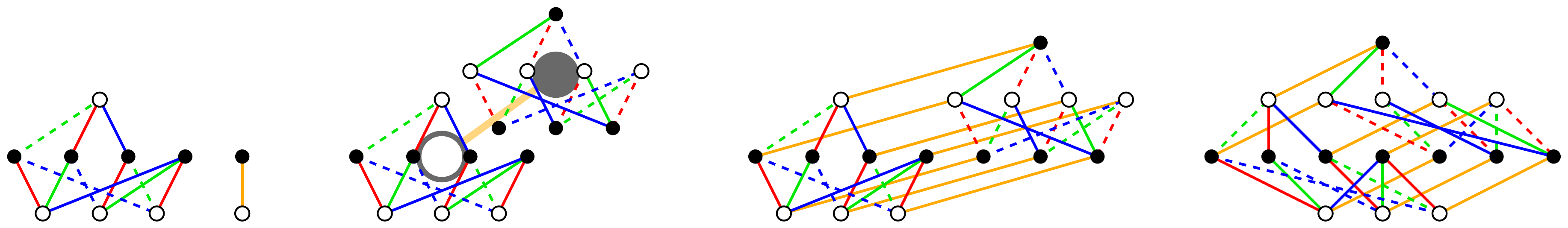}}
       \put(8,20){\small$\cA_+$}
       \put(21,14){\small$\cA_-$}
       \put(9,0){\small Step~0}
       \put(46,0){\small Step~1}
       \put(84,0){\small Step~2}
       \put(127,0){\small Step~3}
     \end{picture}}
 \label{e:11x341}
\end{equation}
and
\begin{equation}
 \vC{\begin{picture}(160,22)(0,-1)
      \put(0,5){\includegraphics[width=150mm]{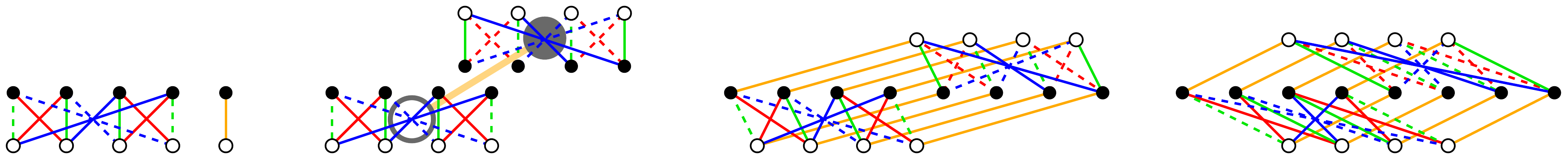}}
       \put(8,14){\small$\cA_+$}
       \put(21,14){\small$\cA_-$}
       \put(9,0){\small Step~0}
       \put(46,0){\small Step~1}
       \put(84,0){\small Step~2}
       \put(127,0){\small Step~3}
     \end{picture}}
 \label{e:11x44}
\end{equation}
We notice that the upside-down boson/fermion-flipped rendition of\eq{e:11x143} is the same as\eq{e:11x341} upon horizontal reshuffling of the nodes and a judicious sign-change in a couple of component (super)fields, \ie, nodes. In a simpler sense, the upside-down renditions of\eq{e:11x1331} and\eq{e:11x44} are equivalent to the originals, and the upside-down rendition of\eq{e:11x242} is equivalent to the boson/fermion flip of the original.

Thus, the five Adinkras\eq{e:11x1331}, \eqs{e:11x143}{e:11x44} and their boson/fermion flips represent ten inequivalent $(3,1)$-Adinkras, and depict ten corresponding, inequivalent off-shell supermultiplets of worldsheet $(3,1)$-supersymmetry. Swapping the roles of $\{\rD_{\a+},\vd_\pp\}$ and $\{\rD_{\ad-},\vd_\mm\}$, each $(3,1)$-Adinkra may be reinterpreted as a $(1,3)$-Adinkra, resulting in the depiction of ten inequivalent off-shell $(1,3)$-supermultiplets.

\subsubsection{Tensor Product (2,2)-Adinkras}
\label{s:(2,2)}
Construction~\ref{C:RxR} is illustrated also by considering the product
\begin{equation}
 \vC{\begin{picture}(160,45)(0,-9)
   \put(0,0){\includegraphics[width=160mm]{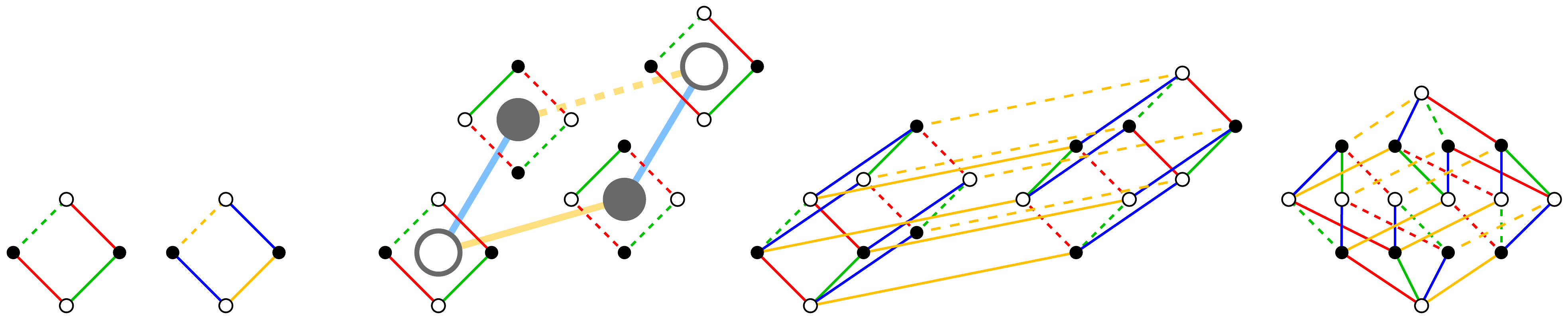}}
    \put(4,6){$\cA_+$}
    \put(20.5,6){$\cA_-$}
    \put(10,-5){Step~0}
    \put(48,-5){Step~1}
    \put(92,-5){Step~2}
    \put(138,-5){Step~3}
 \end{picture}}
 \label{e:121x121}
\end{equation}
The resulting Adinkra\eq{e:121x121} is easily seen to be equivalent to\eq{e:11x1331} by changing the sign of the component (super)fields corresponding to the 2$\text{nd}$, $3^\text{rd}$ and $6^\text{th}$ middle-level node from the left, as well as the top-most node. Its reinterpretation from depicting an off-shell supermultiplet of $(3,1)$-supersymmetry to depicting an off-shell supermultiplet of $(2,2)$-supersymmetry owes to the reassignment of the blue and orange edges from $\rD_{3+}$- and $\rD_-$-action, respectively, in\eq{e:11x1331} to $\rD_{1-}$- and $\rD_{2-}$-action in\eq{e:121x121}.

The (2,2)-Adinkra\eq{e:121x121} thus  (also) depicts the intact off-shell supermultiplet of worldsheet $(2,2)$-supersymmetry:
\begin{equation}
 \vC{\unitlength=1.067mm
     \begin{picture}(140,38)(1,1)
      \put(-5,.5){\includegraphics[height=43mm,width=160mm]{Spindle.pdf}}
      \put(70,37){\cB{$\,\PMB{\cF}\,$}}
      \put(22,30){\cB{$\bX^\mm_{1-}$}}
      \put(54,30){\cB{$\bX^\mm_{2-}$}}
      \put(86.5,30){\cB{$\bX^\pp_{1+}$}}
      \put(118,30){\cB{$\bX^\pp_{2+}$}}
      \put(4.75,20){\cB{${\bf F}_\pp$}}
      \put(31,20){\cB{${\bf F}_{11}$}}
      \put(57.5,20){\cB{${\bf F}_{12}$}}
      \put(83.5,20){\cB{${\bf F}_{21}$}}
      \put(110,20){\cB{${\bf F}_{22}$}}
      \put(135.75,20){\cB{${\bf F}_\mm$}}
      \put(22,10){\cB{$\bJ_{1+}$}}
      \put(53.5,10){\cB{$\bJ_{2+}$}}
      \put(86.5,10){\cB{$\bJ_{1-}$}}
      \put(118,10){\cB{$\bJ_{2-}$}}
      \put(70,2){\cB{$\,\BF\,$}}
     \end{picture}}
 \label{e:22U}
\end{equation}
and also represented by the intact $(2,2)$-superfield with component (super)fields projected {\em\/\`a la\/}\eq{e:CompF}, by means of the tesseract of superderivatives displayed in Figure~\ref{f:22Ds}.
\begin{figure}[t]
 \begin{center}
  \unitlength=1.067mm
     \begin{picture}(140,47)(1,5)
      \put(-5,.5){\includegraphics[width=160mm]{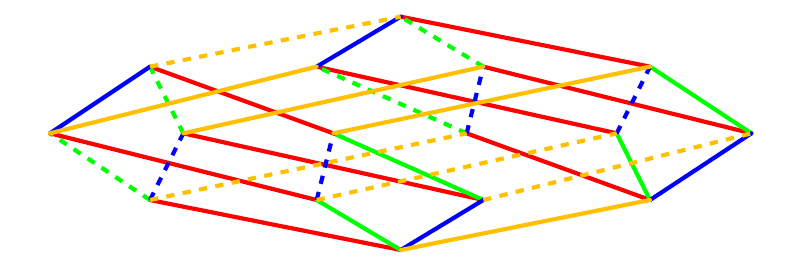}}
      \put(70,47.5){\cB{$\inv4[\rD_{1+},\rD_{2+}][\rD_{1-},\rD_{2-}]$}}
      \put(22,37.5){\cB{$\inv2[\rD_{1+},\rD_{2+}]\rD_{1-}$}}
      \put(54,37.5){\cB{$\inv2[\rD_{1+},\rD_{2+}]\rD_{2-}$}}
      \put(86.5,37.5){\cB{$\inv2[\rD_{1-},\rD_{2-}]\rD_{1+}$}}
      \put(118,37.5){\cB{$\inv2[\rD_{1-},\rD_{2-}]\rD_{2+}$}}
      \put(4.75,25){\cB{$\frc{i}2[\rD_{1+},\rD_{2+}]$}}
      \put(31,24.5){\cB{$i\rD_{1+}\rD_{1-}$}}
      \put(57.5,24.5){\cB{$i\rD_{2+}\rD_{1-}$}}
      \put(83.5,24.5){\cB{$i\rD_{1+}\rD_{2-}$}}
      \put(110,24.5){\cB{$i\rD_{2+}\rD_{2-}$}}
      \put(135.75,25){\cB{$\frc{i}2[\rD_{1-},\rD_{2-}]$}}
      \put(22,12){\cB{$i\rD_{1+}$}}
      \put(53.5,12){\cB{$i\rD_{2+}$}}
      \put(86.5,12){\cB{$i\rD_{1-}$}}
      \put(118,12){\cB{$i\rD_{2-}$}}
      \put(70,3){\cB{$\Ione$}}
     \end{picture}
\end{center}
 \caption{The tesseract of superderivative operators used in projecting component (super)fields of worldsheet $(2,2)$-superfields. Edges are associated with the superderivatives: $\C1{\rD_{1+}\iff\text{red}}$, $\C2{\rD_{2+}\iff\text{green}}$, $\C3{\rD_{1-}\iff\text{blue}}$, $\C4{\rD_{2-}\iff\text{orange}}$; see Table~\ref{t:A} for more details.}
 \label{f:22Ds}
\end{figure}
In addition to\eq{e:121x121}, Construction~\ref{C:pxq>pq} also yields:
\begin{equation}
 \vC{\begin{picture}(160,35)(0,-7)
   \put(0,0){\includegraphics[width=160mm]{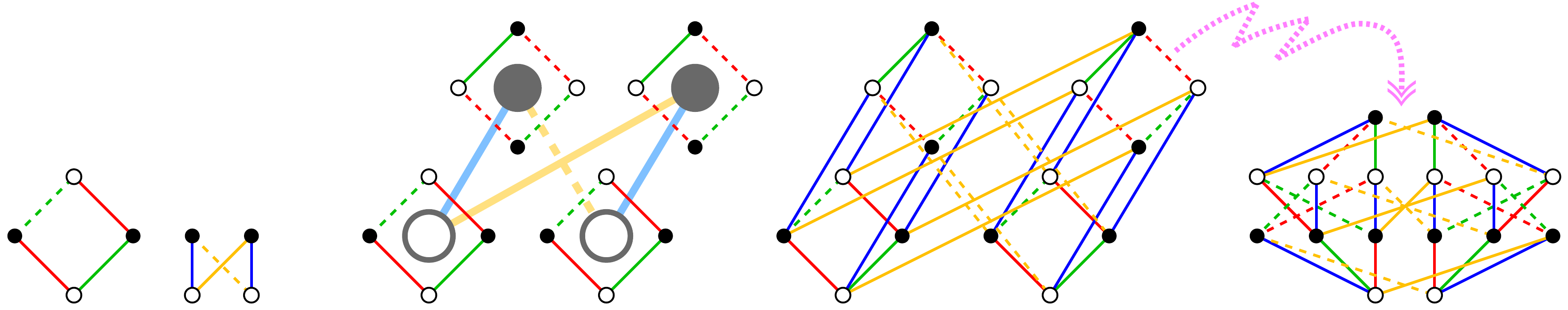}}
    \put(4.5,7){$\cA_+$}
    \put(20,11){$\cA_-$}
    \put(8,-5){Step~0}
    \put(48,-5){Step~1}
    \put(90,-5){Step~2}
    \put(138,-5){Step~3}
 \end{picture}}
 \label{e:121x22}
\end{equation}
where the zig-zagging arrow indicates additional horizontal rearrangement of nodes; see Section~\ref{s:Cpx}. Next, we have
\begin{equation}
 \vC{\begin{picture}(160,34)(0,-7)
   \put(0,0){\includegraphics[width=160mm]{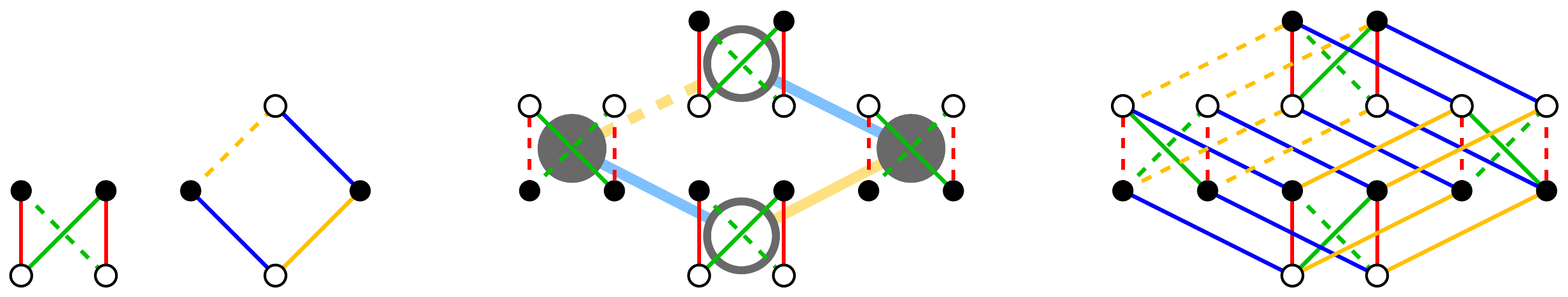}}
    \put(4.5,14){$\cA_+$}
    \put(25,10){$\cA_-$}
    \put(15,-5){Step~0}
    \put(70,-5){Step~1}
    \put(131,-5){Step~2}
 \end{picture}}
 \label{e:22x121}
\end{equation}
where Step~3 was not necessary in this third example.
Note that the $(2,2)$-Adinkra\eq{e:22x121} has the same number of nodes at the same heights as does the $(2,2)$-Adinkra\eq{e:121x22} and they depict isomorphic {\em\/worldline\/} supermultiplets. This may be seen by swapping the edge-colors corresponding to the $\cA_+\iff\cA_-$ swap, horizontally reshuffling the nodes and changing the signs of four of the white nodes in the second row from above, which swaps the solid/dashed parity of the edges incident to those nodes. However, the {\em\/worldsheet\/} supermultiplets depicted by the $(2,2)$-Adinkras\eq{e:121x22} and\eq{e:22x121} are inequivalent: they are each other's $\rD_{\a+}\iff\rD_{\ad-}$ mirror images, {\em\/via\/} the $\cA_+\iff\cA_-$ swap. Alternatively, one may say that the equivalent worldline supermultiplets depicted by the Adinkras\eq{e:121x22} and\eq{e:22x121} extend to inequivalent worldsheet $(2,2)$-supermultiplets.
Finally, we also have
\begin{equation}
 \vC{\begin{picture}(160,30)(0,-8)
   \put(0,0){\includegraphics[width=160mm]{22x22.pdf}}
    \put(3,11){$\cA_+$}
    \put(14,11){$\cA_-$}
    \put(5,-5){Step~0}
    \put(35,-5){Step~1}
    \put(80,-5){Step~2}
    \put(130,-5){Step~3}
 \end{picture}}
 \label{e:22x22}
\end{equation}

The chromotopology of the Adinkras\eqs{e:121x121}{e:22x22} is the same, the 4-cube; the differences between them lie in
 ({\small\bf1})~the height assignments of the nodes|the engineering dimensions of the corresponding component (super)fields,
 ({\small\bf2})~the left/right splitting of the edges between\eq{e:121x22} and\eq{e:22x121}, and
 ({\small\bf3})~some sign-redefinitions of some of the nodes, \ie, component (super)fields.

\subsection{Additional Structures}
As mentioned above, two similar features in Adinkras are of special interest:
\begin{enumerate}\itemsep=-3pt\vspace{-2mm}
 \item A $\ZZ_2$ symmetry, which affords projecting to a $\ZZ_2$ quotient.
 \item A twisted $\ZZ_2$ symmetry, which indicates the admission of a complex structure on the depicted, {\em\/a priori\/} real supermultiplet.

\end{enumerate}

\subsubsection{$\protect\pmb{\protect\ZZ}_2$-Symmetry and Projection}
The Adinkras\eq{e:22x22} and\eq{e:11x44} depict both different worldline supermultiplets and different worldsheet supermultiplets. In fact, and unlike\eq{e:11x44}, the Adinkra\eq{e:22x22} exhibits a $\ZZ_2$ symmetry which is made evident as follows\cite{rGH-obs}:
First, we rearrange the nodes in\eq{e:22x22} horizontally,
\begin{equation}
 \vC{\includegraphics[width=160mm]{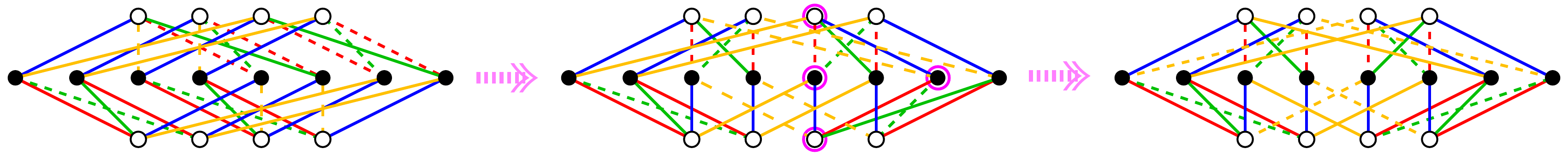}}
 \label{e:484}
\end{equation}
then flip the sign of the component (super)fields represented by the encircled four nodes, in the $(2,2)$-Adinkra obtained in the middle of\eq{e:484}. Of these $(2,2)$-Adinkras, the resulting one makes its $\ZZ_2$ symmetry manifest as a perfect horizontal mirror symmetry, so that its right-hand half may be identified|node-by-node and edge-by-edge|with its left-hand half:
\begin{equation}
 \vC{\begin{picture}(120,28)
   \put(0,0){\includegraphics[height=30mm]{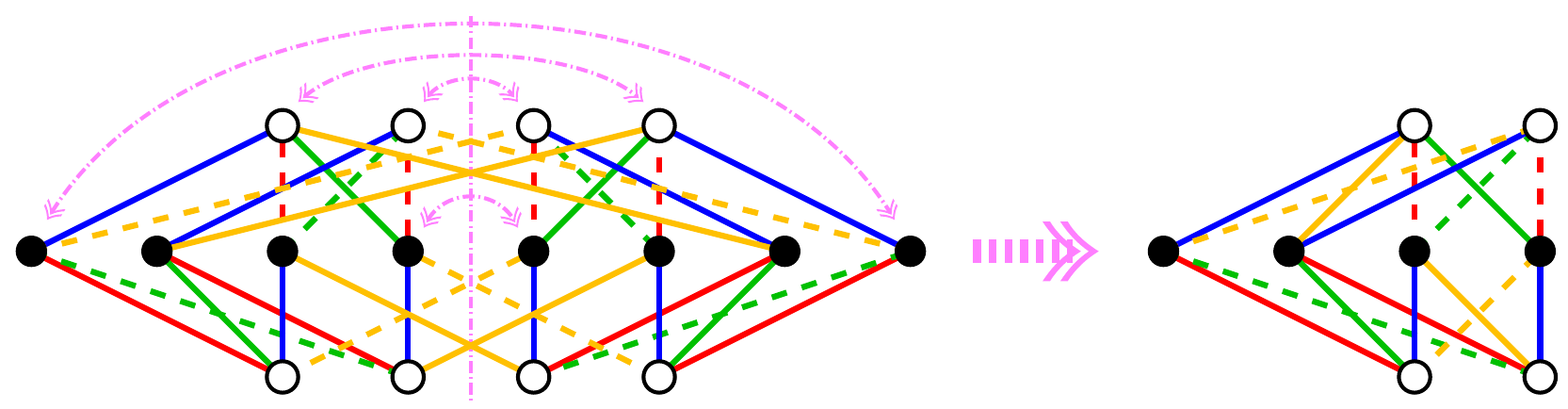}}
 \end{picture}}
 \label{e:B484>242}
\end{equation}
resulting in a half-sized $(2,2)$-Adinkra. By identifying instead the {\em\/negative\/} of each right-hand side node with its corresponding left-hand side node, the orange (left-right crisscrossing) edges flip their solid/dashed parity, and we obtain the twisted version of the half-sized $(2,2)$-Adinkra:
  \begin{equation}
 \vC{\begin{picture}(100,30)
      \put(0,0){\includegraphics[height=30mm]{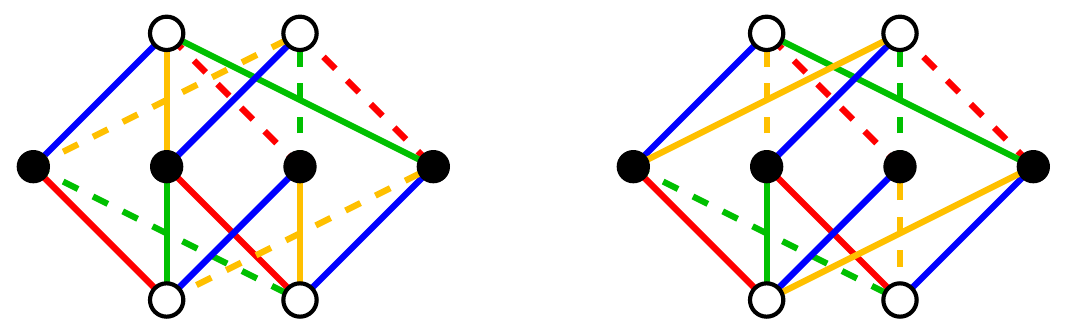}}
       \put(5,5){\Rx{\small(twisted-chiral)}}
       \put(5,25){\Rx{\small($+$)}}
       \put(90,5){\Lx{\small(chiral)}}
       \put(90,25){\Lx{\small($-$)}}
     \end{picture}}
 \label{e:B242}
\end{equation}
The definite identification and naming convention was made\cite{r6-1} comparing with the original definitions of these worldsheet supermultiplets\cite{rGHR}. When depicting {\em\/worldline\/} supermultiplets, these are identical to the pair stacked second from the right in\eq{e:N4half}.

Projected $(p,q)$-Adinkras such as the two depicted in\eq{e:B242} have a {\em\/hallmark\/} that distinguishes them from the unprojected, $N$-cubical ones such as\eq{e:484}: $4k$ distinctly colored edges in every projected Adinkra form closed $4k$-gons, wherein the product of signs associated with dashed edges varies with the order of the permutation of the $4k$ colors\ft{This graphical {\em\/hallmark\/} was recognized in Ref.\cite{r6-1}, generalized for classification purposes and related to certain error-correcting codes in Refs.\cite{r6-3,r6-3.2,r6-3c} and used to define a {\em\/character\/} in Ref.\cite{rUMD09-1}|all for worldline (reductions of) supermultiplets. Herein, these notions are extended to worldsheet supermultiplets.}. For example, beginning with the bottom-left-hand node in the chiral Adinkra in\eq{e:B242} and proceeding clockwise, there is a \C1{red}-\C2{green}-\C3{blue}-\C4{orange} bow-tie shaped tetragon. Associating factors of $(-1)$ with dashed edges, the product along this path is $(+1)(-1)(+1)(+1)=-1$. In the similar tetragon where we permute the colors, say in the last two edges, the \C1{red}-\C2{green}-\C4{orange}-\C3{blue} tetragon has  $(+1)(-1)(-1)(+1)=+1$ associated with it. The same result is obtained starting from any white (bosonic) node (and ending back at it), but the opposite result is obtained when starting and ending at a black (fermionic) node|or when starting from any white (bosonic) node of the twisted-chiral Adinkra.

Being that edges are associated with the supersymmetry and superderivative action, and since these two tetragons both lead back to the same node (as do all others, in such projected Adinkras), we have that in supermultiplets depicted by projected Adinkras there exist operatorial relations such as
\begin{subequations}
 \label{e:d4D+}
\begin{align}
  \text{\bsf Chiral hallmark relation:}\qquad
  (\C4{\rD_{2-}})^{-1}\circ\C3{\rD_{1-}}\circ(\C2{\rD_{2+}})^{-1}\circ\C1{\rD_{1+}}
   &\simeq-(-1)^F\Ione,\\*
  \textit{i.e.}\quad
   \C4{\rD_{2-}}\C3{\rD_{1-}}\C2{\rD_{2+}}\C1{\rD_{1+}}
   &\simeq-(-1)^F(i\vd_\mm)(i\vd_\pp),
\end{align}
\end{subequations}
where $F=0$ for a white (bosonic) initial/final node and $F=1$ for a black (fermionic) initial/final node.
 The color-permutation dependent sign-changes are evidently a consequence of the anticommutivity of the $\rD$'s. Straightforwardly, 
\begin{subequations}
 \label{e:d4D-}
\begin{align}
  \text{\bsf Twisted-chiral hallmark relation:}\qquad
  (\C4{\rD_{2-}})^{-1}\circ\C3{\rD_{1-}}\circ(\C2{\rD_{2+}})^{-1}\circ\C1{\rD_{1+}}
   &\simeq(-1)^F\Ione,\\*
  \textit{i.e.}\quad
   \C4{\rD_{2-}}\C3{\rD_{1-}}\C2{\rD_{2+}}\C1{\rD_{1+}}
   &\simeq(-1)^F(i\vd_\mm)(i\vd_\pp).
\end{align}
\end{subequations}
and the relative sign difference in the right-hand side of\eq{e:d4D+} {\em\/vs.\/} the right-hand side of\eq{e:d4D-} unambiguously detects the relative {\em\/twisting\/}|not only between the chiral and twisted chiral supermultiplets of Ref.\cite{rGHR}, but completely generally\cite{r6-1,r6-3}; see also\cite{rUMD09-1}, where the numerical eigenvalue of an operator closely related to the superdifferential operators on the left-hand side of\eq{e:d4D+} and\eq{e:d4D-} was defined as a character of a worldline (reduction of a) supermultiplet.

It is not difficult to verify the consistency of the $(-1)^F$ factor on the right-hand side of \Eqs{e:d4D+}{e:d4D-}: Suppose that the relation\eq{e:d4D+} holds when applied on a bosonic component (super)field, $\BF$:
\begin{align}
 \P^{11|11}_+\BF&\Defl
   \big[\C4{\rD_{2-}}\C3{\rD_{1-}}\C2{\rD_{2+}}\C1{\rD_{1+}}
        +(i\vd_\mm)(i\vd_\pp)\big]\BF,\nn\\
 &~=\big[\C1{\rD_{1+}}\C2{\rD_{2+}}\C3{\rD_{1-}}\C4{\rD_{2-}}
        +(i\vd_\mm)(i\vd_\pp)\big]\BF\simeq0.
 \label{e:PabF0}
\end{align}
Applying the twisted operator, say, on the fermion $(\C1{\rD_{1+}}\BF)$ produces
\begin{align}
 \P^{11|11}_-(\C1{\rD_{1+}}\BF)
 &=\big[\C1{\rD_{1+}}\C2{\rD_{2+}}\C3{\rD_{1-}}\C4{\rD_{2-}}
        -(i\vd_\mm)(i\vd_\pp)\big](\C1{\rD_{1+}}\BF),\nn\\
 &=-\C1{\rD_{1+}}\big[\C1{\rD_{1+}}\C2{\rD_{2+}}\C3{\rD_{1-}}\C4{\rD_{2-}}
         +(i\vd_\mm)(i\vd_\pp)\big]\BF
  =-\C1{\rD_{1+}}\P^{11|11}_+\BF\simeq0.
 \label{e:PabF1}
\end{align}
Note that by applying the superderivatives\eq{e:MonD} on any one component (super)field one obtains all the other component (super)fields in a supermultiplet or their worldsheet derivatives.
 Then, proceeding in the manner\eqs{e:PabF0}{e:PabF1}, it follows that if a bosonic component (super)field in a supermultiplet is annihilated by $\P^{11|11}_+$, all bosonic component (super)fields are annihilated by the same $\P^{11|11}_+$, whereas all fermionic component (super)fields are annihilated by the complementary $\P^{11|11}_-$.

A comparison of the relations\eq{e:d4D+} and\eq{e:d4D-} implies:
\begin{corl}\label{C:Tw=BF}
Boson/fermion (white/black) node assignment flipping in the Adinkras\eq{e:B242} is,
up to node rearrangement, equivalent to its twisting.
\end{corl}
Projections generalizing\eq{e:B484>242}|and the corresponding {\bsf\/hallmark 4{\slshape k}-gon relations\/} generalizing\eq{e:d4D+} and\eq{e:d4D-}|have been explored and catalogued in Refs.\cite{r6-3.2} for worldline supermultiplets. Using the simple results\eqs{e:B484>242}{e:d4D-} as a template, the general results of Refs.\cite{r6-3,r6-3.2,r6-1.2,r6-3.1} are adapted to worldsheet supermultiplets and explored in more detail in Section~\ref{s:WSCodes}.

Before we turn to that, considering the graphical details of the projection\eq{e:B484>242} we can immediately generalize Corollary~\ref{C:Tw=BF} to conclude:
\begin{corl}\label{C:TwProj}
When an Adinkra is rendered so as to exhibit a (literal) left-right $\ZZ_2$ symmetry, the number of colors of the crisscrossing edges must be odd for the twisted variant of the projection to this $\ZZ_2$ quotient to be inequivalent from the untwisted one.
\end{corl}
If the crisscrossing edges came in an even number of colors, the twisting (identifying the {\em\/negative\/} of the left-hand nodes with the right-hand ones) will flip the solid/dashed parity of the edges in those even number of colors. This can always be compensated by a judicious component (super)field sign-change, whereupon all edges incident to the sign-changing nodes change their solid/dashed parity, and so the twisting is removed.

\subsubsection{Complex Structure}
 \label{s:Cpx}
The Adinkras\eq{e:11x242}, \eq{e:121x22} and\eq{e:22x121} all depict equivalent worldline supermultiplets, but inequivalent worldsheet supermultiplets. The particular arrangement of\eq{e:121x22} makes the horizontal twisted $\ZZ_2$ symmetry in these Adinkras obvious|as per specification in Corollary~\ref{C:Cpx}. In turn, the same structure is evident in the Adinkras\eq{e:11x242} and\eq{e:22x121} by the facts that:
 \begin{enumerate}\itemsep=-3pt\vspace{-2mm}
 \item the nodes and the edges of a chosen pair of colors form multiple copies of\eq{e:22},
 \item nodes in any such copy of\eq{e:22} are connected to the nodes of any other such copy by perfectly like edges (same color, same solid/dashed parity) of the remaining  colors.
\end{enumerate}
When depicting worldsheet supermultiplets, the $(2,2)$-Adinkras\eq{e:121x22} and\eq{e:22x121} are each other's $\rD_{\a+}\iff\rD_{\ad-}$ mirror images, whereas the Adinkra\eq{e:11x242} depicts a $(3,1)$-supermultiplet and its ($1,3)$-supersymmetric mirror-pair.
 Both\eq{e:11x242} and\eq{e:22x121} can be brought into the twisted left-right symmetric form of\eq{e:121x22} by a horizontal repositioning of nodes.
 Using the complex basis {\em\/\`a la\/}\eq{e:Cpx}, these Adinkras may be used to depict the supermultiplets that are also known as semi-chiral superfields\cite{rSChSF0,rSChSF}. For the $(2,2)$-Adinkras\eqs{e:121x22}{e:22x121}, this has been demonstrated explicitly\cite{rGH-obs} by reading off the supersymmetry transformation rules from the Adinkras and comparing them with the complex superfield results.

In the same manner, a complex structure is detected in the first and third $(4,0)$- and $(0,4)$-Adinkras in both their twisted and untwisted versons\eq{e:N4half}. With a little horizontal rearrangement, this can be made evident:
\begin{equation}
 \vC{\begin{picture}(140,45)
   \put(0,0){\includegraphics[width=140mm]{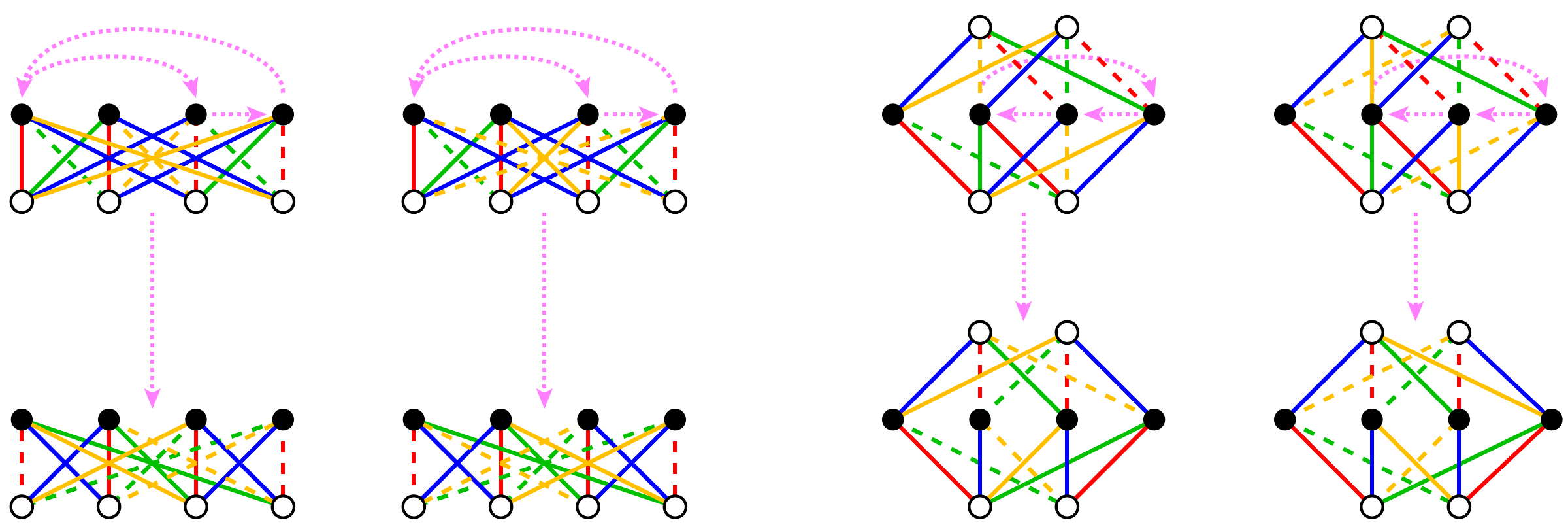}}
    \put(7.5,19){\small(chiral)}
    \put(35.5,19){\small(twisted-chiral)}
    \put(85,23){\small(chiral)}
    \put(115,23){\small(twisted-chiral)}
 \end{picture}}
 \label{e:ChTCh}
\end{equation}
where now the bottom-row Adinkras satisfy the specifications of Corollary~\ref{C:Cpx} and may be used to depict complex supermultiplets of $(4,0)$- or $(0,4)$-supersymmetry. Although these are not the originally so-named $(2,2)$-supermultiplets in Ref.\cite{rTwSJG0,rGHR}, we adopt that terminology, just as has been done for worldline supermultiplets in Refs.\cite{r6-1,r6-3c,r6-3,r6-3.2,r6-1.2}. In fact, the left-hand half of the Adinkras\eq{e:ChTCh} admit a quaternionic structure; see Section~\ref{s:FnF}.

Finally, the $(2,2)$-Adinkras\eq{e:B242} are identical with the right-hand half of the Adinkras\eq{e:ChTCh}, and so admit a conjugate pair of complex structures in just the same way. Thus, the $(2,2)$-Adinkras\eq{e:B242} indeed depict the complex chiral and twisted chiral supermultiplets as well as their conjugates, which are also represented by the superfields of the same name, as introduced in Ref.\cite{rGHR}.

The difference between the (literal) left-right $\ZZ_2$ symmetry, exemplified in\eq{e:B484>242}, and the {\em\/twisted\/} $\ZZ_2$ symmetry, exemplified by the Adinkras in the lower row of\eq{e:ChTCh} or the detailed illustration\eq{e:Ch2D} is highlighted in Table~\ref{t:Z2Z2}.
\begin{table}[htpb]
\begin{center}
  \begin{tabular}{@{} >{\raggedright}p{70mm}|>{\raggedright}p{80mm} @{}c@{}}
  \MC1{c|}{\bsf (literal, left-right) ~$\pmb{\ZZ_2}$ symmetry}
 &\MC1c{\bsf twisted ~$\pmb{\ZZ_2}$ symmetry}&\\ 
    \toprule
 \MC2c{Both types of Adinkras can be drawn to have identical left- and right-hand halves.}&\\
    \midrule
  left-right criss-crossing edges are\break
  pair-wise identical; see, \eg\eq{e:B484>242}
 & left-right criss-crossing edges appear in\newline
   solid-dashed pairs; see, \eg\eq{e:ChTCh} &\\ 
    \midrule
   used to project to a half-sized supermultiplet; see, \eg\eq{e:B484>242}
 & used to indicate that a supermultiplet admits a complex structure; see\eq{e:CpxA}&\\
    \bottomrule
  \end{tabular}
\end{center}\vspace{-3mm}
\caption{A side-by-side comparison between literal and twisted $\ZZ_2$ symmetries}
\label{t:Z2Z2}
\end{table}

\subsubsection{Summary}
Table~\ref{t:N4} summarizes the results of this section.
\begin{table}[htb]
\centering\small
  \begin{tabular}{@{} c|lp{50mm}cc|l @{}}
 \MC1c{}
 & {\bsf({\slshape p},{\slshape q})-Supersymmetry} & {\bsf Adinkras}
 & {\bsf \#} & {\bsf Constr.}& {\bsf Comment} \\ 
    \toprule
 \MR3*{\rotatebox{90}{\bsf off-shell}}
 & $(4,0)~\&~(0,4)$ & all 28 $N=4$ Adinkras$^*$
    & 64 & \ref{C:N>N0} &
 \MR7*{\parbox[c]{30mm}{\rightskip=0pt plus4pc\baselineskip=11pt
 In Ref.\cite{r6-3.2}, each of the Adinkras listed also represents its boson/fermion and upside-down flips, as well as its twisted variant.}} \\ 
 \cline{2-5}
 & $(3,1)~\&~(1,3)$ & (\ref{e:11x1331}), (\ref{e:11x143}), (\ref{e:11x242}),
                    (\ref{e:11x341}), (\ref{e:11x44}) & 10 & \ref{C:pxq>pq} \\
 \cline{2-5}
 & $(2,2)$ & (\ref{e:121x121}), (\ref{e:121x22}), (\ref{e:22x121}), (\ref{e:B242}) & 10 & \ref{C:pxq>pq} \\
 \cline{1-5}
 \MR4*{\rotatebox{90}{\bsf half-shell$^\dag$}}
 & $(4,q)_\mm$ \& $(p,4)_\pp$ & all 28 $N=4$ Adinkras$^*$
   & 64$^\dag$ & \ref{C:N>Nq} & \\
 \cline{2-5}
 & $(3,q)_\mm$ \& $(p,3)_\pp$ & the $N\,{=}\,3$ Adinkras\eq{e:N3list}
   & 10$^\dag$ & \ref{C:N>Nq} \\
 \cline{2-5}
 & $(2,q)_\mm$ \& $(p,2)_\pp$ & the $N\,{=}\,2$ Adinkras\eqs{e:121}{e:22}
   & 4$^\dag$ & \ref{C:N>Nq} \\
 \cline{2-5}
 & $(1,q)_\mm$ \& $(p,1)_\pp$ & the $N\,{=}\,1$ Adinkras in Table~\ref{t:A}
   & 2$^\dag$ & \ref{C:N>Nq} \\
    \bottomrule
 \MC1c{}
 &\MC5l{\footnotesize$^*$\,These Adinkras are listed in tables~6 and~7 in Ref.\cite{r6-3.2}.\qquad $^\dag$\,$(\cdots)_\mm$ = ``annihilated by $\vd_\mm$''}
  \end{tabular}
  \caption{A list of off-shell and on the half-shell adinkraic supermultiplets of worldsheet $(p,q)$-super\-symmetry constructed by tensoring supermultiplets of worldline $N$-extended supersymmetry.}
  \label{t:N4}
\end{table}
In addition, off-shell supermultiplets of worldline $(N\,{<}\,4)$-supersymmetry readily extend to worldsheet supermultiplets on the half-shell. For example by identifying $\rD_I\mapsto\rD_{I+}$ and $\ddt\mapsto\vd_\pp$, the two off-shell supermultiplets of $(N\,{=}\,2)$-extended supersymmetry\eq{e:121} and\eq{e:22} extend to left-moving supermultiplets of worldsheet $(2,q)$-supersymmetry for arbitrary $q>0$; these supermultiplets are on the half-shell, being annihilated by $\rD_{\ad-}$ and $\vd_\mm$. The supermultiplets\eq{e:121} and\eq{e:22} also extend to right-moving supermultiplets of worldsheet $(p,2)$-supersymmetry for arbitrary $p>0$ and are annihilated by $\rD_{\a+}$ and $\vd_\pp$.

\paragraph{Completeness:}
The list presented in Table~\ref{t:N4} provides a complete list of adinkraic off-shell and on the half-shell supermultiplets of various worldsheet $(p,q)$-supersymmetries, all of which are obtained by tensoring worldline $(N\,{\leqslant}\,4)$-extended supersymmetry. The preceding then suggests:
\begin{conj}[completeness]\label{C:Weyl}
The application of Constructions~\ref{C:pxq>pq}, \ref{C:N>Nq} and~\ref{C:N>N0} together with the projections of the kind\eq{e:B484>242}|detailed in Section~\ref{s:WSCodes}|generates {\bsf all} $(p,q)$-Adinkras and so also {\bsf all} adinkraic off-shell worldsheet $(p,q)$-supermultiplets|for all $p,q\geqslant0$.
\end{conj}

\paragraph{Redundancy:}
While Constructions~\ref{C:pxq>pq}, \ref{C:N>Nq} and~\ref{C:N>N0} together with the projections of the kind\eq{e:B484>242} certainly generate a number of $(p,q)$-Adinkras and corresponding worldsheet supermultiplets, some of these may turn out to be equivalent. This phenomenon has been noted in worldline supermultiplets\cite{r6-3.2}, where a criterion for determining when that happens was also given. The phenomenon is likely to also occur amongst worldsheet supermultiplets for large enough $p{+}q$, and is clearly inherited {\em\/verbatim\/} in extensions to unidextrous $(N,0)$- and $(0,N)$-supersymmetry. Section~\ref{s:1} explores a few examples of this phenomenon amongst $(p,q)$-supermultiplets; the full extent to which this equivalence of supermultiplets with distinct Adinkras also extends to ambidextrous $(p,q)$-supersymmetry remains an open question for now.

\section{Supersymmetry and Error-Correcting Codes}
\label{s:WSCodes}
For $N\geqslant4$, certain adinkraic worldline supermultiplets admit $\ZZ_2$ symmetries and corresponding projections to smaller supermultiplets|akin to the projection described in\eqs{e:484}{e:B242}. The action of such symmetries is encoded by error-detecting and error-correcting (binary) {\em\/doubly even\/} linear block codes\cite{r6-3,r6-3.2}, ``\DELB-codes'' for short. Herein, we explore their worldsheet analogues.

\subsection{Encoding Worldsheet Supermultiplets}
\label{s:esDE}
Since component fields within a superfield are defined using the $N$-cube of superderivatives such as in Figure~\ref{f:31Ds} and\eq{f:22Ds}, the component-wise identifications such as made in\eq{e:B484>242} must translate into identification relations among the component-defining superderivatives\eq{e:MonD} and take the general form using the binary exponent notation\eqs{e:MonD}{e:brk}:
\begin{equation}
  \vd_\pp^{n_+}\vd_\mm^{n_-}\rD^{\bf x,y}~\pm~
  \vd_\pp^{n'_+}\vd_{\mm\like{$\SSS\pp$}}^{n'_-}\rD^{\bf x',y'}\simeq0,
 \label{e:D+D}
\end{equation}
where ${\bf x}$ and ${\bf x}'$ have no common bit\ft{Herein, $\4$ denotes bitwise addition (\texttt{Xor}), and $\wedge$ is bitwise multiplication (\texttt{And}).}: $\bf x\wedge x'=0=y\wedge y'$
so the relations\eq{e:D+D} would not induce purely worldsheet differential constraints (with neither $\rD_{\a+}$ nor $\rD_{\ad-}$) on the component (super)fields.
 Applying separately $\rD^{\bf x,y}$ or $\rD^{\bf x',y'}$ from the left, we obtain superderivative relations that are, after clearing common factors, of the general form
\begin{equation}
  \P^{\bf a|b}_\pm\Defl
   \inv2\big[(i\vd_\pp)^{\frac12\bf|a|}(i\vd_\mm)^{\frac12\bf|b|}~\pm~\rD^{\bf a|b}\big]
   ~\simeq~0.
 \label{e:Pab}
\end{equation}
Such operators then provide the generalization of the hallmark $4k$-gon relations generalizing\eq{e:d4D+} and\eq{e:d4D-}. A few remarks are in order:
\begin{enumerate}\itemsep=-3pt\vspace{-2mm}
 \item With engineering dimension homogeneity and $\Spin(1,1)$-covariance, the split binary exponents $\bf a|b$ fully encode the operators\eq{e:Pab} except for the (again binary) choice of the relative sign between the two summands in\eq{e:Pab}.
 \item The choice of the relative sign is called the {\em\/twist\/}, and coincides with the standard terminology such as in chiral {\it vs}.\ twisted-chiral superfields\cite{rGHR}; see below, and Ref.\cite{r6-3} for the worldline variant of the statement.
\end{enumerate}\vspace{-2mm}
We now turn to explore these to features in more detail.

\paragraph{Binary Encoding:}
Superdifferential operators such as\eq{e:Pab} are quite familiar from the superspace formalism in $3{+}1$-dimensional spacetime\cite{r1001,rBK}. These are {\bsf quasi-projection operators}\ft{Refs.\cite{r1001,rBK} construct proper projection operators by formally dividing by spacetime derivatives. This is well defined only when acting on eigenfunctions of those spacetime derivatives with nonzero eigenvalues; herein we refrain from such on-shell restrictions.}, in that they must be quasi-idempotent and mutually orthogonal:
\begin{equation}
  (\P^{\bf a|b}_\pm)^2 \Is
   (i\vd_\pp)^{\frac12|{\bf a}|}(i\vd_\mm)^{\frac12|{\bf b}|}\,\P^{\bf a|b}_\pm,
    \quad\text{and}\quad
  \P^{\bf a|b}_\pm\,\P^{\bf a|b}_\mp \Is 0.
 \label{e:Cond}
\end{equation}
The first of these conditions yields
\begin{equation}
 (\P^{\bf a|b}_\pm)^2
  =\inv4\big[(i\vd_\pp)^{\bf|a|}(i\vd_\mm)^{\bf|b|}+\rD^{\bf a|b}\rD^{\bf a|b}
        \pm2(i\vd_\pp)^{\frac12\bf|a|}(i\vd_\mm)^{\frac12\bf|b|}\rD^{\bf a|b}\big],
\end{equation}
which equals $(i\vd_\pp)^{\frac12\bf|a|}(i\vd_\mm)^{\frac12\bf|b|}\,\P^{\bf a|b}_\pm$ if and only if
\begin{equation}
  \rD^{\bf a|b}\rD^{\bf a|b} =+(i\vd_\pp)^{\bf|a|}(i\vd_\mm)^{\bf|b|}.
\end{equation}
Direct computation yields
\begin{align}
 \rD^{\bf a|b}\rD^{\bf a|b}
  &=(-1)^{\bf|a||b|}\,
     (-1)^{\binom{\bf|a|}2}\,\rD_{1+}^{2a_1}\wedge\cdots\wedge\rD_{p+}^{2a_p}\,{\cdot}\,
      (-1)^{\binom{\bf|b|}2}\,\rD_{1-}^{2b_1}\wedge\cdots\wedge\rD_{q-}^{2b_q},\nn\\
  &=(-1)^{\sss\binom{\bf|a|+|b|}2}\>
     (i\vd_\pp)^{\bf|a|}(i\vd_\mm)^{\bf|b|},
\end{align}
so that the first of the conditions\eq{e:Cond} holds precisely if
\begin{equation}
  \ttt\binom{\bf|a|+|b|}2~\in2\ZZ.
\end{equation}
The second condition\eq{e:Cond} is then satisfied automatically.

In turn, for the (super)differential operators\eq{e:Pab} to be local, $\bf|a|$ and $\bf|b|$ must both be even, whereupon their sum is also even. From this, we have:
\begin{equation}
  \text{both}\quad\ttt\binom{\bf|a|+|b|}2\quad
  \text{and}\quad {\bf|a|+|b|} ~\in2\ZZ \qquad\To\qquad
  {\bf|a|+|b|}=0\pmod4.
\end{equation}
That is, the binary exponent in\eq{e:Pab} with digits $a_1,\cdots,a_p,b_1,\cdots,b_q$ must be doubly-even, and also split into even (not necessarily equal) parts:
\begin{equation}
  |{\bf a}|+|{\bf b}|=0\pmod4,\qquad
  |{\bf a}|,|{\bf b}|=0\pmod2.
 \label{e:DE;E,E}
\end{equation}
For any desired system of identification relations\eq{e:D+D}, the corresponding system of quasi-projection operators\eq{e:Pab} will consist of $k$ independent\ft{A collection of $k$ relations (realized by quasi-projection operators) are independent if the imposition of any $k{-}1$ of them on any supermultiplet does not render the action of the $k^\text{th}$ one trivial.} relations. The corresponding $k$ split binary numbers (codewords) $g_i\Defl({\bf a|b})_i$ then generate an even-split (binary) doubly even linear block (\sDE) code, $\ssC$, that consists of all binary linear combinations $\4_ic_ig_i$; one says that $\ssC$ has rank $k$.

This implies a refining corollary of the results of Ref.\cite{r6-3} and the ``even-split refinement'' of the \DELB\ codes defined therein:
\begin{corl}
When $p,q\neq0$, only the \DELB\ code-encoded $\ZZ_2$ symmetries that admit an even-split (\sDE\ codes) define off-shell $\ZZ_2$ quotient worldsheet $(p,q)$-supermultiplets.
\end{corl}
See Figure~\ref{f:1} below for a depiction of \sDE\ codes for $p{+}q\leqslant8$.

\paragraph{Twisting:}
In the general case, there exist several ($k$) mutually commuting relations of the type\eq{e:Pab}; each defines an even-split binary number (codeword), $g_i=({\bf a}|{\bf b})_i$, which jointly generate an \sDE\ code $\ssC$. Each relation of the type\eq{e:Pab} exhibits a choice of the relative sign, whereupon there exist $2^k$ different combinations of such quasi-projective operators, and correspondingly $2^k$ choices of self-duality type superderivative relations of the type\eq{e:ddD0}.

For {\em\/worldline\/} supermultiplets, the same abundance of sign-choices was shown to nevertheless result in only one untwisted-twisted pair of supermultiplets\cite{r6-3,r6-3.2}|and only in cases where the total number of supersymmetries is $N=0\pmod4$. There exist two separate types of isomorphisms that so effectively reduce the number of inequivalent sign-choices: \begin{description}\itemsep=-3pt\vspace{-2mm}
 \item[\bsf Outer:] On the worldline|all $N$ supersymmetry generators may be freely permuted. Graphically, all $N$ edge-color assignments may be freely permuted. Within a given model, this operation clearly affects all supermultiplets and so is {\em\/global\/}.
 \item[\bsf Inner:] The other employs the fact that changing the sign of a particular component (super)field induces a change in the sign of each superderivative of that component (super)field. Graphically, every edge incident with the node representing the sign-changed component (super)field changes its solid/dashed parity. A judicious application of this operation alone can change the solid/dashed parity of edges of any {\em\/even\/} number of colors\cite{r6-3,r6-3.2}. Within a given model, this operation may be performed on any one supermultiplet at a time and so is {\em\/local\/} to a supermultiplet.
\end{description}

Within worldsheet $(p,q)$-supersymmetry, any permutation of supersymmetry generators must preserve Lorentz $\Spin(1,1)$-covariance: the $\rD_{\a+}$ may be permuted freely amongst themselves, as may the $\rD_{\ad-}$, but there can exist no permutation that would mix the $\rD_{a+}$ with the $\rD_{\ad-}$. This restriction on the possible {\em\/outer\/} equivalence isomorphisms between the $2^k$ sign-choices in\eq{e:Pab} may well, in general, prevent transforming one (un)twisted projected supermultiplet into another.

In turn, however, the {\em\/inner\/} equivalence isomorphisms within a worldsheet $(p,q)$-supermultiplet remain as free as they are within worldline supermultiplets, leading thus to the same conclusion as in Refs.\cite{r6-3,r6-3.2}:
\begin{corl}
Only in case of $p{+}q=0\,\text{mod}\,4$ and only for $\ZZ_2$-projected supermultiplets does  twisting produce inequivalent classes of supermultiplets, and precisely two of them. Their Adinkras differ in the solid/dashed parity of edges of an odd number of colors.
\end{corl}

\paragraph{Extending from Worldline to Worldsheet:}
The above results may be rephrased in terms of extending the worldline constructions and classification of adinkraic off-shell supermultiplets in Ref.\cite{r6-3,r6-3.2,r6-1.2} to worldsheet supersymmetry as follows.
 Let $\sC$ be a \DELB\ $[N;k]$-code, that is, a collection of $N$-digit binary numbers that are all:
\begin{enumerate}\itemsep=-3pt\vspace{-2mm}
 \item doubly even (the sum of digits is divisible by 4),
 \item closed under bitwise binary addition ($\boxplus$, \ie, \texttt{Xor}),
 \item binary linear combinations of some $k$ {\em\/generators\/}.
\end{enumerate}
The ambidextrous extension of $\sC$ and its use in projecting worldsheet supermultiplets|as was the case in\eq{e:B484>242}|requires that we split the $N$ worldline supersymmetries into $p$ left-moving and $q\,{=}\,(N{-}p)$ right-moving supersymmetries in such a way that the corresponding {\em\/left\/} portion and the {\em\/right\/} portion of each codeword in $\sC$ is even.
 It follows that both the left and the right  portions of the codewords separately form (simply) even binary linear block codes. Such a splitting
\begin{equation}
  \text{\DELB\ $[N;k]$-code}~\sC ~~\longrightarrow~~
  \text{\sDE\ $[p,q;k']$-code}~\ssC,\quad k'\leqslant k.
\end{equation}
may turn out to be:
 ({\small\bf1})~impossible,
 ({\small\bf2})~unique, or 
 ({\small\bf3})~multiple,
for any given $[N;k]$-code and any desired extension $\SS1N\to\SS{1,1}{p,N-p}=\SS1p_+\,{\otimes}\,\SS1{N-p}_-$.

Consider now the special case of\eq{e:Pab}, when $\bf b=0$:
\begin{equation}
  \P^{\bf a|0}_\pm\Defl
   \inv2\big[(i\vd_\pp)^{\frac12\bf|a|}~\pm~\rD^{\bf a|0}\big]~\simeq~0.
 \label{e:Pab=0}
\end{equation}
These quasi-projection operators are evidently the unidextrous $\{\rD_I,\ddt\}\mapsto\{\rD_{\a+},\vd_\pp\}$ extension mapping of the worldline quasi-projection operators employed in Ref.\cite{r6-3,r6-3.2,r6-1.2}. Thereby, the classification therein translates {\em\/verbatim\/} into a classification of one of the following two:
\begin{enumerate}\itemsep=-3pt\vspace{-2mm}
 \item Ambidextrous supermultiplets of unidextrous worldsheet $(N,0)$-supersymmetry, where there exist no $\rD_{\ad-}$-superderivatives, so that annihilation by $\vd_\mm$ is not implied and such supermultiplets are free to be off-shell. Such supermultiplets are constructed by means of projecting the result of Construction~\ref{C:N>N0} using the quasi-projection operators\eq{e:Pab=0}.
 \item Unidextrous supermultiplets of ambidextrous worldsheet $(N,q)$-supersymmetry are constructed by means of projecting the result of Construction~\ref{C:N>Nq} using the quasi-projection operators\eq{e:Pab=0}, and for arbitrary $q$. Such supermultiplets are annihilated by the $\rD_{\ad-}$-superderivatives and therefore also by $\vd_\mm$, and so are {\em\/on the half-shell\/}.
\end{enumerate}
The parity mirror-images of these constructions are evidently obtained by means of the unidextrous $\{\rD_I,\ddt\}\mapsto\{\rD_{\ad-},\vd_\mm\}$ extension mapping instead.

\subsection{Supermultiplet Reduction}
\label{s:Red}
While quasi-projection operators\eq{e:Pab} permit reading off the \sDE\ code, the complete and strict identifications that hold on a projected supermultiplet are not generated by the quasi-projectors\eq{e:Pab}, but by self-duality type relations of the form\eq{e:D+D}, where $n_\pm,n'_\pm$ have been chosen to be minimal, typically zero. This subtlety has been detected already for worldline supermultiplets\cite{r6-1.2}, and becomes only more prominent for worldsheet supermultiplets.
 We thus have:
\begin{defn}
Let an even-split doubly even code $\ssC$ be generated by $k$ generators $({\bf a|b})_i$, with $i=1,\cdots,k$. Then, to each generator $({\bf a|b})_i$ $i=1,\cdots,k$ there corresponds a system of ``self-duality'' superderivative operators
\begin{equation}
  \S^{({\bf a|b})_i\,\pm}_{\a\dots|\ad\dots}
   \Defl~\big[\rD_{\a+}{\cdots}\,\rD_{\ad-}{\cdots} ~\pm~
          \inv{({\sss\frac12}|{\bf a}|)!}\inv{({\sss\frac12}|{\bf b}|)!}
           \ve_{\a\dots}{}^{\b\dots}\,\ve_{\ad\dots}{}^{\bd\dots}
            \rD_{\bd+}{\cdots}\,\rD_{\bd-}{\cdots}\big]
 \label{e:Sab}
\end{equation}
where the indices $\a,\b,\dots$ range over those values at which positions the binary number {\bf a} has 1's, and the range of values for $\ad,\bd,\dots$ is similarly determined by the 1's in {\bf b}; see, e.g.,\eq{e:ddD0}.
\end{defn}

To see the need for the operators\eq{e:Sab}, consider the example\eq{e:d4D+}, rewritten in lexicographic order: Applying $\rD_{1+}$ from the left\ft{Since such relations by definition hold when applied from the left on  superfields, any additional operator must be applied from the left.}, we obtain
\begin{alignat}9
 \rD_{1+}\cdot\Big(\rD_{1+}\rD_{2+}\rD_{1-}\rD_{2-} &\simeq \vd_\mm\vd_\pp\Big)\quad
 &\To\quad
 i\vd_\pp\rD_{2+}\rD_{1-}\rD_{2-}
 &\simeq \vd_\mm\vd_\pp\rD_{1+} \label{e:ddD1}.
\intertext{Applying now $\rD_{1-}$ produces:}
 \rD_{1-}\cdot\Big(i\vd_\pp\rD_{2+}\rD_{1-}\rD_{2-} &\simeq \vd_\mm\vd_\pp\rD_{1+}\Big)\quad
 &\To\quad
 -i\vd_\pp\rD_{2+}(i\vd_\mm)\rD_{2-}
 &\simeq -\vd_\mm\vd_\pp\rD_{1+}\rD_{1-},\nn\\
 &&
 \vd_\pp\vd_\mm\rD_{2+}\rD_{2-}
 &\simeq -\vd_\pp\vd_\mm\rD_{1+}\rD_{1-} \label{e:ddD2}.
\end{alignat}
Of these conditions, \eq{e:ddD1} is vacuous on right-moving functions on the worldsheet, and\eq{e:ddD2} is vacuous on harmonic functions. Thus, attempting to reduce a supermultiplet by imposing the hallmark quasi-projections\eq{e:d4D+} would not result in a proper off-shell supermultiplet, being defined only up to fully unrestricted unidextrous and harmonic summands in many of its component (super)fields.

Following\cite{r6-1.2}, the necessary proper conditions are then generated from the ``self-duality'' relations\eq{e:Sab}. For the $d_{2,2}$ even-split doubly even code, which has a single generator, $11|11$, the ``self-duality'' operators are:
\begin{equation}
  \S^{{\sss11|11}+}_{1|1}\Defl\rD_{1+}\rD_{1-}+\rD_{2+}\rD_{2-}\simeq0 \qquad\text{and}\qquad
  \S^{{\sss11|11}+}_{1|2}\Defl\rD_{1+}\rD_{2-}-\rD_{2+}\rD_{1-}\simeq0,
 \label{e:ddD0}
\end{equation}
meaning that these operators annihilate component (super)fields in any $d_{2,2}$-projected supermultiplet.
 Applying $\rD_{1+}$ and then $\rD_{1-}$ on the first of these then results
\begin{alignat}9
 \rD_{1-}{\cdot}\rD_{1+}{\cdot}\htS{11|11}{+}{1|1}
 &=\rD_{1-}{\cdot}\rD_{1+}{\cdot}\big[\rD_{1+}\rD_{1-}+\rD_{2+}\rD_{2-}\big]
  =\rD_{1-}{\cdot}\big[(i\vd_\pp)\rD_{1-}+\rD_{1+}\rD_{2+}\rD_{2-}\big],\nn\\
 &=\big[(i\vd_\pp)(i\vd_\mm)+\rD_{1+}\rD_{2+}\rD_{1-}\rD_{2-}\big]
  =\P^{\sss11|11}_+,
\end{alignat}
the vanishing of which is equivalent to\eq{e:d4D+}. Similar manipulations show that the two operatorial relations\eq{e:ddD0} are both mutually consistent and consistent with\eq{e:d4D+}.

Applying the relations\eq{e:ddD0} on the supermultiplet\eq{e:22U} to reduce it {\em\/does\/} produce an off-shell supermultiplet, albeit in a rather unexpected way. The operators\eq{e:ddD0} evidently produce identification relations within\eq{e:22U} only from the middle level upward. To be precise, by applying one superderivative from left at a time, the generating relations $\S$ produce:
\begin{equation}
  \begin{array}{@{} r@{~\simeq~}l@{~~\Iff~~}r@{~\simeq~}l @{}}
 \MC2{c}{\rD\text{\bsf-relation}}\qquad\qquad &
 \MC2{c}{\text{\bsf Comp. Field Relation}} \\ 
    \toprule
 \rD_{1+}\rD_{1-} & \rD_{2+}\rD_{2-} & {\bf F}_{11} & {\bf F}_{22} \\
 \rD_{1+}\rD_{2-} &-\rD_{2+}\rD_{1-} & {\bf F}_{12} &-{\bf F}_{21} \\
    \midrule
 i\vd_\pp\rD_{1-} & \rD_{1+}\rD_{2+}\rD_{2-} & i\vd_\pp\bJ_{1-} & \bX^\mm_{2-} \\
-i\vd_\pp\rD_{1+} &-\rD_{2+}\rD_{1-}\rD_{2-} & i\vd_\mm\bJ_{1+} & \bX^\pp_{2+} \\
-\rD_{1+}\rD_{2+}\rD_{1-} & i\vd_\pp\rD_{2-} & -\bX^\mm_{1-} & i\vd_\pp\bJ_{2-} \\
 \rD_{1+}\rD_{1-}\rD_{2-} &-i\vd_\pp\rD_{2+} &  \bX^\pp_{1+} &-i\vd_\mm\bJ_{2+} \\
    \midrule
 \vd_\pp\vd_\mm &-\rD_{1+}\rD_{2+}\rD_{1-}\rD_{2-} & \vd_\pp\vd_\mm\BF &-\PMB{\cF} \\
 i\vd_\pp\rD_{1-}\rD_{2-} &-i\vd_\mm\rD_{1+}\rD_{2+} & \vd_\pp{\bf F}_\mm &-\vd_\mm{\bf F}_\pp \\
    \bottomrule
  \end{array}
 \label{e:CFI}
\end{equation}
These identifications may be traced to be body-diagonal within the Adinkra\eq{e:22U}. The first two of the component (super)field identifications\eq{e:CFI} simply identify two pairs of component superfields; the next five express the five component superfields $(\bX^\pp_{\ad-},\bX^\mm_{\ad-};\PMB{\cF})$ in terms of derivatives of component (super)fields of lower engineering dimension.

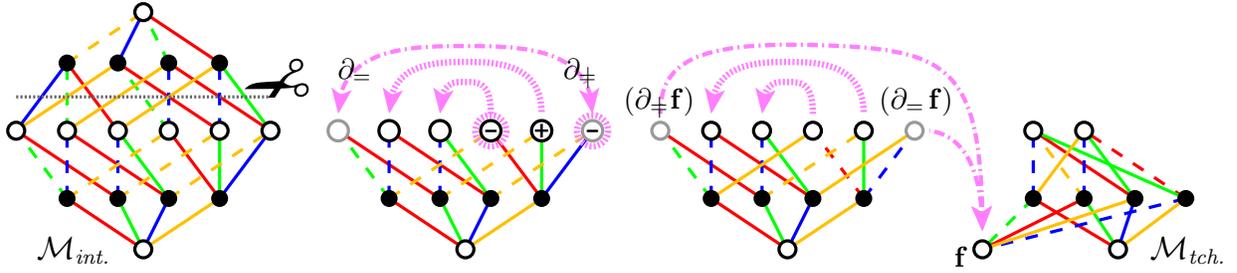
\begin{figure}[htp]
 \begin{center}
  \begin{picture}(160,30)(0,5)
   \put(0,0){\includegraphics[width=160mm]{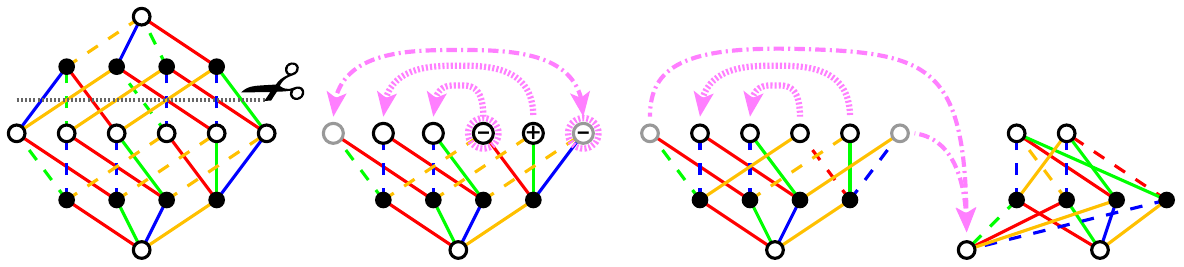}}
    \put(5,1){$\cM_\textit{int.}$}
    \put(45,25){\small$\vd_\mm$}
    \put(75,25){\small$\vd_\pp$}
    \put(83,21){\small$(\vd_\pp{\bf f})$}
    \put(117,21){\small$(\vd_\mm{\bf f})$}
    \put(127,0){\small\bf f}
    \put(153,1){$\cM_\textit{tch.}$}
 \end{picture}
 \end{center}
 \caption{The effects of imposing $\S^{11|11+}_{1+1-}\simeq0\simeq\S^{11|11+}_{1+2-}$ on the intact $(2,2)$-supermultiplet.}
 \label{f:int>tch}
\end{figure}

 However, the last relation, $\vd_\pp{\bf F}_\mm\simeq-\vd_\mm{\bf F}_\pp$|instead of identifying a linear combination of the existing component (super)fields ${\bf F}_\pp$ and ${\bf F}_\mm$|may be ``solved'' in terms of a {\em\/new\/} component (super)field:
\begin{equation}
 \vd_\pp{\bf F}_\mm\simeq-\vd_\mm{\bf F}_\pp\qquad\To\qquad
 {\bf F}_\pp=\vd_\pp{\bf f}\quad\&\quad {\bf F}_\mm=-\vd_\mm{\bf f}.
 \label{e:nlow}
\end{equation}
These identifications are depicted in Figure~\ref{f:int>tch}.
In the second, cut Adinkra, the highlighted $4^\text{th}$ and $6^\text{th}$ (previously) middle-level nodes from the left change signs, whereupon the incident edges change their solid/dashed parity as shown in the third Adinkra. Upon this, the $4^\text{th}$ and $5^\text{th}$ node from left in this row are identified with the $3^\text{rd}$ and $2^\text{nd}$ node, respectively.

 The $1^\text{st}$ and the $6^\text{th}$ (previously) middle-level nodes are shown grayed as they are related by a worldsheet differential condition, rendering both of them a derivative of a boson which is in the final, right-most rendition represented at the {\em\/bottom\/} level, and labeled ``{\bf f}\,''. Thus, part of the self-duality type relations\eq{e:ddD0} in effect imply not an identification of two component (super)fields with each other, but with worldsheet derivatives of a new component (super)field of lower engineering dimension; this is depicted by the simultaneous
 ({\small\bf1})~fusion of two nodes and 
 ({\small\bf2})~lowering of the resulting node.

Therefore, imposing
\begin{equation}
 \text{either}\quad\S^{11|11+}_{1+1-}\simeq0\simeq\S^{11|11+}_{1+2-}\quad
 \text{or}\quad~\S^{11|11-}_{1+1-}\simeq0\simeq\S^{11|11-}_{1+2-}
\end{equation}
on the intact supermultiplet\eq{e:22U} is necessary and sufficient: it generates all the requisite relationships between the component superfields so as to reduce the off-shell supermultiplet\eq{e:22U} into
\begin{equation}
 \vC{\begin{picture}(160,40)
      \put(0,0){\includegraphics[width=160mm]{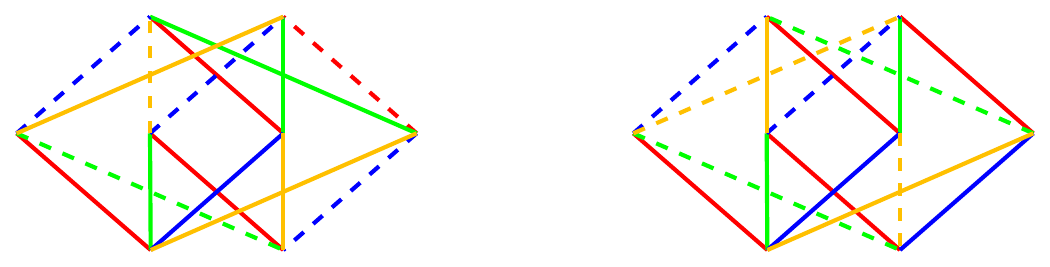}}
      \put(23,36){\cB{${\bf F}_{11}$}}
      \put(43,36){\cB{${\bf F}_{12}$}}
      \put(03,19){\cB{$\bJ_{1+}$}}
      \put(23,19){\cB{$\bJ_{2+}$}}
      \put(43,19){\cB{$\bJ_{1-}$}}
      \put(63,19){\cB{$\bJ_{2-}$}}
      \put(23,1){\cB{$\,\BF\,$}}
      \put(43,1){\cB{$\,\bf f\,$}}
      \small
      \put(0,36){$\S^{11|11+}_{1+1-}$}
      \put(0,30){$\S^{11|11+}_{1+2-}$}
      \put(0,7){twisted}
      \put(0,3){chiral}
      \normalsize
      \put(117,36){\cB{${\bf F}_{11}$}}
      \put(137,36){\cB{${\bf F}_{12}$}}
      \put(97,19){\cB{$\bJ_{1+}$}}
      \put(117,19){\cB{$\bJ_{2+}$}}
      \put(137,19){\cB{$\bJ_{1-}$}}
      \put(157,19){\cB{$\bJ_{2-}$}}
      \put(117,1){\cB{$\,\BF\,$}}
      \put(137,1){\cB{$\,\bf f\,$}}
      \small
      \put(93,36){$\S^{11|11-}_{1+1-}$}
      \put(93,30){$\S^{11|11-}_{1+2-}$}
      \put(93,3){chiral}
 \end{picture}}
 \label{e:ChTwCh}
\end{equation}
Upon flipping the signs of $\bJ_{2-},{\bf F}_{11}$ and ${\bf F}_{12}$ in the twisted chiral Adinkra, and of ${\bf F}_{11}$ and ${\bf F}_{12}$ in the chiral Adinkra, these become identical to those shown in\eq{e:B242}.

In the analogous {\em\/worldline\/} construction, there exist {\em\/three\/} self-duality type relations,
\begin{equation}
 \begin{gathered}
  \S^{1111+}_{12}\Defl \rD_1\rD_2+\rD_3\rD_4\simeq0,
  \Lx{\qquad and}\\
  \S^{1111+}_{13}\Defl \rD_1\rD_3-\rD_2\rD_4\simeq0,\qquad
  \S^{1111+}_{14}\Defl \rD_1\rD_4+\rD_2\rD_3\simeq0.
 \end{gathered}
 \label{e:d4}
\end{equation}
With the mapping $\{\rD_1,\rD_2,\rD_3,\rD_4\}\mapsto\{\rD_{1+},\rD_{2+},\rD_{1-},\rD_{2-}\}$, it is clear that the first of these,
\begin{equation}
  \text{putative}\quad\S^{11|11+}_{1+2+}\Defl\rD_{1+}\rD_{2+}+\rD_{1-}\rD_{2-}
\end{equation}
would violate $\Spin(1,1)$ Lorentz symmetry, and so cannot be used. Nevertheless, as the analysis\eqs{e:ddD0}{e:ChTwCh} shows, the remaining self-duality type constraints that {\em\/are\/} $\Spin(1,1)$-covariant do in fact generate precisely the required identifications to reduce the intact off-shell supermultiplet\eq{e:22U} to a ``half-size'' off-shell projection\eq{e:ChTwCh}. It seems reasonable to expect that this generalizes to all \sDE\ codes:
\begin{conj}
Given an \sDE\ code $\ssC$ with $k$ generators, the corresponding maximal set of linearly independent and $\Spin(1,1)$-covariant self-duality type relations|as given in\eq{e:D+D} and with minimal $n_\pm,n'_\pm$|{\bsf reduce} the intact supermultiplet to one of its $2^{-k}$-sized $\ssC$-encoded $(\ZZ_2)^k$-quotients, together with requisite instances of ``node-lowering,'' as in\eq{e:nlow}.
\end{conj}

Now, on the worldline, the supermultiplet obtained by reducing the intact supermultiplet {\em\/via\/} imposing the self-duality relations\eq{e:d4} on it is indeed a sub-supermultiplet of the original intact supermultiplet:
\begin{equation}
 \vC{\begin{picture}(140,48)
   \put(0,0){\includegraphics[width=120mm]{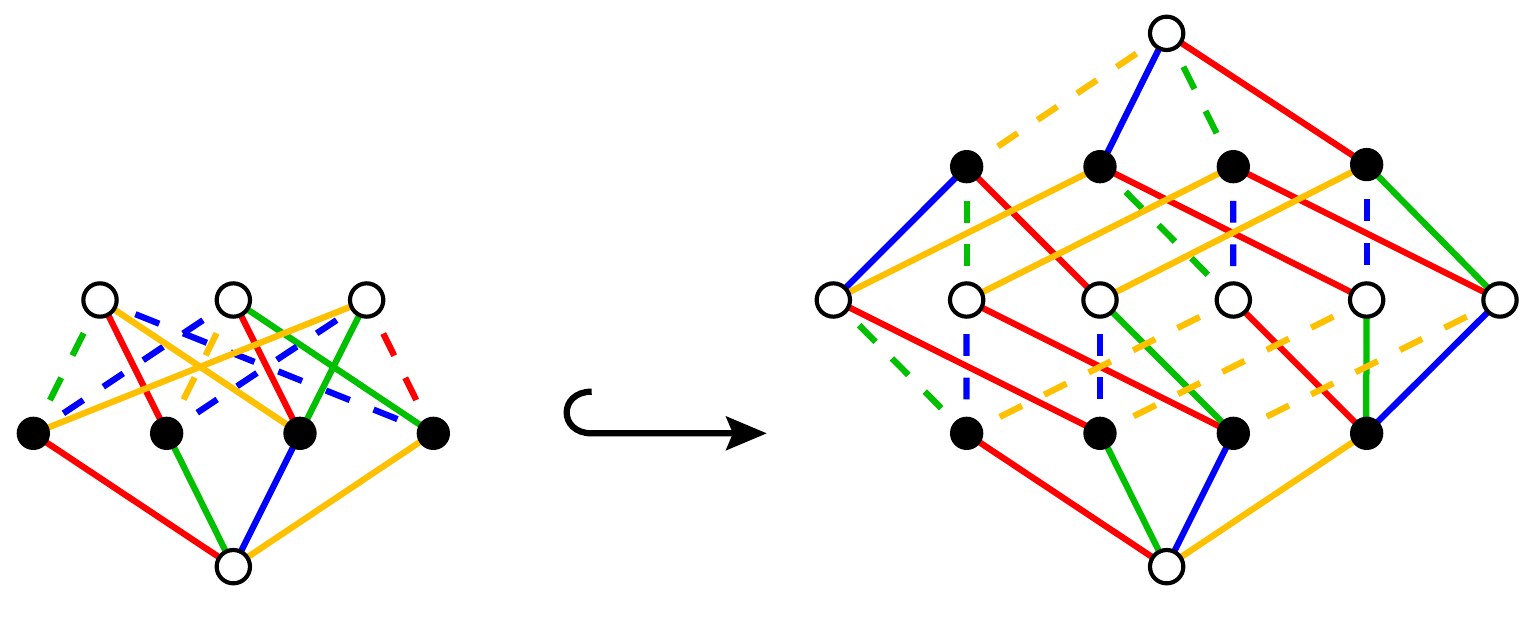}}
    \put(0,38){\small$\S^{1111+}_{12},~\S^{1111+}_{13},~\S^{1111+}_{14}$}
    \put(5,33){\small on the worldline}
    \put(53,18){\Cx{$\k$}}
    \put(54,11){\Cx{\footnotesize(sub-supermultiplet)}}
    \put(0,3){$\cM'_\textit{tch.}$}
    \put(100,3){$\cV_\textit{int.}$}
 \end{picture}}
\end{equation}
The quotient $\cV_\textit{int.}/\k(\cM'_\textit{tch.})$ is well known to represent the 1-dimensional dimensional reduction of the off-shell vector supermultiplet of simple $({\cal N}=1)$ supersymmetry in $(3{+}1)$-dimensional spacetime, and in the Wess-Zumino gauge\cite{r6-1.2}.

However, in stark contrast with this worldline result, the worldsheet off-shell supermultiplet obtained by reducing a supermultiplet by means of imposing self-duality constraints of the type\eq{e:Sab} need not be a {\em\/strict\/} sub-supermultiplet of the initial off-shell supermultiplet, in the sense of the definition\cite{r6-3.2}.
It is evident from considering the initial and final Adinkra in Figure~\ref{f:int>tch}, that the mapping from the reduced (twisted chiral) supermultiplet to the intact supermultiplet
\begin{equation}
 \vC{\begin{picture}(140,45)
   \put(0,0){\includegraphics[width=120mm]{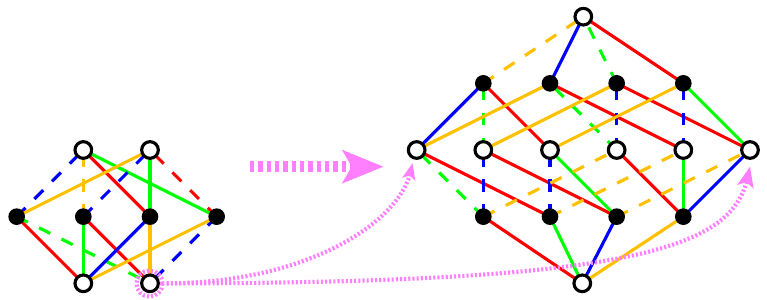}}
    \put(0,1){$\cM_\textit{tch.}$}
    \put(100,1){$\cM_\textit{int.}$}
    \put(47,23){$\k$}
    \put(60,10){$\vd_\pp$}
    \put(112,10){$-\vd_\mm$}
    \put(27,1.5){{\bf f}}
    \put(60,27){${\bf F}_\pp$}
    \put(117,27){${\bf F}_\mm$}
    \put(-5,38){${\bf f}\mapsto\{{\bf F}_\pp=(\vd_\pp{\bf f}),
                                {\bf F}_\mm=(\vd_\mm{\bf f})\}$}
    \put(6,33){so that $\vd_\mm{\bf F}_\pp=\vd_\pp{\bf F}_\mm$}
 \end{picture}}
\end{equation}
is local, but its inverse, shown in Figure~\ref{f:int>tch} is not. The quotient $\cM_\textit{int.}/\k(\cM_\textit{tch.})$ is then evidently not an off-shell supermultiplet, the mapping $\k$ is not a strict homomorphism of off-shell supermultiplets, and $\cM_\textit{tch.}\not\subset\cM_\textit{int.}$.

\subsection{Some Low-$(p,q)$ Split Codes}
\label{s:LowSpC}
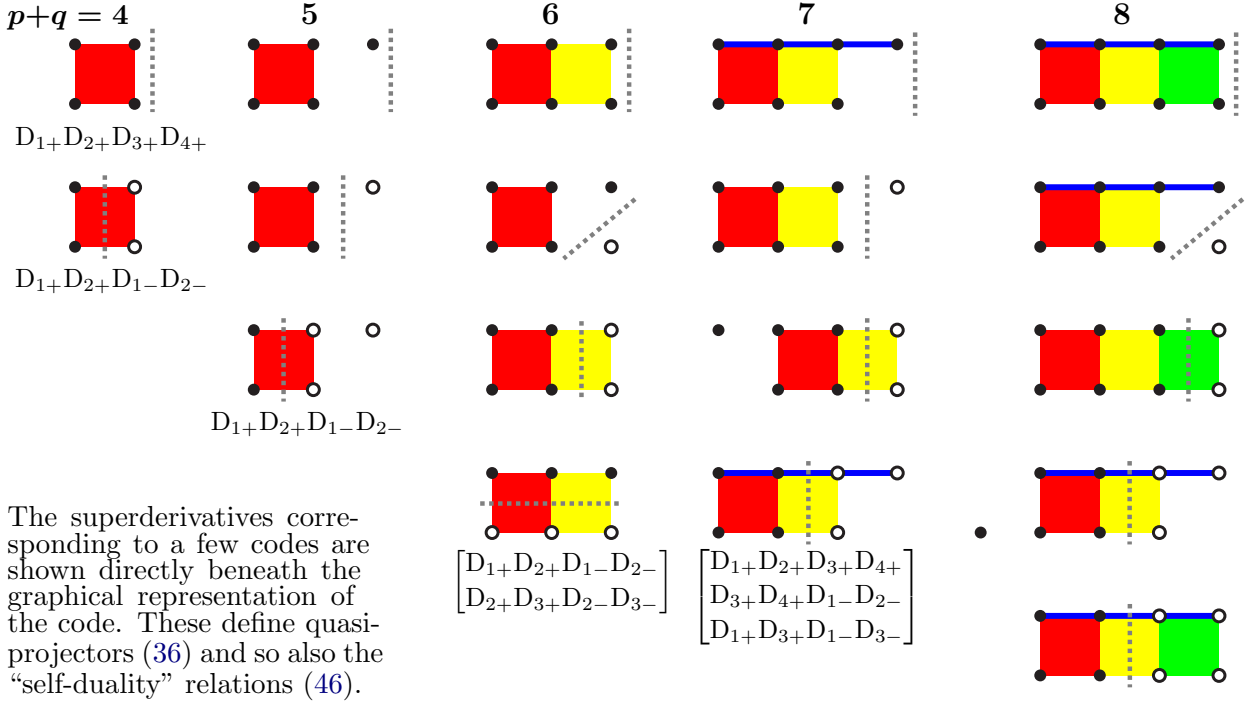
\begin{figure}[htp]
 \begin{center}
  \begin{picture}(160,90)(0,3)
   \put(0,0){\includegraphics[width=160mm]{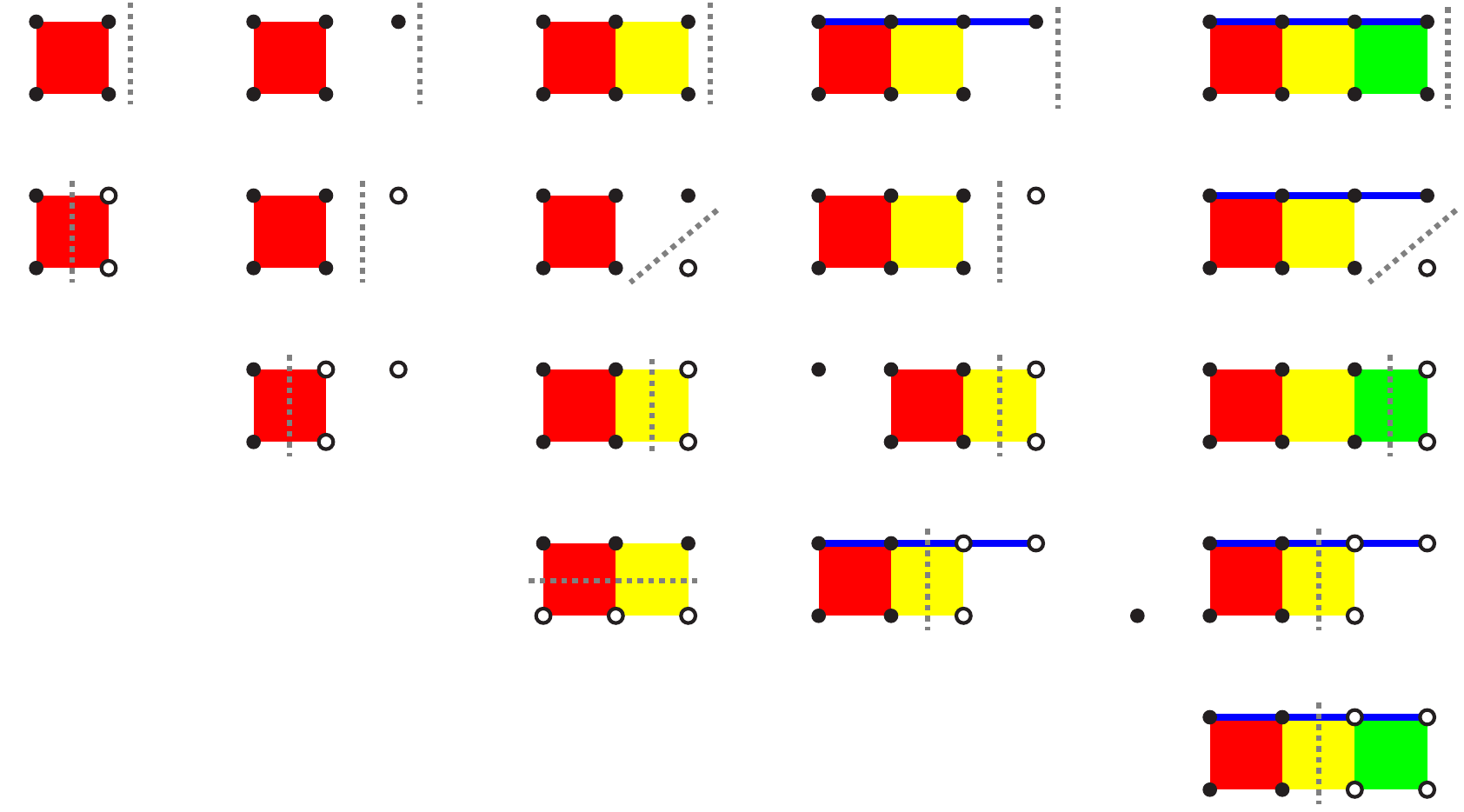}}
    \put(-5,89){\small\boldmath$p{+}q=4$}
    \put(34,89){\bf5}
    \put(66,89){\bf6}
    \put(100,89){\bf7}
    \put(142,89){\bf8}
    \put(-4,73){\footnotesize$\rD_{1+}\rD_{2+}\rD_{3+}\rD_{4+}$}
    \put(-4,54){\footnotesize$\rD_{1+}\rD_{2+}\rD_{1-}\rD_{2-}$}
    \put(22,35){\footnotesize$\rD_{1+}\rD_{2+}\rD_{1-}\rD_{2-}$}
    \put(54,14){\footnotesize$\Bm{\rD_{1+}\rD_{2+}\rD_{1-}\rD_{2-}\\
                                  \rD_{2+}\rD_{3+}\rD_{2-}\rD_{3-}}$}
    \put(86,12){\footnotesize$\Bm{\rD_{1+}\rD_{2+}\rD_{3+}\rD_{4+}\\
                                  \rD_{3+}\rD_{4+}\rD_{1-}\rD_{2-}\\
                                  \rD_{1+}\rD_{3+}\rD_{1-}\rD_{3-}}$}
    \put(-5,0){\parbox[b]{50mm}{\small\rightskip=0pt plus1pc\baselineskip=10pt
     The superderivatives corresponding to a few codes are shown directly beneath the graphical representation of the code. These define quasi-projectors\eq{e:Pab} and so also the ``self-duality'' relations\eq{e:Sab}.}}
  \end{picture}
 \end{center}
 \caption{A graphical method (depicting\eq{e:Pab}-type hallmark $4k$-gon relations) for finding maximal split doubly even binary linear block codes for $4\leqslant(p{+}q)\leqslant8$. {\em\/Warning:\/}~The $4k$-gon colors are independent of edge-colors in Adinkras and serve merely to distinguish the hallmark $4k$-gon relations: here, black vertices depict the $\rD_{\a+}$'s and the white ones the $\rD_{\protect\ad-}$'s|or the other way around.}
 \label{f:1}
\end{figure}

We consider some of the lower values of $p{+}q$, and the possible extension of the worldline supermultiplet projections to their analogue within worldsheet $(p,q)$-supersymmetry. The $4k$-gon graphical method of Ref.\cite{r6-3c} may be adapted to determine the possible ways of splitting the \DELB\ codes, and the result for $p{+}q\leqslant8$ is shown in Figure~\ref{f:1}.

\paragraph{Maximal Projections:}
Already the $p{+}q\leqslant8$ listing, presented graphically in Figure~\ref{f:1} and detailed below, reveals a feature of maximal \sDE\ codes that is unlike the maximal \DELB\ codes as used in Refs.\cite{r6-3,r6-3.2,r6-1.2}! The number of generators of maximal \DELB\ codes equals
\begin{equation}
 \vk(N):=
  \begin{cases}
   0 &\text{for $N<4$},\\
   \big\lfloor\frac{(N-4)^2}{4}\big\rfloor+1 &\text{for $N=4,5,6,7$},\\
   \vk(N{-}8)+4 &\text{for $N>7$, recursively},
  \end{cases}
 \label{e:Kmax}
\end{equation}
and depends only on $N$, the length of the codewords\cite{r6-3}. In turn, for any \DELB\ $[N,k]$-code $\sC$, the chromotopology of a $\sC$-projected supermultiplet is $I^N/(\ZZ_2)^k$ and it has $2^{N-k}$ component (super)fields, half bosonic and half fermionic. Consequently, supermultiplets projected by {\em\/maximal\/} \DELB-codes have $2^{N-\vk(N)}$ component (super)fields and so are {\em\/minimal\/} off-shell supermultiplets of $N$-extended worldline supersymmetry.

In contrast, Figure~\ref{f:1} and the listing\eqs{e:SpC6}{e:SpC8} below show that the number of generators in maximal \sDE\ codes varies for a fixed $p{+}q$, and depends on the $(p,q)$-split.
 In particular, for a specified $(p,q)$-supersymmetry, there exist maximal \sDE\ codes which are not a split of a maximal \DELB\ code, but of a sub-code. Consequently, the total dimension of a minimal, $\ssC$-projected off-shell {\em\/worldsheet\/} supermultiplet is on several occasions strictly larger than $2^{p+q-\vk(p+q)}$.

\paragraph{Decomposing:}
In the projection\eq{e:B484>242}, the graph identification may be taken to either identify the corresponding component (super)fields on the left-hand half with the component (super)fields on the right-hand half, or the negatives thereof. The two resulting ``half-sized'' supermultiplets are distinct|see\eq{e:B242+t} below.
 In general, the so-obtained ``half-sized'' supermultiplets may in fact be inequivalent as\eq{e:B242+t} are, or may turn out to be equivalent through a redefinition of the basis for the component (super)fields and/or the superderivatives, \ie, the supersymmetry generators. For worldline supersymmetry, Ref.\cite{r6-3.2} provides an algorithm to resolve this question on a case-by-case basis; this may have to be revised for application to supermultiplets of worldsheet $(p,q)$-supersymmetry with $p,q\neq0$.

Indeed, a supermultiplet that can be so projected to two ``half-sized'' supermultiplets is said to be {\em\/decomposable\/}, and\eq{e:B484>242} demonstrates that this is equally possible for worldsheet supermultiplets. See the Appendix for the details of this decomposition.

We now read the \sDE\ codes from the graphics in Fig.~\ref{f:1} in turn, and discuss the implications for worldsheet supermultiplets, and so provide a listing of them for worldsheet $(p,q)$-supersymmetry with $p{+}q\leqslant8$, employing to the above-derive constraints.

 \Pgf{4} For the worldline $4$-extended supersymmetry, there is only the $d_4$ code, generated by the single codeword, $1111$. There are then only two possibilities:
 \begin{enumerate}\itemsep=-3pt\vspace{-2mm}

  \item The unidextrously split $d_{4,0}$, for the unidextrous worldsheet $(4,0)$-supersymmetry or its parity-reflection, $d_{0,4}$ for worldsheet $(0,4)$-supersymmetry. A $d_{4,0}$-projected $(4,0)$-supermultiplet must be annihilated by one of the two operators:
\begin{align}
    \big[(i\vd_\pp)^2+\rD_{1+}\rD_{2+}\rD_{3+}\rD_{4+}\big]
  \quad&\text{or}\quad
    \big[(i\vd_\pp^2)-\rD_{1+}\rD_{2+}\rD_{3+}\rD_{4+}\big],
\intertext{or, equivalently, by the systems}
  \big[\rD_{\a+}\rD_{\b+}-\inv2\ve_{\a\b}{}^{\g\d}\rD_{\g+}\rD_{b+}\big]
  \quad&\text{or}\quad
  \big[\rD_{\a+}\rD_{\b+}+\inv2\ve_{\a\b}{}^{\g\d}\rD_{\g+}\rD_{b+}\big].
\end{align}
  Stated another way, on any of the components of a $d_{4,0}$-projected supermultiplet, the action of $\rD_{1+}\rD_{2+}\rD_{3+}\rD_{4+}$ is indistinguishable from either $+\vd_\pp^2$ or $-\vd_\pp^2$. This then imposes component (super)field identifications of the type
  \begin{equation}
    \rD_{1+}\rD_{2+}\rD_{3+}\rD_{4+}\bs\F|=+\vd_\pp^2\bs\F|,\quad\text{or}\quad
    \rD_{1+}\rD_{2+}\rD_{3+}\rD_{4+}\bs\F|=-\vd_\pp^2\bs\F|,\quad\textit{etc.}
  \end{equation}
  The analogous holds for a $d_{0,4}$-projected $(0,4)$-supermultiplet. Thus, the $d_4$-projected worldline $4$-extended supermultiplets can extend both to unidextrous worldsheet $(4,0)$-super\-sym\-metry and to $(0,4)$-supersym\-met\-ry simply by reinterpreting, say, $\ddt\to\vd_\pp$ and $\rD_I\to\rD_{\a+}$.
  
  \item The ambidextrous split, $d_{2,2}$ for worldsheet $(2,2)$-supersymmetry. $d_{2,2}$-projected $(2,2)$-super\-mul\-ti\-plets must be annihilated by one of the two operators:
\begin{align}
    \big[\vd_\pp\vd_\mm+\rD_{1+}\rD_{2+}\rD_{1-}\rD_{2-}\big]
  \quad&\text{or}\quad
    \big[\vd_\pp\vd_\mm-\rD_{1+}\rD_{2+}\rD_{1-}\rD_{2-}\big],
   \label{e:d4Ps}
\intertext{or, equivalently, by the systems}
  \big[\rD_{1+}\rD_{\ad+}+\inv2\ve_{\ad}{}^{\bd}\rD_{2+}\rD_{\bd-}\big]
  \quad&\text{or}\quad
  \big[\rD_{1+}\rD_{\ad+}-\inv2\ve_{\ad}{}^{\bd}\rD_{2+}\rD_{\bd-}\big].
\end{align}
  Stated another way, on any of the components of a $d_{2,2}$-projected supermultiplet|such as\eq{e:B242} \ie,\eq{e:ChTwCh}, the action of $\rD_{1+}\rD_{2+}\rD_{1-}\rD_{2-}$ is indistinguishable from one of $\mp\vd_\pp\vd_\mm$. This then imposes component (super)field identifications of the type
  \begin{equation}
    \rD_{1+}\rD_{2+}\rD_{1-}\rD_{2-}\bs\F|=-\vd_\pp\vd_\mm\bs\F|,\quad\text{or}\quad
    \rD_{1+}\rD_{2+}\rD_{1-}\rD_{2-}\bs\F|=+\vd_\pp\vd_\mm\bs\F|,\quad\textit{etc.}
  \end{equation}
The $(2,2)$-Adinkra\eq{e:B242} depicts such multiplets. As discussed in Ref.\cite{r6-3.2} and above, this graph admits a ``twist,'' whereby the solid/dashed parity assignments of the edges of an odd number of colors is flipped; \eq{e:B242} and its so twisted version are equivalent (by sending $\rD_{2-}\to-\rD_{2-}$):
\begin{equation}
 \vC{\begin{picture}(140,24)
       \put(20,0){\includegraphics[height=25mm]{B242c.pdf}}
         \put(23,3){\Rx{chiral}}
         \put(60,12){\textit{vs.}}
        \put(70,0){\includegraphics[height=25mm]{B242t.pdf}}
         \put(98,3){twisted chiral}
     \end{picture}}
 \label{e:B242+t}
\end{equation}
but {\bsf usefully distinct}: although the transformation $\rD_{2-}\to-\rD_{2-}$ maps one into the other, when used jointly in a model, their coupling provides for features not describable with only one or only the other\cite{rGHR}.
  
  \item None of the $(1,3)$- and $(3,1)$-supermultiplets admits a projection. Conversely, none of the $d_4$-projected supermultiplets of worldline $4$-extended supersymmetry extend to worldsheet either $(1,3)$- or $(3,1)$-supersymmetry.
 \end{enumerate}

To summarize, the minimal off-shell supermultiplets
 \begin{enumerate}\itemsep=-3pt\vspace{-2mm}
 \item of $(4,0)$- $(0,4)$--supersymmetry have 4+4 components and are depicted in\eq{e:N4half};
 \item of $(3,1)$- and $(1,3)$-supersymmetry have 8+8 components;
 \item of $(2,2)$-supersymmetry have 4+4 components and are depicted in\eq{e:B242}.
\end{enumerate}\vspace{-2mm}
It is gratifying that the last case recovers the well-known chiral and twisted chiral supermultiplets|and their complex conjugates, as specified in Corollary~\ref{C:Cpx}.

\Pgf{5} For the worldline $N=5$-extended supersymmetry, there is only the $d_4\oplus t_1$ \DELB\ code, where the $t_n$ summand denotes the trivial (empty) code of length $n$, \ie, the binary codeword $0\cdots0$ with $n$ zeros. This denotes the fact that the fifth supersymmetry generator is not involved in any hallmark $4k$-gon relation. For $p{+}q=5$, there are three maximal split codes\ft{The trivial code $t_{p,q}$ consists of only the split binary codeword with $p{+}q$ zeros: $0\cdots0|0\cdots0$.} (see Figure~\ref{f:1}):
\begin{equation}
  d_{4,0}\oplus t_{1,0}=\pC{11110&},\qquad
  d_{4,0}\oplus t_{0,1}=\pC{1111&0},\qquad
  d_{2,2}\oplus t_{1,0}=\pC{11&110},
  \label{e:d5}
\end{equation}
respectively, for worldsheet $(5,0)$,- $(4,1)$,- and $(3,2)$-supersymmetry.
These are all depicted by means of a tensor product Adinkra such as:
\begin{equation}
 \vC{\begin{picture}(160,35)(0,-1)
   \put(0,0){\includegraphics[width=160mm]{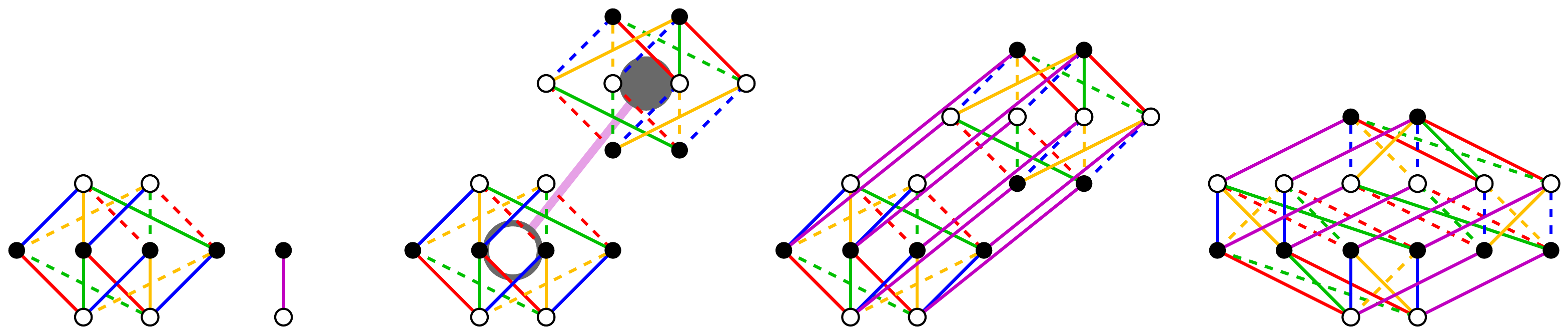}}
    \put(10,19){$\cA_1$}
    \put(27,12){$\cA_2$}
 \end{picture}}
 \label{e:242x11}
\end{equation}
The first two of\eq{e:d5} are straightforward extensions of the $N=5$ worldline $d_4$-projected supermultiplet to the worldsheet $(5,0)$- and $(4,1)$-supersymmetry, respectively, where:
\begin{enumerate}\itemsep=-3pt\vspace{-2mm}
 \item the Adinkra $\cA_1$ depicts a representation of $\SS14_+$ (could be any one of the 8 $N=4$ inequivalent Adinkras listed in\eq{e:N4half} and their boson/fermion flips),
 \item the fifth edge-color and $\cA_2$ depict either the $\rD_{5+}$- or the $\rD_{1-}$-action for either the unidextrous $(5,0)$- or the ambidextrous $(4,1)$-supersymmetry, respectively.
\end{enumerate}

 In the third case, $d_{2,2}\oplus t_{1,0}$ for $(3,2)$-supersymmetry, $\cA_1$ depicts a representation of $\SS{1,1}{2,2}$ (could be any one of the Adinkras\eq{e:121x121}, \eq{e:121x22}, \eq{e:22x121}, \eq{e:B242} and their boson/fermion flips) with the assignments, say: \C1{red\,=\,$\rD_{1+}$}, \C2{green\,=\,$\rD_{2+}$}, \C3{blue\,=\,$\rD_{1-}$}, \C4{orange\,=\,$\rD_{2-}$}, whereupon, say, \C5{purple\,=\,$\rD_{3+}$}.

 Lastly, any off-shell supermultiplet of the unidextrous $(5,0)$-supersymmetry may always be extended to a left-moving (unidextrous) supermultiplet of the ambidextrous worldsheet $(5,q)$-supersymmetry, for arbitrary $q$; all such supermultiplets are on the half-shell, \ie, are annihilated by $\vd_\mm$. As this can always be done with off-shell representations of $(p,0)$- and $(0,q)$-supersymmetry, it will no longer be pointed out explicitly.

Thus, minimal supermultiplets of $(p,5{-}p)$-supersymmetry all have 8+8 components, for all choices $0\leqslant p\leqslant5$. Recall however that there do exist adinkraic supermultiplets with 16 bosonic and 16 fermionic component (super)fields that do not decompose into direct sums of minimal supermultiplets, the prime example being the straightforward, $(p,5{-}p)$-supersymmetric generalization of\eq{e:121x121}.

It is quite evident that the resulting Adinkra\eq{e:242x11} is 1-color-decomposable, since deleting the purple, $\rD_{3-}$-edges decomposes the Adinkra. However, when deleting edges of any color other than the fifth one (purple), one must delete edges of {\em\/two\/} colors for the Adinkra to decompose|since the the factor-Adinkra $\cA_1$ is 2-color-decomposable.

\Pgf{6}For the worldline $N=6$-extended supersymmetry, $d_6$ is the maximal \DELB\ code. For $p{+}q=6$, there are four maximal split codes (see Figure~\ref{f:1}):
\begin{equation}\itemsep=-3pt\vspace{-2mm}
  \begin{array}{@{} rc@{\qquad}c@{\qquad}c@{\qquad}c @{}}
 \text{\bsf Supersymmetry:}
  & \bs{(6,0)} & \bs{(5,1)} & \bs{(4,2)} & \bs{(3,3)}\\[1mm]
  \midrule
 \MR2*{{\bsf Maximal Code:}}
  & d_{6,0} & d_{4,0}\oplus t_{1,1} & d_{4,2} & d_{3,3} \\
  &\pC{111100&\\001111&}
  &\pC{11110&0}
  &\pC{1111&00\\[-1pt]0011&11}
  &\pC{110&110\\[-1pt]011&011}\\[2mm]
  \midrule
 \text{\bsf Minimal Dim.:} & (4|8|4) & (8|16|8) & (4|8|4) & (4|8|4) 
  \end{array}
 \label{e:SpC6}
\end{equation}
Notice that the minimal supermultiplets of the unidextrous $(5,1)$- and $(1,5)$-supersymmetry have $16+16$ component (super)fields, as opposed to the half as large, $(8|8)$-dimensional (super)fields of the other $(p,6{-}p)$-supersymmetry cases.

For illustration, a minimal $d_{3,3}$-projected off-shell supermultiplet of worldsheet $(3,3)$-super\-sym\-metry may be constructed by decomposing, in the manner of\eq{e:B484>242}, the tensor product of two $N=3$ so-called {\em\/valise\/} supermultiplets:
\begin{equation}
 \vC{\begin{picture}(140,22)(0,-2)
   \put(0,0){\includegraphics[width=140mm]{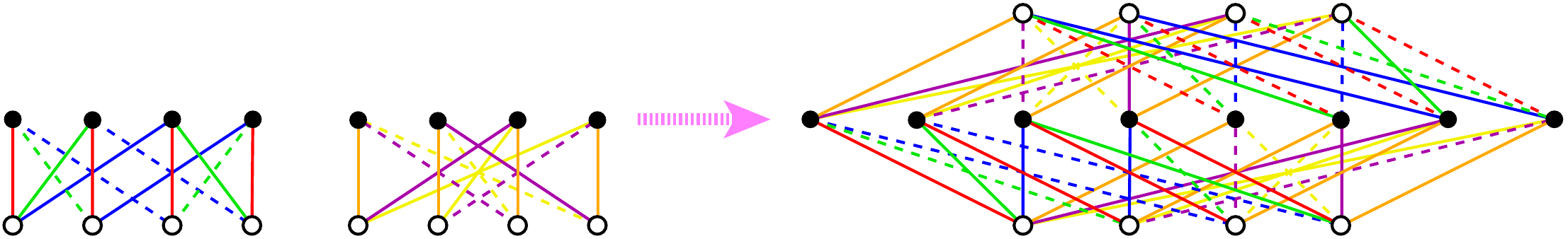}}
    \put(10,13){$\cA_+$}
    \put(41,13){$\cA_-$}
    \put(25,4.5){\Large$\otimes$}
    \put(57,13){\small$/d_{3,3}$}
 \end{picture}}
 \label{e:D33-484}
\end{equation}
Since the $d_{3,3}$ code has two generators, the ``$/d_{3,3}$'' annotation denotes a $d_{3,3}$-encoded $(\ZZ_2)^2$-quotient, which reduces the $(4|4)\otimes(4|4)=(16|32|16)$-dimensional tensor product $(3,3)$-Adinkra (not shown in\eq{e:D33-484}) to the $(4|8|4)$-dimensional one shown. Conceptually, this is the $(3,3)$-supersymmetric generalization of the $(2,2)$-supersymmetric construction\eqs{e:22x22}{e:B242}. The analogous $d_{4,2}$-quotient $(4,2)$-supermultiplet is obtained in a similar way:
\begin{equation}
 \vC{\begin{picture}(140,22)(0,-2)
   \put(0,0){\includegraphics[width=140mm]{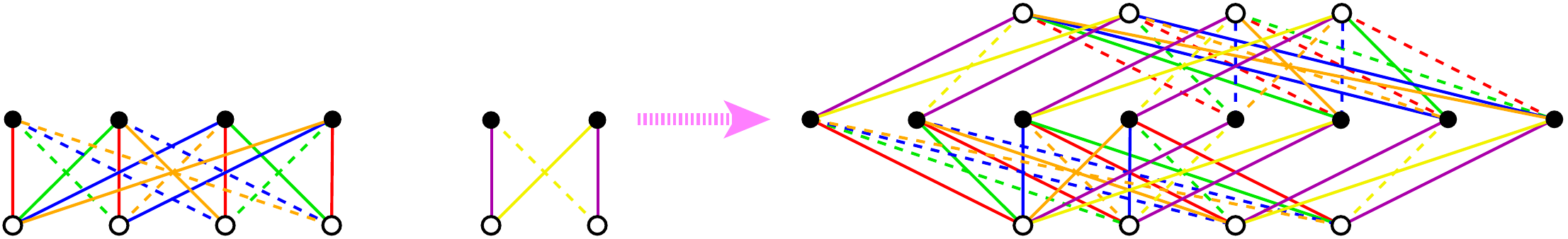}}
    \put(13,13){$\cA_+$}
    \put(47,13){$\cA_-$}
    \put(34,4.5){\Large$\otimes$}
    \put(57,13){\small$/d_{4,2}$}
 \end{picture}}
 \label{e:D42-484}
\end{equation}
Note that the left-hand factor in the tensor product is already a $d_{4,0}$-quotient, so that this subcode $d_{4,0}\subset d_{6,2}$ acts trivially when passing to the final result in\eq{e:D42-484}.

 Clearly, there exist many $(16+16)$- and $(32+32)$-dimensional representations which do not exhibit two commuting $\ZZ_2$ symmetries|because of the vertical positioning of the nodes|so as to be so decomposed. The simplest example is the intact supermultiplets, which are the straightforward, $(p,6{-}p)$-supersymmetric generalizations of\eq{e:121x121}.
 These supermultiplets are {\em\/reducible\/}, in that they may be reduced to smaller supermultiplets by means of (now {\em\/two\/} mutually commuting sets of) self-dual type relations such as\eq{e:Sab} and akin to the procedure shown in Figure~\ref{f:int>tch}. These smaller, reduced supermultiplets always have a higher $n$-color-decomposability than the bigger ones, prior to the reduction.

\Pgf{7}For the worldline $N=7$-extended supersymmetry, $e_7$ is the maximal \DELB\ code. For $p{+}q=7$, there are four maximal split codes (see Figure~\ref{f:1}):
\begin{equation}
  \begin{array}{@{} rc@{\qquad}c@{\qquad}c@{\qquad}c @{}}
 \text{\bsf Supersymmetry:}
  & \bs{(7,0)} & \bs{(6,1)} & \bs{(5,2)} & \bs{(4,3)}\\[1mm]
  \midrule
 \MR2*{{\bsf Maximal Code:}}
  & d_{7,0} & d_{6,0}\oplus t_{0,1} & d_{4,2}\oplus t_{1,0} & e_{4,3} \\
  &\pC{1111000&\\[-1pt]0011110&\\[-1pt]1010101}
  &\pC{111100&0\\[-1pt]001111&0}
  &\pC{01111&00\\[-1pt]00011&11}
  &\pC{1111&000\\[-1pt]0011&110\\[-1pt]1010&101}\\[2mm]
  \midrule
 \text{\bsf Minimal Dim.:} & (8|8) & (16|16) & (16|16) & (8|8) 
  \end{array}\qquad\qquad
 \label{e:SpC7}
\end{equation}
Recall that there do exist adinkraic supermultiplets with up to 64 bosonic and 64 fermionic component (super)fields that do not decompose into direct sums of minimal supermultiplets. The simplest example is the intact supermultiplets, which are the straightforward, $(p,7{-}p)$-supersymmetric generalizations of\eq{e:121x121}.
 These supermultiplets {\em\/can\/} be reduced by means of the self-duality type equations such as\eq{e:Sab}.

\Pgf{8}For the worldline $N=8$-extended supersymmetry, $e_8$ is the maximal \DELB\ code. For $p{+}q=8$, there are five maximal split codes (see Figure~\ref{f:1}):
\begin{equation}
  \begin{array}{@{} rc@{\qquad}c@{\qquad}c@{\qquad}c@{\qquad}c @{}}
 \text{\bsf Supersymmetry:}
  &\bs{(8,0)} & \bs{(7,1)} & \bs{(6,2)} & \bs{(5,3)} & \bs{(4,4)}\\[1mm]
  \midrule
 \MR2*{{\bsf Maximal Code:}}
  & d_{8,0} & e_{7,0}\oplus t_{0,1} & d_{6,2} & e_{4,3}\oplus t_{1,0} & e_{4,4} \\
  &\pC{11110000&\\[-1pt]00111100&\\[-1pt]00001111\\[-1pt]10101010}
  &\pC{1111000&0\\[-1pt]0011110&0\\[-1pt]1010101&0}
  &\pC{111100&00\\[-1pt]001111&00\\[-1pt]000011&11}
  &\pC{11110&000\\[-1pt]01010&101\\[-1pt]00110&110}
  &\pC{1111&0000\\[-1pt]0011&1100\\[-1pt]0000&1111\\[-1pt]1010&1010}\\[2mm]
  \midrule
 \text{\bsf Minimal Dim.:} & (8|8) & (16|16) & (16|16) & (16|16) & (8|8) 
  \end{array}
 \label{e:SpC8}
\end{equation}
For illustration, a minimal $e_{4,4}$-projected off-shell supermultiplet of worldsheet $(4,4)$-supersymmetry may be constructed by decomposing, in the manner of\eq{e:B484>242}, the tensor product of two $N=4$ valises:
\begin{equation}
 \vC{\begin{picture}(140,22)(0,-2)
   \put(0,0){\includegraphics[width=140mm]{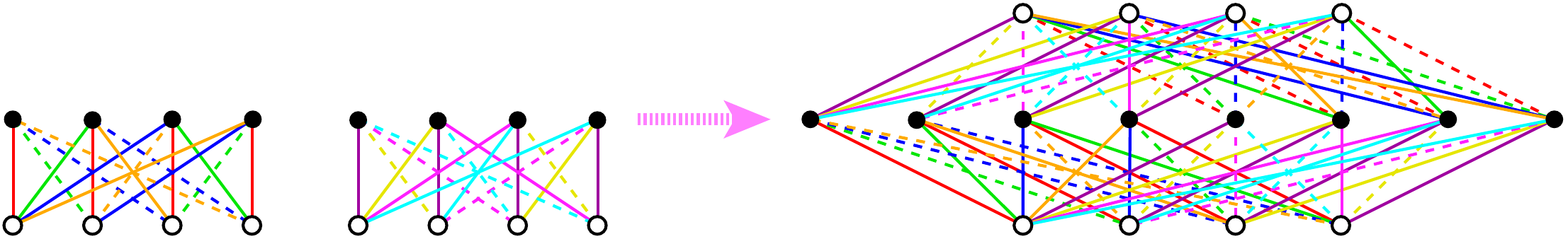}}
    \put(10,13){$\cA_+$}
    \put(41,13){$\cA_-$}
    \put(25,4.5){\Large$\otimes$}
    \put(57,13){\small$/e_{4,4}$}
 \end{picture}}
 \label{e:D44-484}
\end{equation}
Since the $e_{4,4}$ code has four generators, the ``$/e_{4,4}$'' quotient denotes a $e_{4,4}$-encoded $(\ZZ_2)^4$-quotient. However, both factors in the tensor product are already $d_{4,0}$,- \ie, $d_{0,4}$-quotients, respectively. These subcodes
\begin{equation}
 d_{4,0}, d_{0,4}\subset e_{4,4}:\qquad
 \pC{1111&0000}\oplus\pC{0000&1111}\subset
 \pC{1111&0000\\[-1pt]0011&1100\\[-1pt]0000&1111\\[-1pt]1010&1010}
\end{equation}
therefore act trivially in passing to the final quotient in\eq{e:D44-484}. Thus, the  $(4|4)\otimes(4|4)=(16|32|16)$-dimensional tensor product $(4,4)$-Adinkra (not shown in\eq{e:D44-484}) is reduced only by the $(\ZZ_2)^2$-quotient encoded by the two generators of $e_{4,4}/(d_{4,0}+d_{0,4})$, producing the $(4|8|4)$-dimensional Adinkra shown.

It is worth noting that in the Adinkra\eq{e:D44-484}, the factor $\cA_+$ is chiral and $\cA_-$ is twisted-chiral, as can be verified by checking the hallmark $4k$-gon relations\eq{e:d4D+} and\eq{e:d4D-}; these relations remain valid in the final quotient of the tensor product $(4,4)$-Adinkra. By transforming $\rD_{4-}\to-\rD_{4-}$, $\cA_-$ is turned into a chiral Adinkra, and the tensor product $(4,4$)-Adinkra also changes into its twisted variant; an equivalent result is obtained by transforming $\rD_{4+}\to-\rD_{4+}$, which turns $\cA_+$ into a twisted-chiral Adinkra. In fact, the simultaneous change $(\rD_{4+},\rD_{4-})\to(-\rD_{4+},-\rD_{4-})$ can always be compensated by a judicious sign-change in some component (super)fields, and so leads to an equivalent supermultiplet.

As before, there exist adinkraic supermultiplets with up to 128 bosonic and 128 fermionic component (super)fields that do not decompose into direct sums of minimal supermultiplets, the prime example being the intact supermultiplet, which is the straightforward, $(p,8{-}p)$-supersymmetric generalization of\eq{e:121x121}.
 These supermultiplets {\em\/can\/} be reduced by means of the self-duality type equations such as\eq{e:Sab}. As before, iterated $\ZZ_2$ projection increases $n$-color-decomposability, and it is not hard to see that the end result in\eq{e:D44-484} is 4-color-decomposable.

\subsection{Ground Fields {\protect\itshape vs.}\ Tensor Products}
\label{s:FnF}
In the discussion\eqs{e:Cpx1}{e:CpxSuSy} leading to Corollary~\ref{C:Cpx}, it was shown that certain supermultiplets admit a complex structure. This means that:
\begin{enumerate}\itemsep=-3pt\vspace{-2mm}
 \item pairs of nodes at the same height may be combined into a ``complex node,'' which depict pairs of component fields combined as real and imaginary parts of a complex component field,
 \item two pairs of edges (of two colors) connecting two ``complex nodes'' at adjacent heights combine into a single ``complex edge,'' depicting the complex supersymmetry transformation from one complex component field to another, as depicted in\eq{e:CpxA}.
\end{enumerate}
Combining nodes and edges in this manner in the chiral Adinkra shown in the left of\eq{e:B242}, we obtain that the real $(2|4|2)$-dimensional supermultiplet of real $(2,2)$-supersymmetry may also be thought of as a complex $(1|2|1)$-dimensional supermultiplet of complex $(1,1)$-supersymmetry. In turn, reverse-engineering Construction~\ref{C:RxR}, this complex $(1|2|1)$-dimensional $(1,1)$-supermultiplet is obtainable as the (external) tensor product of two complex $(1|1)$-dimensional supermultiplets:
\begin{equation}
  (2|4|2)^{\sss(2,2)}_\IR = (1|2|1)^{\sss(1,1)}_\IC
  = (1|1)^1_{\IC+} \otimes (1|1)^1_{\IC-}.
 \label{e:C11Sq}
\end{equation}
Note that one transforms this into the twisted chiral supermultiplet by changing the sign of an odd number of real $\rD_I$-actions. By choosing this to be the one identified as the imaginary part of the complex $\BD_-$-action, we effectively conjugate the complex structure of the $(1|1)^1_{\IC-}$ factor.

 On the other hand, we have that
\begin{enumerate}\itemsep=-3pt\vspace{-2mm}
 \item The $(2|4|2)^{\sss(2,2)}_\IR$ supermultiplet is the $d_{2,2}$-specified $\ZZ_2$ quotient of the $(4|8|4)^{\sss(2,2)}_\IR$ supermultiplet; see\eq{e:B484>242}.
 \item The $(4|8|4)^{\sss(2,2)}_\IR$ supermultiplet is the (real) tensor product of a left- and a right-handed copy of the $(2|2)^2_\IR$ supermultiplet; see\eq{e:22x22}.
\end{enumerate}
This proves that
\begin{alignat}9
  (2|2)^2_{\IR+} \otimes (2|2)^2_{\IR-}
 &\too{~/d_{2,2}~} \5{\sss\text{chiral\qquad}}{(2|4|2)^{\sss(2,2)}_\IR}
 &= (1|2|1)^{\sss(1,1)}_\IC
 &= (1|1)^1_{\IC+} \otimes (1|1)^1_{\IC-}, \label{e:Ch>CC}\\
  (2|2)^2_{\IR+} \otimes (2|2)^2_{\IR-}
 &\too{~/d_{2,2}~} \5{\sss\text{tw.-chiral\qquad}}{(2|4|2)^{\sss(2,2)}_\IR}
 &= (1|2|1)^{\sss(1,\bar1)}_\IC
 &= (1|1)^1_{\IC+} \otimes \ba{(1|1)^1_{\IC-}},
 \label{e:R<C}
\end{alignat}
where over-bar indicates complex conjugation; recall that left- and right-handed objects can be handled independently on the worldsheet.

 That is, the chiral Adinkra shown in the left of\eq{e:B242} and indicated in the middle of\eq{e:Ch>CC} is not a real tensor product of real supermultiplets, but {\em\/is\/} a complex tensor product of complex supermultiplets. For brevity, we will say that the chiral Adinkra shown in the left of\eq{e:B242} does not factorize over $\IR$, but does factorize over $\IC$. Moreover, it follows that the 2-color decomposability and other details of the connectivity specified by the $d_{2,2}$ \sDE\ code in the chiral Adinkra stem from the complex tensor product\eq{e:R<C}.

It is then reasonable to ask:
 ({\small\bf1})~which real quotients of real tensor products turn out to factorize over $\IC$ or $\IH$, and
 ({\small\bf2})~if there exist real quotients of real tensor products that factorize over no ground field.
 We now turn to answer these, at least within the scope of the examples presented herein.

\paragraph{Complex Tensor Products:}
Modeling on the extended equality\eq{e:R<C} considered in reverse, considering here only a dimension-count and assuming the supermultiplets to admit the indicated complex structures, we compute:
\begin{equation}
    (d_1|d_1)^p_{\IC+} \otimes (d_2|d_2)^q_{\IC-}
  = (d_1{\cdot}d_2|2{\cdot}d_1{\cdot}d_2|d_1{\cdot}d_2)^{p,q}_\IC
  = (2{\cdot}d_1{\cdot}d_2|4{\cdot}d_1{\cdot}d_2|2{\cdot}d_1{\cdot}d_2)^{2p,2q}_\IR.
 \label{e:CC>R0}
\end{equation}
In turn, we know that, for any particular number $N$ of edge-colors, Adinkras are largest when not projected, and have $2^{N-1}+2^{N-1}$ nodes. Selecting $d_1=2^{p-1}$ and $d_2=2^{q-1}$, we have
\begin{equation}
    (2^{p-1}|2^{p-1})^p_{\IC+} \otimes (2^{q-1}|2^{q-1})^q_{\IC-}
  = (2^{p+q-2}|2^{p+q-1}|2^{p+q-2})^{p,q}_\IC
  = (2^{p+q-1}|2^{p+q}|2^{p+q-1})^{2p,2q}_\IR.
 \label{e:CC>R}
\end{equation}
Since an intact, real $2(p{+}q)$-supermultiplet is $(2^{2p+2q-1}|2^{2p+2q-1})$-dimensional and the result\eq{e:C>R} is only $(2^{p+q}|2^{p+q})$-dimensional, it must be that\eq{e:CC>R} is a $(\ZZ_2)^{p{+}q{-}1}$-quotient, which had to have been specified by a $(p,q)$-split \sDE\ code with $p{+}q{-}1$ generators. If the quotient admits a complex structure and factorizes over $\IC$, it must be that
\begin{equation}
 \begin{aligned}
  (2^{2p-1}|2^{2p-1})^{2p}_{\IR+} \otimes (2^{2q-1}|2^{2q-1})^{2q}_{\IR+}
  \too{~/\ssC_{\!2p,2q}~}~&
  (2^{p+q-1}|2^{p+q}|2^{p+q-1})^{2p,2q}_\IR\\
  &=(2^{p-1}|2^{p-1})^p_{\IC+} \otimes (2^{q-1}|2^{q-1})^q_{\IC-}
 \end{aligned}
 \label{e:R>CC}
\end{equation}
for $0\leqslant p,q\in\ZZ$, and for some rank-$(p{+}q{-}1)$ \sDE\ code $\ssC_{2p,2q}$.

 In this sense, such \sDE\ codes $\ssC_{\!2p,2q}$ may be said to stem from complex tensor products, and in all the cases considered herein, $\ssC_{\!2p,2q}=d_{2p,2q}$ in fact. Since the rank-$(p{+}q{-}1)$ \sDE\ codes $d_{2p,2q}$ are maximal except when $p{+}q\id0\,\text{(mod~8)}$, and Adinkras with $N$ edge-colors have at most $(2^{N-1}|2^{N-1})$ nodes, it follows that the relation\eq{e:R>CC} is saturated except when $p{+}q\id0\,\text{(mod~8)}$:
 On the right-hand sideof\eq{e:R>CC}, factor Adinkras with more nodes would have to decompose into direct sums, and Adinkras with fewer nodes would require a quotient code violating\eq{e:Kmax}.
 On the left-hand side of\eq{e:R>CC}, the factor Adinkras can be smaller only if they have been ``pre-quotiented'' by some \sDE\ code $Z'$. Then, the \sDE\ code $\ssC_{\!2p,2q}$ could always be made to subsume this $Z'$, reverting the left-hand side factor Adinkras into the intact ones used in\eq{e:R>CC}. Since the indicated projection code is maximal, the left-hand side of\eq{e:R>CC} is also saturated. The exceptional cases with $p{+}q\id0\,\textrm{(mod~8)}$ turn out to involve quaternions; see\eq{e:H+-}.

Suffice it here to recall that the conditions of Corollary~\ref{C:Cpx} and relations\eqs{e:CC>R}{e:R>CC} are necessary. While finding the sufficient conditions is of considerable interest, it turns out to be rather involved (see below), and will have to be deferred to a later effort.

\paragraph{Quaternionic Tensor Products:}
Although satisfying both the conditions of  Corollary~\ref{C:Cpx} and relations\eqs{e:CC>R}{e:R>CC}, the $(4|4$)-dimenssional (valise) supermultiplets in the left-hand half of\eq{e:ChTCh} do not admit a simple complex structure, but a quaternionic structure instead. For example, by judicious identification of the nodes in the lower left-most (chiral valise) Adinkra
\begin{equation}
  \vC{\begin{picture}(60,20)(0,4)
       \put(-5,.5){\includegraphics[width=60mm]{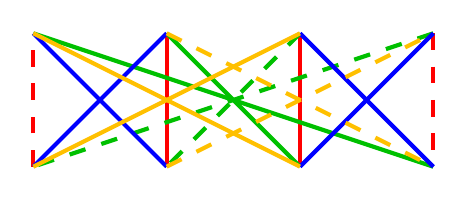}}
       \put(0,21){\cB{$\j_2$}}
       \put(17,21){\cB{$\j_3$}}
       \put(34,21){\cB{$\j_4$}}
       \put(50,21){\cB{$\j_1$}}
       \put(0,5){\cB{$\f_3$}}
       \put(17,5){\cB{$\f_2$}}
       \put(34,5){\cB{$\f_1$}}
       \put(50,5){\cB{$\f_4$}}
      \end{picture}}
  \label{e:ChV}
\end{equation}
the supersymmetry transformation rules are completely captured by the quaternionic equation
\begin{equation}
   \inv4[\C3{\rD_1}+i\C4{\rD_2}+j\C2{\rD_3}+k\C1{\rD_4}]
    \,{\times}\,(\f_1+i\f_2+j\f_3+k\f_4)
   =(i\j_1)+i(i\j_2)+j(i\j_3)+k(i\j_4).
 \label{e:ChVQ}
\end{equation}
The imaginary unit $i$ plays a double role: both as one of the three quaternionic units, and as the imaginary unit from the right-hand side of\eq{e:pqSuSy}. This can be rearranged \begin{equation}
   \inv4[(\C3{\rD_1}{+}i\C4{\rD_2})+k(\C1{\rD_4}{+}i\C2{\rD_3})]
    \,{\times}\,[(\f_1{+}i\f_2)+k(\f_4{+}i\f_3)]
   =\big((i\j_1){+}i(i\j_2)\big)+k\big((i\j_4){+}i(i\j_3)\big)
 \label{e:ChV2c}
\end{equation}
to indicate the $i$-complex combinations,
\begin{equation}
  [\C3{\bs{D}_{12}}{+}k\C1{\bs{D}_{43}}]\,{\times}\,(\bs\f_{12}{+}k\bs\f_{43})
  =(i\bs\j_{12}){+}k(i\bs\j_{43}),
 \label{e:ChViC}
\end{equation}
where $\C4{\rD_2}$ and $\C2{\rD_3}$ play the roles of $i$-imaginary parts and correspond to the edges criss-crossing between the two sides in\eq{e:ChV}. Expanding the $k$-real and $k$-imaginary parts of\eq{e:ChViC}, we obtain a $k$-complex $(2|2)$-dimensional supermultiplet effectively depicted by the Adinkra\eq{e:22}, where however both the nodes and the edges are already $i$-complexified. The two ($i$- and $k$-)complex structures\eq{e:ChV2c} admitted by the real Adinkra\eq{e:ChV} are independent and generate the quaternionic structure\eq{e:ChVQ}:
\begin{equation}
  (4|4)^4_\IR = (2|2)_\IC^2 = (1|1)_\IH^1.
\end{equation}

Considering again only a dimension-count and assuming the supermultiplets to admit the indicated quaternionic structures, we retrace the equality\eq{e:C>R} with quaternionic tensor products on the left-hand side instead of complex ones:
\begin{equation}
    (2^{p-1}|2^{p-1})^p_{\IH+} \otimes (2^{q-1}|2^{q-1})^q_{\IH-}
  = (2^{p+q-2}|2^{p+q-1}|2^{p+q-2})^{p,q}_\IH
  = (2^{p+q}|2^{p+q+1}|2^{p+q})^{4p,4q}_\IR.
 \label{e:C>R}
\end{equation}
Since a real Adinkra with $N$ edge-colors can have no fewer than $2^{N-\vk(N)-1}$ white (and as many black) nodes, and must decompose into a direct sum of Adinkras if it has more than $2^{N-1}$ white (and as many black) nodes, we impose the extended condition
\begin{equation}
  2^{4p+4q-\vk(4p+4q)-1} \leqslant 2^{p+q+1} \leqslant 2^{4p+4q-1},
\end{equation}
which is satisfied only when $p{+}q=1,2$. This produces only one ambidextrous case:
\begin{equation}
  (8|8)^4_{\IR+} \otimes (8|8)^4_{\IR+}
  \too{~e_{4,4}~} (4|8|4)^{\sss(4,4)}_\IR
  =(1|1)^1_{\IH+} \otimes (1|1)^1_{\IH-}
 \label{e:H+-}
\end{equation}
and several unidextrous cases. Allowing for the quaternionic Adinkras to decompose as direct products again permits many more cases, and we defer their study to a subsequent effort.

\paragraph{Read Quotients {\slshape vs.}\ (Hyper-)Complex Tensor Products:}
Summarizing the results\eqs{e:C11Sq}{e:H+-}, Table~\ref{t:R=C} lists the nontrivial real quotients $(\fR_+\otimes\fR_-)/(\ZZ_2)^k$ discussed above, which factorize as (hyper-)complex tensor products.
\begin{table}[htbp]
  \centering
  \begin{tabular}{@{} rl@{~}c@{~}l @{}}
\bsf Real Adinkra & \MC1l{\bsf Listed}  && \bsf Factorization \\ 
    \toprule
 chiral $(2|4|2)^{\sss(2,2)}_\IR$
 & (\ref{e:B242})-right, (\ref{e:ChTwCh})-right, (\ref{e:B242+t})-left
 &=& $(1|1)^1_{\IC+}\otimes(1|1)^1_{\IC-}$ \\[1mm]
 twisted-chiral $(2|4|2)^{\sss(2,2)}_\IR$
 & (\ref{e:B242})-left, (\ref{e:ChTwCh})-left, (\ref{e:B242+t})-right
 &=& $(1|1)^1_{\IC+}\otimes\ba{(1|1)^1_{\IC-}}$ \\
    \midrule
 $(4|8|4)^{\sss(4,2)}_\IR$
 & (\ref{e:D42-484})
 &=& $(2|2)^2_{\IC+}\otimes(1|1)^1_{\IC-}$ \\
    \midrule
 \MC2l{$\big((32|32)^6_{\IR+}\otimes(2|2)^2_{\IR-}\big)/d_{6,2}
        =(8|16|8)^{\sss(6,2)}_\IR$;~~ see\eq{e:SpC8}}
 &=& $(4|4)^3_{\IC+}\otimes(1|1)^1_{\IC-}$ \\
    \midrule
 ultra multiplet\cite{rFGH}$^\ddag$
 & (\ref{e:D44-484})
 &=& $(1|1)^1_{\IH+}\otimes(1|1)^1_{\IH-}$ \\
    \bottomrule
 \MC4l{\parbox{140mm}{\raggedright\footnotesize\baselineskip=9pt$^\ddag$\,As shown in\eq{e:D44-484}, the worldline ultra multiplet of Ref.\cite{rFGH} extends to the worldsheet.}}
  \end{tabular}\vspace*{-1mm}
  \caption{Non-trivial real quotient Adinkras that are unprojected (hyper-)complex tensor products}
  \label{t:R=C}
\end{table}

More importantly however, note that the supermultiplet\eq{e:D33-484}
\begin{equation}
  (4|4)^3_{\IR+} \otimes (4|4)^3_{\IR-} \too{/d_{3,3}} (4|8|4)^{\sss(3,3)}_\IR
  \tag{$\ref{e:D33-484}'$}
 \label{e:?}
\end{equation}
cannot factorize over (hyper-)complex numbers, for the simple reason that the odd numbers (three each) of $\rD_{\a+}$- and $\rD_{\ad-}$-actions cannot be combined into (hyper-)complex $\BD_\pm$-actions. The same reasoning shows that the examples
\begin{alignat}9
  (16|16)^5_{\IR+} \otimes (2|2)^2_{\IR-}
 & \too{/(d_{4,2}\oplus t_{1,0})}&~& (8|16|8)^{\sss(5,2)}_\IR, \label{e:d42}\\
  (8|8)^4_{\IR+} \otimes (4|4)^3_{\IR-}
 & \quad\too{/e_{4,3}}&~& (4|8|4)^{\sss(4,3)}_\IR,\\
  (16|16)^5_{\IR+} \otimes (4|4)^3_{\IR-}
 & \too{/(e_{4,3}\oplus t_{1,0})}&~& (8|16|8)^{\sss(5,3)}_\IR, \label{e:e43}
\end{alignat}
are all real quotients of real tensor products of real representations that are not themselves tensor products over either of ground fields $\IR,\IC,\IH$. In fact, five of the nine ambidextrous codes for $4\leqslant p{+}q\leqslant8$ give rise to such non-factorizable real quotients of real tensor products, indicating that a seizable fraction of worldsheet supermultiplets obtained by Construction~\ref{C:pxq>pq} do not factorize over either of ground fields $\IR,\IC,\IH$.

The explicit examples\eq{e:?} and\eqs{e:d42}{e:e43} then suffice to prove:
\begin{corl}
There exist real, \sDE-code $\ssC$-encoded quotients $(\fR_+\otimes\fR_-)/\ssC$ of tensor products of left- and right-handed representations, $\fR_+$ of $\SS1p_+$ and $\fR_-$ of $\SS1q_-$, which are not themselves tensor products over any of the ground fields $\IR,\IC,\IH$, and are off-shell representations of $\SS{1,1}{p,q}$.
\end{corl}
As compared with the standard representation theory of Lie algebras, this result may well come as a surprise.

\subsection{Worldsheet Adinkra Degeneracy 1}
\label{s:1}
Ref.\cite{r6-3.2} shows that worldline supermultiplets with different chromotopology may nevertheless be equivalent, and provides both a criterion for this to happen and an explicit isomorphism. Whereas this type of equivalence evidently extends to chiral worldsheet $(N,0)$- and $(0,N)$-supersymmetry, it is nontrivial to deterimine under what circumstances|and if at all|this type of equivalence can extend to ambidextrous worldsheet $(p,q)$-supersymmetry.

As an example, consider the worldsheet $(8,2)$-supermultiplets with the chromotopology
\begin{equation}
  I^{8,2}/d_{8,2}\qquad\textit{vs.}\qquad I^{8,2}/(e_{8,0}\oplus t_{0,2}).
\end{equation}
where $t_{0,2}=\bf0$ is the trivial code of length 2, \ie, $00$.
The explicit proof of the supermultiplet equivalence\cite{r6-3.2} starts with valise supermultiplets, which are in 1--1 correspondence with the known representations of Clifford algebras.
 Since in such Adinkras all bosonic nodes are on one level and all fermionic on another level, any attempt at an extension to any ambidextrous worldsheet supersymmetry would be ruled out by the twin theorems of Ref.\cite{rGH-obs} and the consequent extension criterion. Furthermore, we also have that
\begin{equation}
  I^{8,2}/(e_{8,0}\oplus t_{0,2}) = (I^8/e_8)_+\otimes(I^2/t_2)_-
   = (I^8/e_8)_+\otimes(I^2)_- = I^{8,2}/e_{8,0}
\end{equation}
is a left-right tensor product worldsheet representation as obtained in Construction~\ref{C:pxq>pq}, whereas $I^{8,2}/d_{8,2}$ is not, making an isomorphism unlikely. Nevertheless, in view of the somewhat surprising equivalence mapping discovered in Ref.\cite{r6-3.2}, it behoves to explore this a little further. First, akin to\eq{e:22x22}, we construct
\begin{equation}
 (I^8/e_8)_+\otimes(I^2)_- =~
 \vC{\begin{picture}(100,26)(3,0)
   \put(1,1.2){\includegraphics[width=100mm]{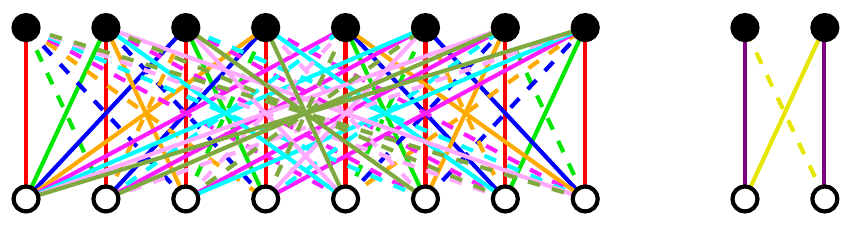}}
    \put(77,13){\Large$\otimes$}
 \end{picture}}
 \label{e:E8xxI2}
\end{equation}
where generators of the $e_8$ code in the left factor are found by tracing (closed) hallmark 4-color tetragons; a convenient basis is given by:
\begin{equation}
  e_{8,0}\oplus t_{0,2} =
  e_{8,0} =\pC{1100\,1100&00\\[-2pt]0110\,0110&00\\[-2pt]
               0011\,0011&00\\[-2pt]1111\,0000&00\\[-2pt]}
  =\vC{\includegraphics[width=30mm]{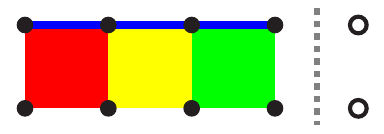}},
 \label{e:e8,0}
\end{equation}
and the product results in
\begin{equation}
 \vC{\begin{picture}(160,30)
   \put(0,0){\includegraphics[width=160mm]{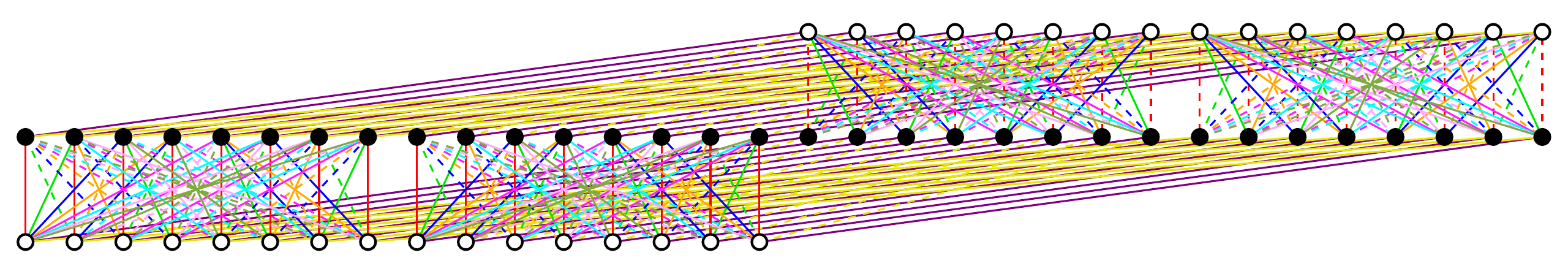}}
    \put(3,22){$I^{8,2}/e_{8,0}$}
 \end{picture}}
 \label{e:E8xI2}
\end{equation}
where the $e_8$ code encodes relations entirely amongst the $\rD_{\a+}$.

On the other hand, $I^{8,2}/d_{8,2}$ is not a tensor product but a quotient thereof, since the $d_{8,2}$ code involves all the $\rD_{\a+},\rD_{\ad-}$.
 However, a $d_{8,0}$ subcode encodes relations entirely amongst the $\rD_{\a+}$, whereupon a fourth generator must encode the mixed $d_{2,2}$-type relations such as\eq{e:ddD0}. Thus, we can construct
\begin{equation}
 (I^8/d_8)_+\otimes(I^2)_- =~
 \vC{\begin{picture}(130,22)(3,0)
   \put(1,1.2){\includegraphics[width=130mm]{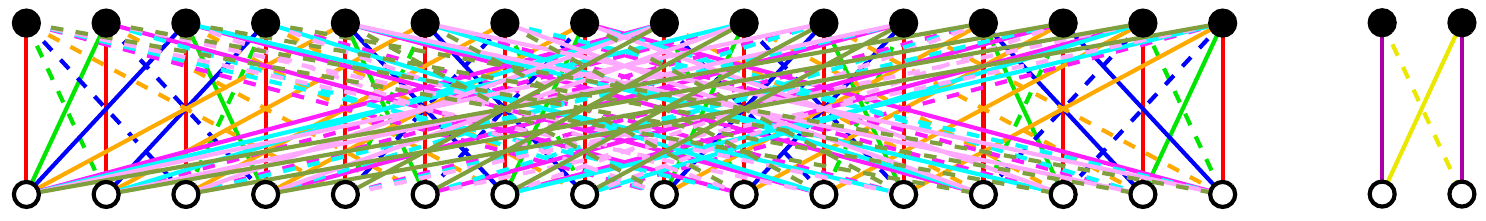}}
    \put(113,9){\Large$\otimes$}
 \end{picture}}
 \label{e:D8xI2}
\end{equation}
and then impose that final, mixed relation to obtain
\begin{equation}
 \vC{\begin{picture}(160,30)
   \put(0,0){\includegraphics[width=160mm]{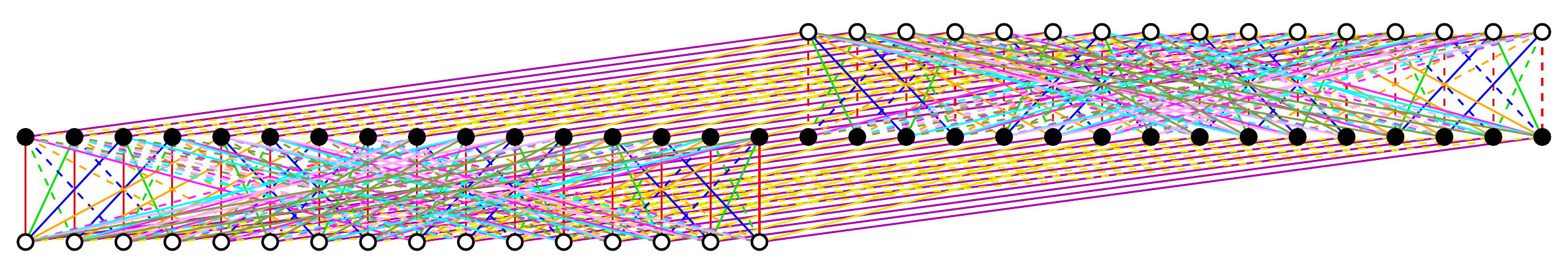}}
    \put(3,22){$I^{8,2}/d_{8,2}$}
 \end{picture}}
 \label{e:D82}
\end{equation}
in direct analogy with\eq{e:D33-484}, \eq{e:D42-484} and\eq{e:D44-484}: the $d_{8,0}$ subcode of $d_{8,2}$ acts trivially on the factors, but the $d_{2,2}\simeq d_{8,2}/d_{8,0}$ part acts non-trivially and produces the resulting Adinkra. Tracing (closed) hallmark 4-color tetragons in the left-hand factor of the product\eq{e:D8xI2} a convenient basis is given by:
\begin{equation}
   \vC{\includegraphics[width=30mm]{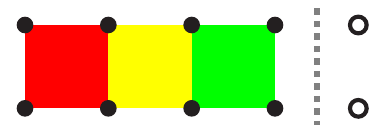}}
  =\bm{1100\,1100\\0110\,0110\\0011\,0011}=(d_{8,0}\oplus t_{0,2})
  ~\subset~\e_{8,0} =\pC{1100\,1100&00\\[-2pt]0110\,0110&00\\[-2pt]
                               0011\,0011&00\\[-2pt]1111\,0000&00\\[-2pt]}
  =\vC{\includegraphics[width=30mm]{E80.pdf}},
 \label{e:d80<e80}
\end{equation}
where the subcode relationship is easily spotted by comparing the $4k$-gon diagrams at the far ends.
 A final, fourth closed hallmark 4-color tetragon may be found in\eq{e:D82} to correspond to $10001000|11$, thus giving a convenient basis
\begin{equation}
  d_{8,2}=\pC{1100\,1100&00\\[-2pt]0110\,0110&00\\[-2pt]
              0011\,0011&00\\[-2pt]0001\,0001&11\\[-2pt]}
  =\vC{\includegraphics[width=30mm]{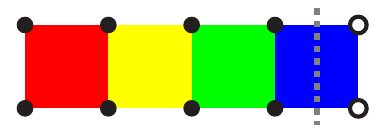}}.
 \label{e:d8,2}
\end{equation}
Again, the relationship between $e_{8,0}$, $d_{8,2}$ and the common subcode $d_{8,0}\oplus t_{0,2}$ is easily spotted on comparing the $4k$-gon diagrams in\eq{e:d80<e80} and\eq{e:d8,2}.
 In terms of superderivatives, we have that
\begin{equation}
  \bm{\vd_\pp^2&-&\rD_{1+}\rD_{2+}\rD_{5+}\rD_{6+}\\
      \vd_\pp^2&-&\rD_{2+}\rD_{3+}\rD_{6+}\rD_{7+}\\
      \vd_\pp^2&-&\rD_{3+}\rD_{4+}\rD_{7+}\rD_{8+}\\
      \vd_\pp^2&-&\rD_{1+}\rD_{2+}\rD_{3+}\rD_{4+}\\}_{e_{8,0}}\mkern-15mu
       \simeq0~\text{in\eq{e:E8xI2}},
  \quad\textit{vs.}\quad
  \bm{~~\,\vd_\pp^2&-&\rD_{1+}\rD_{2+}\rD_{5+}\rD_{6+}\\
      ~~\,\vd_\pp^2&-&\rD_{2+}\rD_{3+}\rD_{6+}\rD_{7+}\\
      ~~\,\vd_\pp^2&-&\rD_{3+}\rD_{4+}\rD_{7+}\rD_{8+}\\
      \vd_\pp\vd_\mm&-&\rD_{4+}\rD_{8+}\rD_{1-}\rD_{2-}\\}_{d_{8,2}}\mkern-15mu
       \simeq0~\text{in\eq{e:D82}},
 \label{e:e80vd82}
\end{equation}
so that the two are seen to differ only in the fourth generating hallmark superderivative relation, depicted in blue in the $4k$-gon diagrams\eq{e:d80<e80} and\eq{e:d8,2}.

The isomorphism between\eq{e:E8xI2} and\eq{e:D82} is now constructed\cite{r6-3.2} by changing the component (super)field basis of\eq{e:E8xI2}: Seeing that the fourth generator of the $e_{8,0}$ code, $[\vd_\pp^2{-}\rD_{1+}\rD_{2+}\rD_{3+}\rD_{4+}]$, does not produce a closed quadrangle in the Adinkra\eq{e:D82}, but maps any component (super)field into the one obtained by following the formal action of the {\em\/formal\/} operator
\begin{equation}
  \ha\rD^{1111\,0000|00}\Defl(\rD_{1+})^{\pm1}\circ(\rD_{2+})^{\pm1}\circ(\rD_{3+})^{\pm1}\circ(\rD_{4+})^{\pm1},
 \label{e:Dhat}
\end{equation}
where, say, $(\rD_{1+})^{-1}$ indicates following $\rD_{1+}$ ``in reverse,'' \ie, finding the {\em\/pre-image\/} of $\rD_{1+}$. That is $(\rD_{1+})^{-1}\f$ denotes the component (super)field|or a linear combination thereof|upon which the application of $\rD_{1+}$ produces $\f$.
 The powers in the definition\eq{e:Dhat} are chosen depending on the component (super)field upon which the operator is acting, and so that the path of the corresponding edges remains in the given Adinkra.

For example, if we start with the leftmost lower node, applying $\ha\rD^{1111\,0000|00}$ one factor at a time, we identify the highlighted path:
\begin{equation}
 \vC{\begin{picture}(140,24)
   \put(0,0){\includegraphics[width=140mm]{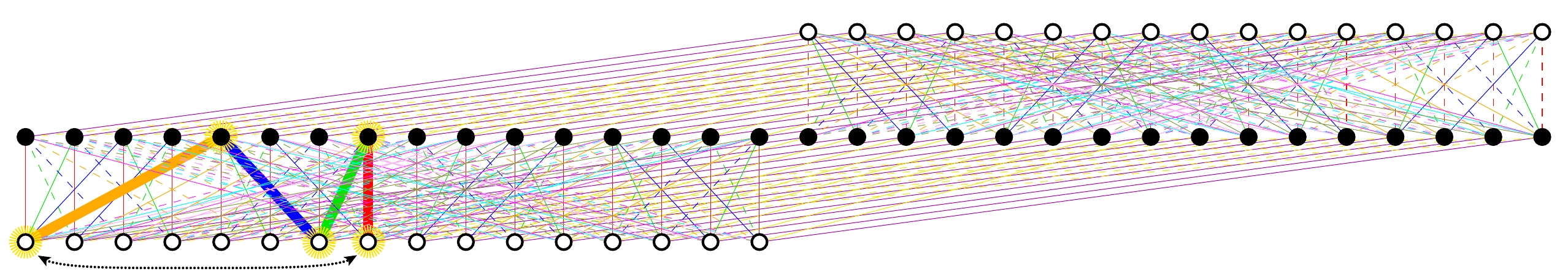}}
    \put(3,18){$I^{8,2}/d_{8,2}$}
    \put(-2,1){\footnotesize$\f$}
 \end{picture}}
 \label{e:d=e}
\end{equation}
Notice that the initial and the final node are at the same height, \ie, the corresponding component (super)fields have the same engineering dimension. They also have the same spin, since
\begin{equation}
  \spin\big[\ha\rD^{1111\,0000|00}(\f)\big]
  =(-\inv2)+(+\inv2)+(-\inv2)+(+\inv2)+\spin[\f]=\spin[\f].
\end{equation}
It is not hard to verify that the same is true for any other starting node in\eq{e:D82}.
This makes the linear combination of any component (super)field and its $\ha\rD^{1111\,0000|00}$-image consistent with both engineering dimension homogeneity and $\Spin(1,1)$ Lorentz-covariance. Finally, since the $\rD$'s anticommute with supersymmetry generators, the mapping defined by $\ha\rD^{1111\,0000|00}$ is manifestly supersymmetry-covariant.

The construction of the new basis starts with defining $\f'_+\Defl\fc12[\Ione+\ha\rD^{1111\,0000|00}](\f)$|indicated in\eq{e:d=e} by the dotted arrow|as a new component (super)field. It is not hard to ascertain that the formal operators $\fc12[\Ione\pm\ha\rD^{1111\,0000|00}]$ act as complementary projection operators, are closely related to\eq{e:Pab}, and
\begin{equation}
  \ha\P^{1111\,0000|00}_-\f'_+\Defl
   \fc12[\Ione-\ha\rD^{1111\,0000|00}]\f'_+=0.
 \label{e:P-f+=0}
\end{equation}
The remainder of this new basis of component (super)fields is obtained by applying the $10$-cube of $(8,2)$-superderivatives\eq{e:MonD} on $\f'_{\sss\pm}$. The resulting collection of component (super)fields is then manifestly just a (super)field redefinition of the supermultiplet depicted by\eq{e:D82}. However, in this new basis, the path corresponding to $\ha\rD^{1111\,0000|00}$ is a closed hallmark 4-color tetragon, thus manifesting the $e_{8,0}\oplus t_{0,2}$ rather than the $d_{8,2}$ code and ensuring that the resulting Adinkra must take the form of\eq{e:E8xI2}. Equivalently and owing to\eq{e:P-f+=0}, each component (super)field in this new basis is annihilated by $\ha\P_-^{1111\,0000|00}$.

Needless to say, the converse can be done as well: in\eq{e:E8xI2}, we start from a component (super)field and now identify its image under the action of $\ha\rD^{0001\,0001|11}$ (where $0001\,0001|11$ is a generator of $d_{8,2}$ that is not also in $e_{8,0}\oplus t_{0,2}$),
\begin{equation}
 \vC{\begin{picture}(160,28)
   \put(0,0){\includegraphics[width=160mm]{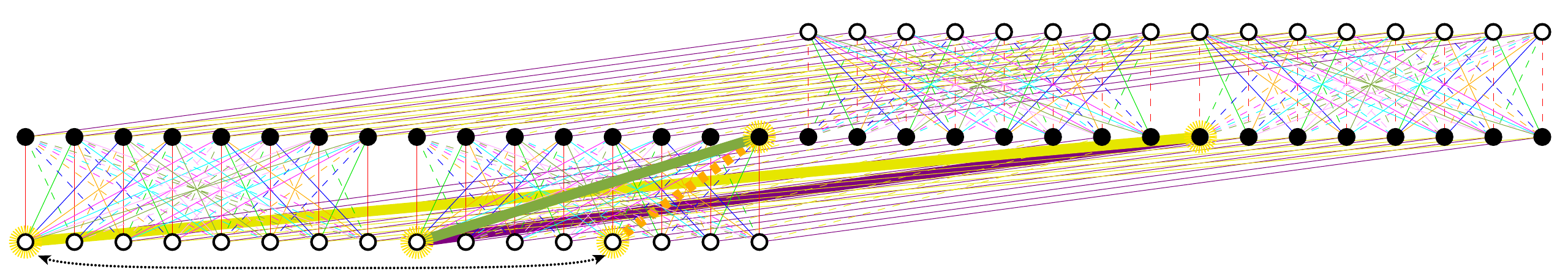}}
    \put(3,22){$I^{8,2}/(e_{8,0}\oplus t_{0,2})$}
    \put(-2,1){\footnotesize$\vf$}
 \end{picture}}
 \label{e:e=d}
\end{equation}
The new component (super)field basis is defined by starting with the linear combination component (super)field $[\Ione+\ha\rD^{0001\,0001|11}]\vf$.
 The remaining component (super)fields are then obtained by applying the $10$-cube of $(8,2)$-superderivatives\eq{e:MonD} on $\vf'_+$. The resulting Adinkra will then have $\rD^{0001\,0001|11}$ generate a hallmark $4k$-gon relation, and so will be depicted by the Adinkra\eq{e:D82}.
\ping

The necessary and sufficient criterion to determine if two {\em\/worldline\/} Adinkras depict isomorphic supermultiplets involves the definition of the ``node choice group'' (NCG)\cite{r6-3.2}. This is the symmetry generated by the horizontal permutations of nodes that result in the same Adinkra. NCG is encoded by the binary exponents of the formal $\rD$-monomials required to connect the component (super)fields which correspond to the permuted nodes, and these exponents form a binary (not necessarily doubly) even linear block code, $\sN$. For two Adinkras $\cA_1\simeq I^N/\sC_1$ and $\cA_2\simeq I^N/\sC_2$ to depict isomorphic supermultiplets supermultiplets, it is necessary and sufficient for both adinkras must have the same NCG encoded by $\sN$, and that $\sC_1\subset\sN$ as well as $\sC_2\subset\sN$.

 Clearly, this criterion translates to {\em\/worldsheet\/} Adinkras, but the node choice group is now encoded by a {\em\/split\/} binary even linear block code. Thereupon, the criterion is virtually the same:
\begin{corl}\label{c:EquivA}
Let $\ssH$ denote the split even linear block code encoding the horizontal permutation of nodes in a given worldsheet Adinkra, and let two Adinkras, $\cA_i$, have the split chromotopology $I^{p,q}/\ssC_i$, $i=1,2$. They depict supermultiplets that are isomorphic, and by (super)field redefinitions only, precisely if:
\begin{enumerate}\itemsep=-3pt\vspace{-2mm}\leftskip2pc
 \item both Adinkras have the same node choice group of symmetries, encoded by $\ssH$, and
 \item both $\ssC_1\subset\ssH$ and $\ssC_2\subset\ssH$,
 \item $\spin[\rD^{\bf a|b}]=0$ for both $\rD^{\bf a|b}\in(\ssC_1\smallsetminus\ssC_2)$ and
        $\rD^{\bf a|b}\in(\ssC_2\smallsetminus\ssC_1)$.
\end{enumerate}

\end{corl}

\subsection{Worldsheet Adinkra Degeneracy 2}
\label{s:2}
Whereas section~\ref{s:1} shows that there exist inequivalent Adinkras that nevertheless depict equivalent worldsheet supermultiplets, we now show that some \DELB\ codes have more than one inequivalent splits. Consequently, an Adinkra with the chromotopology $I^N/\sC$ may be used to depict two inequivalent worldsheet supermultiplets, one with the split chromotopology $I^{p,q}/\ssC_1$ the other with $I^{p,q}/\ssC_2$.

The simplest example is constructed by splitting the $d_8$ code in two distinct ways, as
\begin{table}[ht]
$$
   \begin{array}{ccclc}
 \text{\bsf Code} & \MC2c{\text{\bsf Generators}} & \MC2c{\text{\bsf Adinkra}}\\
 &\text{\bsf 4{\slshape k}-gon} &\text{\bsf Codewords} & \MC1c{\text{\bsf Top.}}
 &\MR2*{\kern-10pt\includegraphics[width=55mm]{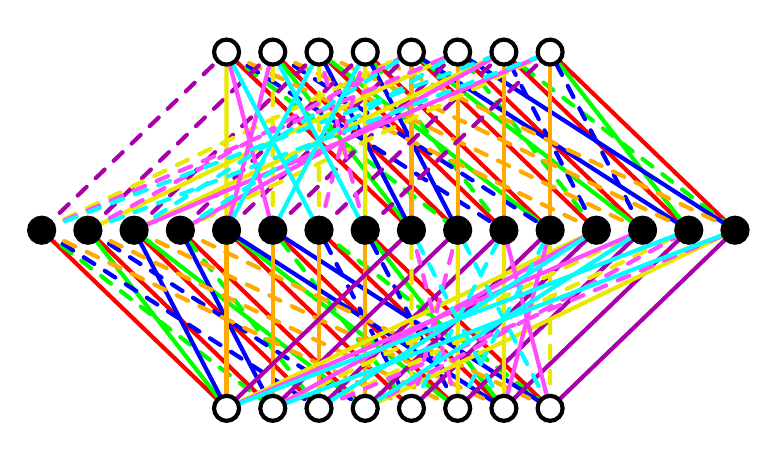}}\\ 
 d_8
 &\vC{\includegraphics[width=25mm]{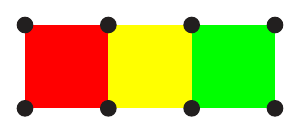}}
 &\text{\scriptsize$\left[\begin{array}{@{\,}c@{\,}}
                           11001100\\[-1pt]01100110\\[-1pt]00110011\\[-1pt]
                          \end{array}\right]$}
 & I^8/d_8
 &\\
 d_{4,4}
 &\vC{\includegraphics[width=25mm]{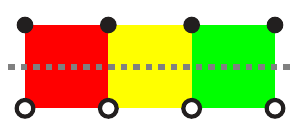}}
 &\pC{1100&1100\\[-2pt]0110&0110\\[-2pt]0011&0011\\[-2pt]}
 & I^{4,4}/d_{4,4}\\
 d_{4,4}'
 &\vC{\includegraphics[width=25mm]{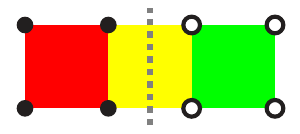}}
 &\pC{1111&0000\\[-2pt]0011&1100\\[-2pt]0000&1111\\[-2pt]}
 & \MC2l{I^{4,4}/d_{4,4}'=(I^{4,0}/d_{4,0}\,{\times}\,I^{0,4}/d_{0,4})/d_{2,2}}\\
  \end{array}
$$
  \caption{Two inequivalent splits of the $d_8$ \DELB-code. The same Adinkra depicts inequivalent supermultiplets via inequivalent assignment of edge-colors to the $(4,4)$-supersymmetry generators.}
  \label{t:d8}
\end{table}
shown in Table~\ref{t:d8}. The inequivalence $d_{4,4}\not\simeq d_{4,4}'$ is easy to see as follows. Unlike for $d_{4,4}'$, all three generators (and therefore also all elements) of $d_{4,4}$ are split symmetrically: each codeword has the same weight on the left- and the right-hand side of the partition. In turn, $d_{4,4}'$ contains the $d_{4,0}$ and $d_{0,4}$ subcodes, while $d_{4,4}$ contains no (nontrivial) unidextrous subcode.

Now, neither is $d_8$ a maximal \DELB-code, nor are $d_{4,4}$ and $d_{4,4}'$ maximal $(4,4)$-split \sDE-codes. Indeed, $d_8\subset e_8$, and $d_{4,4}\subset e_{4,4}$, and $d_{4,4}'\subset e_{4,4}'$:
\begin{equation}
  \vC{\includegraphics[width=30mm]{d44.pdf}}\subset
  \vC{\includegraphics[width=30mm]{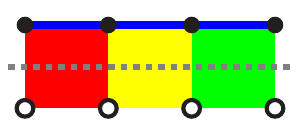}},\quad\text{and}\quad
  \vC{\includegraphics[width=30mm]{d44a.pdf}}\subset
  \vC{\includegraphics[width=30mm]{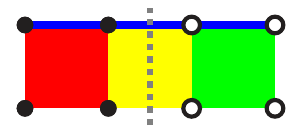}}.
 \label{e:d>e44<d'}
\end{equation}
Finally, it is not hard to show that the two versions of $e_{4,4}$ are in fact equivalent, so that $d_{4,4}$ and $d_{4,4}'$ are inequivalent \sDE-subcodes of the same maximal \sDE-code $e_{4,4}$.

Projections by the inequivalent \sDE-codes such as $d_{4,4},d'_{4,4}$ depicted in Table~\ref{t:d8}, clearly define inequivalent Adinkras, each of which depicts a distinct supermultiplet. An isomorphism between two such supermultiplets may again be constructed in the manner described in\eqs{e:E8xxI2}{e:e=d}. Consider, for example the two worldsheet supermultiplets depicted by the Adinkra in Table~\ref{t:d8}, and projected, respectively, by $d_{4,4}$ and $d'_{4,4}$. A comparison of the $4k$-gon diagrams of the two codes indicates one common generator (the middle one):
\begin{equation}
  d_{4,4}=\vC{\includegraphics[width=25mm]{d44.pdf}}
  =\pC{1100&1100\\[-2pt]0110&0110\\[-2pt]0011&0011\\[-2pt]}
  \qquad\textit{vs.}\qquad
  \pC{1111&0000\\[-2pt]0011&1100\\[-2pt]0000&1111\\[-2pt]}
  =\vC{\includegraphics[width=25mm]{d44a.pdf}}=d'_{4,4},
\end{equation}
which suggests reordering $\rD_{2+}\mapsto\rD_{3+}\mapsto\rD_{4+}\mapsto\rD_{2+}$ and $\rD_{2-}\mapsto\rD_{3-}\mapsto\rD_{4-}\mapsto\rD_{2-}$ in $d'_{4,4}$, which results in
\begin{equation}
  d_{4,4}
  =\vC{\begin{picture}(25,11)
        \put(0,0){\includegraphics[width=25mm]{d44.pdf}}
         \put(.5,10){$\sss1+$}
         \put(7.5,10){$\sss2+$}
         \put(14.5,10){$\sss3+$}
         \put(21.5,10){$\sss4+$}
         \put(.5,-.5){$\sss1-$}
         \put(7.5,-.5){$\sss2-$}
         \put(14.5,-.5){$\sss3-$}
         \put(21.5,-.5){$\sss4-$}
       \end{picture}}
  =\pC{1100&1100\\[-2pt]0110&0110\\[-2pt]0011&0011\\[-2pt]}
  \qquad\textit{vs.}\qquad
  \pC{1111&0000\\[-2pt]0110&0110\\[-2pt]0000&1111\\[-2pt]}
  =\vC{\begin{picture}(25,11)
        \put(0,0){\includegraphics[width=25mm]{d44a.pdf}}
         \put(.5,10){$\sss1+$}
         \put(7.5,10){$\sss2+$}
         \put(14.5,10){$\sss2-$}
         \put(21.5,10){$\sss1-$}
         \put(.5,-.5){$\sss4+$}
         \put(7.5,-.5){$\sss3+$}
         \put(14.5,-.5){$\sss3-$}
         \put(21.5,-.5){$\sss4-$}
       \end{picture}}
  =d'_{4,4},
\end{equation}
which makes the middle generator, $0110|0110$, common to both. In addition, the $\4$-sum of the first and the third generator in both codes, $1111|1111$, is also a common codeword and may itself be used as a generator. In both $d_{4,4}$ and $d'_{4,4}$, one more codeword is needed to act as the third generator, the requirement being only that it be linearly independent\ft{In the context of binary codes, ``linear independence'' refers to $\4$-addition of binary multiples.} from the common generators $0110|0110$ and $1111|1111$. To this end, we may well use the bases
\begin{equation}
  d_{4,4}=\pC{1111&1111\\[-2pt]0110&0110\\[-2pt]0011&0011\\[-2pt]}
  \qquad\textit{vs.}\qquad
  d'_{4,4}=\pC{1111&1111\\[-2pt]0110&0110\\[-2pt]0000&1111\\[-2pt]},
\end{equation}
which now has a single differing generator, and is in this respect in the same situation as were $e_{8,0}$ and $d_{8,2}$ in\eq{e:e80vd82}. We then start from a particular component (super)field, $\f$, in a $d_{4,4}$-projected supermultiplet and apply the superdifferential operator $\ha\rD^{0000|1111}$ encoded by the differing generator from $d'_{4,4}$ in such a way that the result has the same spin as the initial field, and define:
\begin{equation}
  \f'_{\sss\pm}\Defl
  \inv2\big[\Ione-(\rD_{1-})^-1\circ(\rD_{2-})\circ(\rD_{3-})^-1\circ(\rD_{4-})\big]\f.
 \label{e:f>f'}
\end{equation}
Using either of $\f'_{\sss\pm}$ as a starting point, we reconstruct the remainder of the supermultiplet by applying all superdifferential operators from the $(4,4)$-basis\eq{e:MonD}. In the Adinkra depicting the so reconstructed supermultiplet, the hallmark $4k$-gon relations encoded by the \sDE-code $d'_{4,4}$ will all trace closed hallmark $4k$-gons, rather than the ones encoded by $d_{4,4}$ and which were closed before the component (super)field basis redefinition started with\eq{e:f>f'}.

We have thus constructed an isomorphism between the component (super)field basis for the supermultiplet depicted by the $d_{4,4}$-projected Adinkra to the component (super)field basis for the supermultiplet depicted by the $d'_{4,4}$-projected Adinkra, proving that the two are merely two distinct bases for the same supermultiplet.

A few comments are in order: First, the \sDE-codes $d_{4,4}$ and $d'_{4,4}$ are not maximal \sDE-codes: they are both distinct sub-codes of the $e_{4,4}$ \sDE-code, as shown in the display\eq{e:d>e44<d'} and the subsequent text. One may suspect that the above isomorphism is in fact due to this non-maximality, and dismiss the distinction $d_{4,4}\neq d'_{4,4}$ as irrelevant for constructing supermultiplets that are not equivalent by (super)field redefinitions.

However, there do exist {\em\/maximal\/} \sDE-codes that are inequivalent even-splits of the same \DELB-code. For example,
\begin{equation}
  \vC{\begin{picture}(50,14)
       \put(.5,0){\includegraphics[width=50mm]{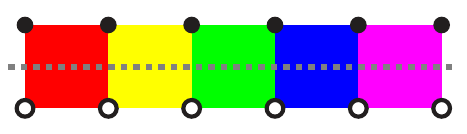}}
        \put(6,3.5){\small A}
        \put(15,3.5){\small B}
        \put(24,3.5){\small C}
        \put(33,3.5){\small\C8D}
        \put(42,3.5){\small E}
      \end{picture}}
       \quad\text{and}\quad
  \vC{\begin{picture}(50,14)
       \put(.5,0){\includegraphics[width=50mm]{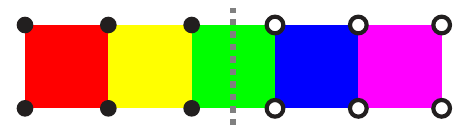}}
        \put(6,3.5){\small A}
        \put(15,3.5){\small B}
        \put(24,3.5){\small C}
        \put(33,3.5){\small\C8D}
        \put(42,3.5){\small E}
      \end{picture}}
 \label{e:d66x2}
\end{equation}
are evidently $(p,q)=(6,6)$ even-splits of $d_{12}$, and are inequivalent in precisely the same manner as are $d_{4,4}$ and $d_{4,4}'$. It is clear that every \DELB-code $d_{4k}$ for $k=2,3,4\dots$ has such two inequivalent even-splits, $d_{2k,2k}$ and $d'_{2k,2k}$. Of these, the \DELB-codes $d_{4k}$ and the \sDE-codes $d_{2k,2k}$ and $d'_{2k,2k}$ with odd $k$|starting with $d_{6,6}$ and $d'_{6,6}$ in\eq{e:d66x2}|are also maximal.

In this case, the bases for the $d_{6,6}$ and $d'_{6,6}$ may again be changed so as to exhibit a maximum (three) of common generators,
\begin{equation}
  \pC{001100&001100\\[-2pt]011110&011110\\[-2pt]110011&110011\\[-2pt]}
\end{equation}
corresponding in turn to the $4k$-gons \C2C, \C6B$\boxplus$\C3D and \C1A$\boxplus$\C7E on both sides of\eq{e:d66x2}. Let $\ha\rD_1,\ha\rD_2$ be the formal superderivative operators corresponding to two generators of $d_{6,6}$ that are not in $d'_{6,6}$, and $\ha\rD'_1,\ha\rD'_2$ be the formal superderivative operators corresponding to two generators of $d'_{6,6}$ that are not in $d_{6,6}$. The supermultiplet isomorphism is then constructed by starting with $\f$ a component (super)field from a $d_{6,6}$-projected supermultiplet, identifying
\begin{equation}
  \f'\Defl\inv2\big[\Ione-\ha\rD'_1\big]\,\inv2\big[\Ione-\ha\rD'_2\big]\f
\end{equation}
with a starting component (super)field in the new basis, and reconstructing the remainder of the supermultiplet by acting with the $(6,6)$-superderivatives\eq{e:MonD} upon $\f'$. In the so-constructed basis and starting with any (new) component (super)field, both $\ha\rD'_1$ and $\ha\rD'_2$ will sweep out closed hallmark $4k$-gons; therefore, the Adinkra depicting this new basis for the $d_{6,6}$-projected supermultiplet will have the topology of $I^{6,6}/d'_{6,6}$ rather than $I^{6,6}/d_{6,6}$ from which we started. This then constructs the isomorphism between the $d_{6,6}$-projected supermultiplet and the $d'_{6,6}$-projected one.

We have thus demonstrated that there exist \DELB-codes that have inequivalent \sDE-code splits, and some of which are maximal, but that the Adinkras projected by at least some of those inequivalent \sDE-code splits of \DELB-codes in fact depict isomorphic supermultiplets.

As the number for inequivalent \sDE-codes grows combinatorially with $(p,q)$, a computer-aided listing of the type done for \DELB-codes\cite{r6-3,r6-3.2,rRLM-codes} is clearly necessary for their classification, and for a consequent classification of all off-shell supermultiplets of worldline $(p,q)$-supersymmetry, e.g.\ for $p{+}q\leqslant32$, which limit is expected from $M$-theory considerations\cite{rBBS}.

\section{Conclusions}
\label{e:coda}
In the foregoing analysis, the classification efforts of Refs.\cite{r6-3,r6-3.2,r6-1.2} are generalized so as to outline the analogous classification of off-shell supermultiplets of $(p,q)$-extended worldsheet supersymmetry. In particular, the main results are as follows.
\begin{enumerate}\itemsep=-3pt\vspace{-2mm}

 \item Section~\ref{s:WSSm} provides three constructions, \ref{C:pxq>pq}, \ref{C:N>Nq} and~\ref{C:N>N0}, by which off-shell and on the half-shell representations of worldsheet $(p,q)$-supersymmetry are obtained as tensor products of left- and right-moving worldline supermultiplets. A complete listing of such supermultiplets for $p{+}q\leqslant8$ is given in Section~\ref{s:LowSpC}.

 \item Generalizing the situation with worldline supermultiplets, certain worldsheet off-shell supermultiplets {\em\/decompose\/} into a direct sum of two half-sized sized supermultiplets, while others {\em\/reduce\/} to half-sized supermultiplets. Possible iteration of such $\ZZ_2$ decompositions and reductions is encoded by even-split doubly even linear block (\sDE) codes, which are discussed and classified for $p+q\leqslant8$ in Section~\ref{s:WSCodes} and depicted in Figure~\ref{f:1}. Such decompositions and reductions produce the minimal supermultiplets for given $(p,q)$-supersymmetry.

 \item Corollary~\ref{C:Cpx} identifies a type of {\em\/twisted\/} $\ZZ_2$ symmetry that signals the existence of a complex structure. Section~\ref{s:Cpx} verifies this amongst off-shell worldsheet $(p,q)$-supermultiplets for $p{+}q=4$. Section~\ref{s:FnF} shows that some \sDE-quotients of real tensor product supermultiplets are (hyper)complex tensor products of (hyper)complex supermultiplets, but that many are not. Therefore, the results listed above under \#1 and \#2 (Construction~\ref{C:pxq>pq}) produces some genuinely novel off-shell supermultiplets of worldsheet $(p,q)$-supersymmetry.

 \item Sections~\ref{s:1} demonstrates that some worldsheet supermultiplets depicted by topologically inequivalent Adinkras are nevertheless equivalent, by adapting the analogous worldline result of Ref.\cite{r6-3}. Corollary~\ref{c:EquivA} specifies the appropriate conditions for this isomorphism.
 
 \item In turn, Section~\ref{s:2} constructs inequivalent splits of the same doubly even linear block code, producing inequivalent \sDE-codes, and whereby the same $(p,q)$-Adinkra is made to depict distinct worldsheet supermultiplets. 
 At least some of such distinct supermultiplets however may be shown to be equivalent by adapting the analogous worldline result of Ref.\cite{r6-3}.
 
 \item Ref.\cite{rGH-obs} observes that, as a necessary avoidance of the obstruction defined for its ``twin theorems~2.1 and~2.2'', ambidextrous off-shell supermultiplets of ambidextrous supersymmetry must have at least three levels\ft{Ref.\cite{rGH-obs} also notes that the only fully off-shell supermultiplets with only two levels may exist: ({\bsf1})~in worldline supersymmetric systems where this obstruction is void, ({\bsf2})~in {\em\/unidextrous\/} $(p,0)$- and $(0,q)$-supersymmetric systems on the worldline, and {\em\/perhaps\/}, ({\bsf3})~in higher-dimensional systems with $s$ space and $t$ time dimensions where $(s{-}t)=0\,\text{mod}\,8$. The analysis herein fully agrees.}, \ie, their component (super)fields must have at least three distinct, adjacent engineering dimensions\cite{rCRT,r6-1}.
 Herein, we see this to follow as an elementary consequence of Adinkra tensor products, as defined in Constructions~\ref{C:pxq>pq} and~\ref{C:RxR}, and exemplified in Sections~\ref{s:(3,1)} and~\ref{s:(2,2)}; for and ambidextrous supersymmetry, $p\neq0\neq q$ and $\fR_+\neq\Ione\neq\fR_-$ in these constructions. Since the minimal level of $\fR_\pm\neq\Ione$ is two, the minimal level of $(\fR_+\otimes\fR_-)/\ssC$ cannot, by construction, be less than three; see\eqs{e:22x22}{e:B242} for a simple illustration.

\end{enumerate}

Owing to the combinatorial growth of these tasks with $p{+}q$, a mechanization of the methods presented herein would be welcome, perhaps in synergy with those reported in Ref.\cite{rGH-obs}, so as to extend the classification of worldsheet supermultiplets beyond $p{+}q\leqslant8$ through $p{+}q\leqslant32$.

\bigskip\paragraph{\bfseries Acknowledgments:}
I am indebted to C.~Doran, M.~Faux, K.~Iga, G.~Landweber and R.~Miller for prior extensive collaboration on the classification of {\em\/worldline\/} off-shell supermultiplets, of which the present work is a generalization,
 to S.J.~Gates, Jr.\ for continued and invaluable collaborative discussions,
 and to the anonymous Referee for invaluable constructive criticism.
I am grateful to the Department of Energy for the generous support through the grant DE-FG02-94ER-40854, as well as the
 Department of Physics, University of Central Florida, Orlando FL,
and the
 Physics Department of the Faculty of Natural Sciences of the University of Novi Sad, Serbia,
 for recurring hospitality and resources.
 Many of the Adinkras were drawn with the help of the {\em Adinkramat\/}~\copyright\,2008 by G.~Landweber.

\appendix
\section{Details of the $d_{2,2}$-Encoded $\PMB{\ZZ}_2$-Symmetry}
 %
To help with translating the Adinkra manipulations that turn\eq{e:121x121} into\eq{e:484} and then this into\eq{e:B484>242}, let us revisit the same depictions, but annotated with corresponding superderivatives, as taken from\eq{f:22Ds}. To save space, we use the binary exponent notation\eq{e:MonD}, modified so as to absorb $\vd_\pp$ and $\vd_\mm$ factors, so for example
\begin{equation}
  \rD^{00|00}=\Ione,\quad
  \rD^{01|00}=\rD_{2+},\quad
  \rD^{20|10}_\pp\Defl i\vd_\pp\rD_{1-},\quad
  \rD^{11|20}_\mm\Defl i\vd_\mm\rD_{1+}\rD_{2+},\quad \textit{etc.}
\end{equation}
We start with\eq{f:22Ds} and define a supermultiplet depicted by the Adinkras in\eq{e:484}:
\begin{equation}
 \vC{\begin{picture}(160,55)
   \put(0,0){\includegraphics[width=160mm]{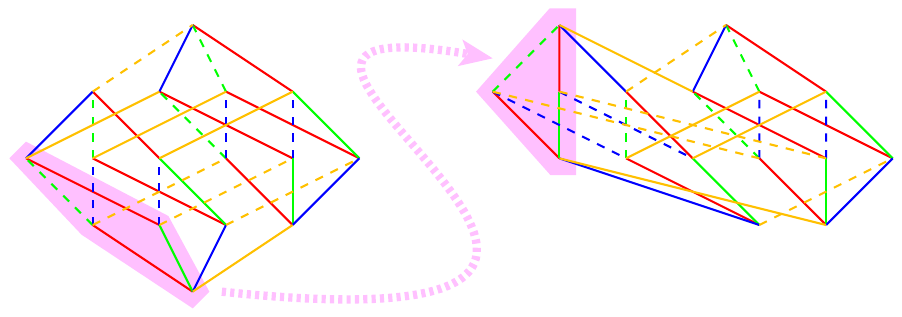}}
    \put(10,5){$\BF$}
    \put(120,5){$\bJ_{\ad-}\Defl(\rD_{\ad-}\BF)$}
    \put(34,51){\scriptsize\cB{$\rD^{1111}$}}
    \put(16,39){\scriptsize\cB{$\rD^{1110}$}}
    \put(28,39){\scriptsize\cB{$\rD^{1101}$}}
    \put(40,39){\scriptsize\cB{$\rD^{1011}$}}
    \put(54,39){\scriptsize\cB{$\rD^{0111}$}}
    \put(4,27){\scriptsize\cB{$\rD^{1100}$}}
    \put(16,27){\scriptsize\cB{$\rD^{1010}$}}
    \put(28,27){\scriptsize\cB{$\rD^{0110}$}}
    \put(40,27){\scriptsize\cB{$\rD^{1001}$}}
    \put(52,27){\scriptsize\cB{$\rD^{0101}$}}
    \put(64,27){\scriptsize\cB{$\rD^{0011}$}}
    \put(16,15){\scriptsize\cB{$\rD^{1000}$}}
    \put(28,15){\scriptsize\cB{$\rD^{0100}$}}
    \put(40,15){\scriptsize\cB{$\rD^{0010}$}}
    \put(52,15){\scriptsize\cB{$\rD^{0001}$}}
    \put(34,3){\scriptsize\cB{$\rD^{0000}$}}
    \put(56,5){$(\rD_{1-})^2=i\vd_\mm$}
    \put(129,51){\scriptsize\cB{$\rD^{1111}$}}
    \put(110,39){\scriptsize\cB{$\rD^{1110}$}}
    \put(122,39){\scriptsize\cB{$\rD^{1101}$}}
    \put(134,39){\scriptsize\cB{$\rD^{1011}$}}
    \put(148,39){\scriptsize\cB{$\rD^{0111}$}}
    \put(98,51){\scriptsize\cB{$\rD^{1120}$}}
    \put(110,27){\scriptsize\cB{$\rD^{1010}$}}
    \put(122,27){\scriptsize\cB{$\rD^{0110}$}}
    \put(134,27){\scriptsize\cB{$\rD^{1001}$}}
    \put(148,27){\scriptsize\cB{$\rD^{0101}$}}
    \put(158,27){\scriptsize\cB{$\rD^{0011}$}}
    \put(86,39){\scriptsize\cB{$\rD^{1020}$}}
    \put(98,39){\scriptsize\cB{$\rD^{0120}$}}
    \put(134,15){\scriptsize\cB{$\rD^{0010}$}}
    \put(148,15){\scriptsize\cB{$\rD^{0001}$}}
    \put(98,27){\scriptsize\cB{$\rD^{0020}$}}
 \end{picture}}
 \label{e:int>vas}
\end{equation}
The maneuver indicated by the lilac dashed arrow resembles ``node raising'' of Refs.\cite{r6-1,r6-3c,r6-3,r6-3.2}. However, since individual nodes of an Adinkra cannot be raised if it is to continue depicting an off-shell {\em\/worldsheet\/} $(p,q)$-supermultiplet with $p,q\neq0$\cite{rGH-obs}, the indicated sub-Adinkra is the minimal contiguous portion that can be consistently raised. The maneuver depicts the consequence of defining a superfield $\bJ_{\ad-}$ to be a superderivative of an intact superfield $\BF$: the highlighted component (super)fields of $\bJ_{\ad-}$ (on the right) are identified with the $\vd_\mm$-derivatives of the highlighted component (super)fields of $\BF$ (on the left).

A similar maneuver produces (we display only the exponents for brevity):
\begin{equation}
 \vC{\begin{picture}(160,40)
   \put(0,0){\includegraphics[width=160mm]{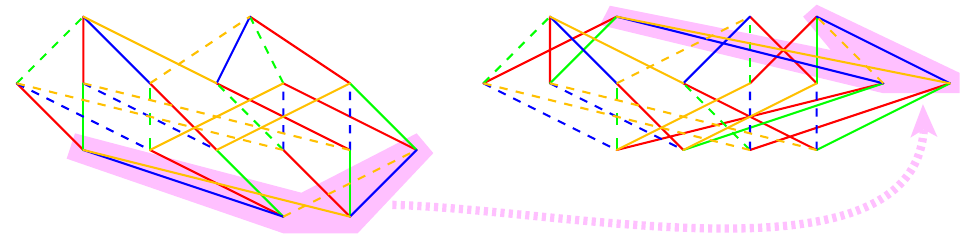}}
    \put(10,5){$\bJ_{\ad-}$}
    \put(110,10){${\bf F}_{\a\ad}\Defl(\rD_{\a+}\bJ_{\ad-})$}
    \put(41,38){\tiny\cB{1111}}
    \put(25,27){\tiny\cB{1110}}
    \put(36,27){\tiny\cB{1101}}
    \put(47,27){\tiny\cB{1011}}
    \put(58,27){\tiny\cB{0111}}
    \put(15,38){\tiny\cB{1120}}
    \put(25,15){\tiny\cB{1010}}
    \put(36,15){\tiny\cB{0110}}
    \put(47,15){\tiny\cB{1001}}
    \put(58,15){\tiny\cB{0101}}
    \put(69,15){\tiny\cB{0011}}
    \put(3,27){\tiny\cB{1020}}
    \put(14,27){\tiny\cB{0120}}
    \put(47,4){\tiny\cB{0010}}
    \put(58,4){\tiny\cB{0001}}
    \put(14,15){\tiny\cB{0020}}
    \put(75,9){$(\rD_{1+})^2=i\vd_\pp$}
    \put(124,38){\tiny\cB{1111}}
    \put(102,27){\tiny\cB{1110}}
    \put(113,27){\tiny\cB{1101}}
    \put(124,27){\tiny\cB{1011}}
    \put(135,27){\tiny\cB{0111}}
    \put(91,38){\tiny\cB{1120}}
    \put(102,15){\tiny\cB{1010}}
    \put(113,15){\tiny\cB{0110}}
    \put(124,15){\tiny\cB{1001}}
    \put(135,15){\tiny\cB{0101}}
    \put(135,38){\tiny\cB{2011}}
    \put(80,27){\tiny\cB{1020}}
    \put(91,27){\tiny\cB{0120}}
    \put(146,27){\tiny\cB{2010}}
    \put(157,27){\tiny\cB{2001}}
    \put(102,38){\tiny\cB{2020}}
 \end{picture}}
\end{equation}
To bring the result into the shape shown in\eq{e:B484>242}, some horizontal reshuffling is needed, and a subsequent change in the signs of a six highlighted superderivatives:
\begin{equation}
 \vC{\begin{picture}(160,27)
   \put(-5,0){\includegraphics[width=170mm]{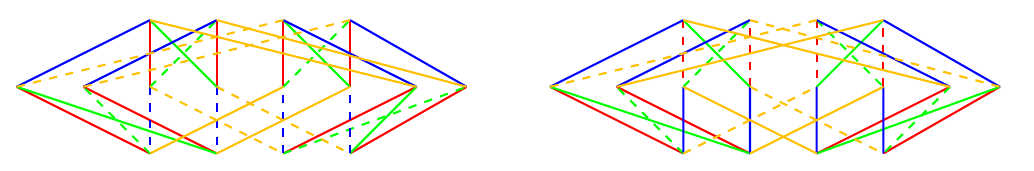}}
    \put(52,24){\tiny\cB{1111}}
    \put(9,13){\tiny\cB{\cb{yellow}{1110}}}
    \put(72,13){\tiny\cB{1101}}
    \put(41,13){\tiny\cB{\cb{yellow}{1011}}}
    \put(52,13){\tiny\cB{\cb{yellow}{0111}}}
    \put(30,24){\tiny\cB{\cb{yellow}{1120}}}
    \put(19,2){\tiny\cB{1010}}
    \put(30,2){\tiny\cB{\cb{yellow}{0110}}}
    \put(41,2){\tiny\cB{1001}}
    \put(52,2){\tiny\cB{0101}}
    \put(41,24){\tiny\cB{2011}}
    \put(19,13){\tiny\cB{\cb{yellow}{1020}}}
    \put(30,13){\tiny\cB{0120}}
    \put(-1,13){\tiny\cB{2010}}
    \put(62,13){\tiny\cB{2001}}
    \put(19,24){\tiny\cB{2020}}
    \put(71,3){${\bf F}_{\a\ad}\to\Tw{{\bf F}}_{\a\ad}$}
    \put(140,24){\tiny\cB{1111}}
    \put(97,13){\tiny\cB{$-1110$}}
    \put(160,13){\tiny\cB{1101}}
    \put(129,13){\tiny\cB{$-1011$}}
    \put(140,13){\tiny\cB{0111}}
    \put(118,24){\tiny\cB{$-1120$}}
    \put(107,2){\tiny\cB{1010}}
    \put(118,2){\tiny\cB{$-0110$}}
    \put(129,2){\tiny\cB{1001}}
    \put(140,2){\tiny\cB{0101}}
    \put(129,24){\tiny\cB{2011}}
    \put(107,13){\tiny\cB{$-1020$}}
    \put(118,13){\tiny\cB{0120}}
    \put(87,13){\tiny\cB{2010}}
    \put(150,13){\tiny\cB{$-2001$}}
    \put(107,24){\tiny\cB{2020}}
 \end{picture}}
 \label{e:SgCh}
\end{equation}
This is finally the Adinkra shown on the left-hand side in\eq{e:B484>242}. To see what this symmetry implies, we revert from the cryptic annotations to the implied (super)derivatives:
\begin{equation}
 \vC{\begin{picture}(160,48)(0,0)\unitlength=.975mm
   \put(0,0){\includegraphics[width=156mm]{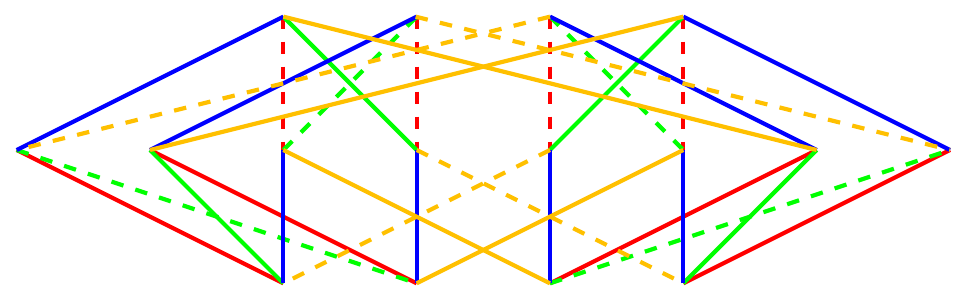}}
    \put(80,0){\rule[-1mm]{.4pt}{49mm}}
    \put(22,3){\large$\Tw{\bf F}_{\a\ad}$}
    \put(46,46){\scriptsize\cB{$-\vd_\mm\vd_\pp$}}
    \put(68,46){\scriptsize\cB{$-i\vd_\mm\rD_{1+}\rD_{2+}$}}
    \put(90,46){\scriptsize\cB{$\vd_\pp\rD_{1-}\rD_{2-}$}}
    \put(114,46){\scriptsize\cB{$\rD_{1+}\rD_{2+}\rD_{1-}\rD_{2-}$}}
    \put(4,24){\scriptsize\cB{$i\vd_\pp\rD_{1-}$}}
    \put(24,24){\scriptsize\cB{$-\rD_{1+}\rD_{2+}\rD_{1-}$}}
    \put(46,24){\scriptsize\cB{$-i\vd_\mm\rD_{1+}$}}
    \put(68,24){\scriptsize\cB{$i\vd_\mm\rD_{2+}$}}
    \put(90,24){\scriptsize\cB{$-\rD_{1+}\rD_{1-}\rD_{2-}$}}
    \put(112,24){\scriptsize\cB{$\rD_{2+}\rD_{1-}\rD_{2-}$}}
    \put(134,24){\scriptsize\cB{$-i\vd_\pp\rD_{2-}$}}
    \put(156,24){\scriptsize\cB{$\rD_{1+}\rD_{2+}\rD_{2-}$}}
    \put(46,2){\scriptsize\cB{$\rD_{1+}\rD_{1-}$}}
    \put(68,2){\scriptsize\cB{$-\rD_{2+}\rD_{1-}$}}
    \put(90,2){\scriptsize\cB{$\rD_{1+}\rD_{2-}$}}
    \put(112,2){\scriptsize\cB{$\rD_{2+}\rD_{2-}$}}
 \end{picture}}
 \label{e:Ch2D}
\end{equation}
where the tilde on $\Tw{\bf F}_{\a\ad}$ denotes the sign-changes performed in\eq{e:SgCh}.

The horizontal (literal) mirror identifications across the vertical divide indicated in\eq{e:B484>242} are now seen as identifications of superderivatives that are complementary within $\rD_{1+}\rD_{2+}\rD_{1-}\rD_{2-}$: literally so in the bottom row:
\begin{equation}
  (\rD_{1+}\rD_{1-})(\rD_{2+}\rD_{2-}) = -\rD_{1+}\rD_{2+}\rD_{1-}\rD_{2-}
   = (-\rD_{2+}\rD_{1-})(\rD{1+}\rD_{2-}),
\end{equation}
and padded with appropriate $\vd_\pp$- and $\vd_\mm$-factors in the middle and top row to match the engineering dimension and spin. Note that the particular assignment of superderivatives to the Adinkra nodes\eq{e:Ch2D} is the only one (up to sign-changes) that permits the horizontal (literal) mirror symmetry of the Adinkra to reflect in the superderivatives.

Thus, the formal identifications in\eq{e:B484>242} imply that, when acting on the components of the mirror-identified ``half-sized'' supermultiplets, the superderivatives satisfy relations such as
\begin{subequations}
\begin{gather}
  [\rD_{1+}\rD_{1-}\pm\rD_{2+}\rD_{2-}]\simeq0,\quad
  [\rD_{2+}\rD_{1-}\mp\rD_{1+}\rD_{2-}]\simeq0,\\[0mm]
  [i\vd_\pp\rD_{1-}\pm\rD_{1+}\rD_{2+}\rD_{2-}]\simeq0,\quad
  [\rD_{1+}\rD_{2+}\rD_{1-}\pm i\vd_\pp\rD_{2-}]\simeq0,\\[0mm]
  [-\vd_\pp\vd_\mm\pm\rD_{1+}\rD_{2+}\rD_{1-}\rD_{2-}]\simeq0,\quad\textit{etc.}
\end{gather}
\end{subequations}
fully consistent with\eq{e:d4Ps}.

We may thus define a supermultiplet in terms of an Adinkra of intact superfields:
\begin{equation}
 \vC{\begin{picture}(160,50)(3,0)
   \put(0,0){\includegraphics[width=160mm]{Cayak.pdf}}
    \put(3,10){\large$\cA_{2,2}$}
    \put(3,3){\large$\{{\bf Y};\bJ,\bX;{\bf Z}\}$}
    \put(47,46){\footnotesize\cB{${\bf Z}^{\mm\pp}$}}
    \put(68,46){\footnotesize\cB{${\bf Z}^\pp_\pp$}}
    \put(91,46){\footnotesize\cB{${\bf Z}^\mm_\mm$}}
    \put(113,46){\footnotesize\cB{${\bf Z}_{\pp\mm}$}}
    \put(4,24){\footnotesize\cB{$\bJ^\mm_{1-}$}}
    \put(24,24){\footnotesize\cB{$\bX^\mm_{1-}$}}
    \put(46,24){\footnotesize\cB{$\bJ^\pp_{1+}$}}
    \put(68,24){\footnotesize\cB{$\bJ^\pp_{2+}$}}
    \put(90,24){\footnotesize\cB{$\bX^\pp_{1+}$}}
    \put(112,24){\footnotesize\cB{$\bX^\pp_{2+}$}}
    \put(134,24){\footnotesize\cB{$\bJ^\mm_{2-}$}}
    \put(156,24){\footnotesize\cB{$\bX^\mm_{2-}$}}
    \put(46,2){\footnotesize\cB{${\bf Y}_{1+1-}$}}
    \put(68,2){\footnotesize\cB{${\bf Y}_{2+1-}$}}
    \put(90,2){\footnotesize\cB{${\bf Y}_{1+2-}$}}
    \put(112,2){\footnotesize\cB{${\bf Y}_{2+2-}$}}
 \end{picture}}
 \label{e:Ch2YJZ}
\end{equation}
where, directly generalizing\eqs{e:121}{e:22}, the edges specify the superdifferential relations:
{\small
\begin{subequations}
\begin{alignat}9
 \C1{\rD_{1+}}{\bf Y}_{1+1-}&=i\bJ^\mm_{1-},&\quad
 \C2{\rD_{2+}}{\bf Y}_{1+1-}&=i\bX^\mm_{1-},&\quad
 \C3{\rD_{1-}}{\bf Y}_{1+1-}&=i\bJ^\pp_{1+},&\quad
 \C4{\rD_{2-}}{\bf Y}_{1+1-}&=-i\bX^\pp_{1+},\\
 \C1{\rD_{1+}}{\bf Y}_{2+1-}&=i\bX^\mm_{1-},&\quad
 \C2{\rD_{2+}}{\bf Y}_{2+1-}&=-i\bJ^\mm_{1-},&\quad
 \C3{\rD_{1-}}{\bf Y}_{2+1-}&=i\bJ^\pp_{2+},&\quad
 \C4{\rD_{2-}}{\bf Y}_{2+1-}&=i\bX^\pp_{2+},\\
 \C1{\rD_{1+}}{\bf Y}_{1+2-}&=i\bJ^\mm_{2-},&\quad
 \C2{\rD_{2+}}{\bf Y}_{1+2-}&=-i\bX^\mm_{2-},&\quad
 \C3{\rD_{1-}}{\bf Y}_{1+2-}&=i\bX^\pp_{1+},&\quad
 \C4{\rD_{2-}}{\bf Y}_{1+2-}&=i\bJ^\pp_{1+},\\
 \C1{\rD_{1+}}{\bf Y}_{2+2-}&=i\bX^\mm_{2-},&\quad
 \C2{\rD_{2+}}{\bf Y}_{2+2-}&=i\bJ^\mm_{2-},&\quad
 \C3{\rD_{1-}}{\bf Y}_{2+2-}&=i\bX^\pp_{2+},&\quad
 \C4{\rD_{2-}}{\bf Y}_{2+2-}&=-i\bJ^\pp_{2+},\\
 \C1{\rD_{1+}}\bJ^\mm_{1-}&=\7{\sss\pp}{\bf Y}_{1+1-},&\quad
 \C2{\rD_{2+}}\bJ^\mm_{1-}&=-\7{\sss\pp}{\bf Y}_{2+1-},&\quad
 \C3{\rD_{1-}}\bJ^\mm_{1-}&={\bf Z}^{\mm\pp},&\quad
 \C4{\rD_{2-}}\bJ^\mm_{1-}&=-{\bf Z}^\pp_{1+2+},\\
 \C1{\rD_{1+}}\bX^\mm_{1-}&=\7{\sss\pp}{\bf Y}_{2+1-},&\quad
 \C2{\rD_{2+}}\bX^\mm_{1-}&=\7{\sss\pp}{\bf Y}_{1+1-},&\quad
 \C3{\rD_{1-}}\bX^\mm_{1-}&={\bf Z}^\pp_{1+2+},&\quad
 \C4{\rD_{2-}}\bX^\mm_{1-}&={\bf Z}_{\pp\mm},
\end{alignat}
\end{subequations}
}%
and so on for a total of 64 relations; $\7{\sss\pp}{\bf Y}\Defl\vd_\pp{\bf Y}$ (in the likeness of $\Dt{\bf Y}\Defl\ddt{\bf Y}$) to save space.

The projection\eq{e:B484>242} is then seen as the imposition of one of the (anti-)self-duality constraints:
\begin{equation}
  {\bf Y}_{\a\ad}~=~\ve_\a{}^\b\,\ve_\ad{}^\bd\,{\bf Y}_{\b\bd},
   \qquad\text{or}\qquad
  {\bf Y}_{\a\ad}~=-~\ve_\a{}^\b\,\ve_\ad{}^\bd\,{\bf Y}_{\b\bd}.
 \label{e:d4Y}
\end{equation}
The corresponding projection relations between the $\bJ$'s and $\bX$'s, and the ${\bf Z}$'s then follow by combining\eq{e:d4Y} and\eq{e:Ch2YJZ}.
 Each of the sign-choices in\eq{e:d4Y} reduces the number of independent component (super)fields in $\{{\bf Y};\bJ,\bX;{\bf Z}\}$ by a factor of two, and the two possible sign-choices produce the supermultiplets
\begin{equation}
  \big\{\,({\bf Y}_{\a\ad}+\ve_\a{}^\b\ve_\ad{}^\bd{\bf Y}_{\b\bd});\cdots\}
   \qquad\textit{vs}.\qquad
  \big\{\,({\bf Y}_{\a\ad}-\ve_\a{}^\b\ve_\ad{}^\bd{\bf Y}_{\b\bd});\cdots\}
 \label{e:Ch2Y+}
\end{equation}
as depicted in\eq{e:B242}; the particular linear combinations of the $\bJ$'s and $\bX$'s, and the {\bf Z}'s that were omitted are recovered by comparing the notation defined by\eq{e:Ch2YJZ} with the operators\eq{e:Ch2D}.
\ping

The supermultiplet\eq{e:Ch2YJZ} {\em\/decomposes\/} into a direct sum of the chiral and twisted-chiral supermultiplets\eq{e:B242}. In turn, the supermultiplet\eq{e:22U} is not decomposable, but may be {\em\/reduced\/} to either the chiral or the twisted-chiral supermultiplet; see\eqs{e:ddD0}{e:ChTwCh} and Figure~\ref{f:int>tch}. In retrospect, the fact that the sequence of transformations\eqs{e:int>vas}{e:Ch2D} is local in one direction but not in the other indicates the inequivalence of the non-decomposable\eq{e:22U} and the decomposable\eq{e:Ch2YJZ}.

\section{Solving Superdifferential Relations}
Since the superfields $\{{\bf Y}_{\a+\ad-};\bJ^\mm_{\ad-},\bX^\mm_{\ad-},\bJ^\pp_{\a+},\bX^\pp_{\a+};{\bf Z}^{\mm\pp},{\bf Z}^\pp_\pp,{\bf Z}^\mm_\mm,{\bf Z}_{\pp\mm}\}$|and so also their complementary linear combinations\eq{e:Ch2Y+}|are otherwise completely free, they may be used in a completely unrestricted path-integration to define a partition functional. Admittedly, however, this is an unwieldy description.

Instead, following such well-known practices\cite{r1001,rPW,rWB,rBK} and Theorem~7.6 of Ref.\cite{r6-1} in particular, the supermultiplet\eq{e:Ch2D} may be ``solved'' (and simplified) in terms of a single intact (unconstrained, ungauged, unprojected...) superfield, $\IU$, by identifying the array
\begin{equation}
   \{{\bf Y};\bJ,\bX;{\bf Z}:\text{\eq{e:Ch2YJZ}}\}\mapsto\IU:~~
   \IY_{\a\ad}\Defl(\rD_{\a+}\rD_{\ad-}\IU),
 \label{e:Y=U}
\end{equation}
and where the remaining component superfields are obtained by applying the tesseract of superderivatives in Figure~\ref{f:22Ds} on the relations\eq{e:Y=U}.

The worldsheet evaluations of\eq{e:Y=U} themselves simply give the component (super)fields depicted by the bottom four nodes. Application of additional superderivatives followed by worldsheet evaluation yields either the component (super)fields depicted by the middle- and top-level nodes, or produces $\vd_\pp$- and $\vd_\mm$-derivatives of these component (super)fields.

However, such ``solutions'' of superdifferential relations typically exhibit gauge invariances. In the case at hand, the component $U_\pp\Defl\frc{i}2[\rD_{1+},\rD_{2+}]\IU|$ does not occur directly in the supermultiplet $\IY_{\a\ad}$, as defined in\eq{e:Y=U}. Instead, $\IY_{\a\ad}$ contains $(\vd_\mm U_\pp)$. Consequently, we are free to replace
\begin{equation}
  U_\pp ~\to~ U_\pp ~+~ V_\pp,\qquad\text{where}\qquad \vd_\mm V_\pp=0.
\end{equation}
The analogous is true of $U_\mm$, which is {\em\/undefined\/} up to the addition of a right-moving summand: $U_\mm\simeq U_\mm+W_\mm$ with $\vd_\pp W_\mm=0$.
 Similarly, only $\vd_\mm$- and $\vd_\pp$-derivatives of the fermions $\j_{\a+}\Defl i\rD_{\a+}\IU|$ and  $\j_{\ad-}\Defl i\rD_{\ad-}\IU|$, respectively, occur within $\IY_{\a\ad}$. These fermions are thus {\em\/undefined\/} up to the addition of a unidextrous fermion: $\j_{\a+}\simeq\j_{\a+}+\c_{\a+}$ and $\j_{\ad-}\simeq\j_{\ad-}+\w_{\ad-}$ where $\vd_\mm\c_{1+}=0$ and $\vd_\pp\w_{\ad-}=0$. Finally, only the D'Alembertian of $u\Defl\IU|$ occurs within $\IY_{\a\ad}$ and so is undefined up to the addition of an arbitrary harmonic function, $u\simeq u+(v+w)$ where $\vd_\mm v=0$ and $\vd_\pp w=0$.

Whereas for worldline supermultiplets such ``gauge'' degrees of freedom were merely the first few terms in a Taylor series|a few constants|in ``solving'' the superderivative relations amongst worldsheet superfields, the gauge degrees of freedom form entire unidextrous supermultiplets:
\begin{equation}
  \vd_\mm\{v;\c_{\a+};V_\pp\}=0\qquad\text{and}\qquad\vd_\pp\{w;\w_{\ad-};W_\mm\}=0.
 \label{e:udif}
\end{equation}
Hence, the array of quadratic superderivatives $\rD_{\a+}\rD_{\ad-}$ provides a {\em\/surjection\/}:
\begin{equation}
  \rD_{\a+}\rD_{\ad-}:\, \IU ~\onto~
  \IY_{\a\ad}=\Big(\rD_{\a+}\rD_{\ad-}(\IU\simeq\IU
                                       +\{v;\c_{\a+};V_\pp\}+\{w;\w_{\ad-};W_\mm\})\Big)
 \label{e:Y=U/}
\end{equation}
Were it not that the supermultiplets\eq{e:udif} are unidextrous and were it not that a harmonic function on the worldsheet is a sum of two unidextrous functions, these would span precisely the degrees of freedom gauged away in the Wess-Zumino gauge from the Hermitian ``vector'' superfield in ${\cal N}{=}1$ supersymmetry in $3{+}1$-dimensional spacetime\cite{r1001,rPW,rWB,rBK}. In that 3+1-dimensional case, the gauge degrees of freedom are in fact themselves off-shell.

 In this sense, worldsheet ``solving'' superderivative relations is ``half-way'' between the worldline and the higher-dimensional spacetime cases:
\begin{equation}
  \begin{array}{@{} cp{40mm}p{65mm} @{}}
 \text{\bsf Dim.} & \MC2l{\text{\bsf Gauge Degrees of Freedom and Their Nature/Occurrence }} \\ 
    \toprule
 1+0 & constants         & first few terms in Taylor series \\ 
 1+1 & unidextrous       & supermultiplets on the half-shell \\ 
 1+n & off-shell $(n>1)$ & off-shell supermultiplets; \\ 
    \bottomrule
  \end{array}
\end{equation}

This analogous adaptation of Theorem~7.6 of Ref.\cite{r6-1} to Adinkras that depict worldsheet $(p,q)$-supermultiplets then guarantees:
\begin{corl}\label{c:Th7.6}
Each Adinkra depicting an off-shell supermultiplet of worldsheet $(p,q)$-super\-sym\-metry without central charges admits|up to unidextrous gauge degrees of freedom|a superfield representation in terms of an intact $(p,q)$-superfield modifying the steps in Theorem~7.6 of Ref.\cite{r6-1} by judiciously replacing $\ddt$ by $\vd_\pp$ or $\vd_\mm$ so as to insure $\Spin(1,1)$-covariance.
\end{corl}
Closely related to Theorem~7.6 of Ref.\cite{r6-1} is also the general construction of ghost-free kinetic Lagrangian terms for worldline supermultiplets reported in Ref.\cite{r6-2}. The close relationship of these to Corollary~\ref{c:Th7.6} would then seem to suggest the existence of an adaptation of this construction of ghost-free kinetic Lagrangian terms for all worldsheet supermultiplets.
\ping

Consistent with the conclusion of the previous appendix is the fact that the off-shell supermultiplets $\{{\bf Y};\bJ,\bX;{\bf Z}\}$ depicted in\eq{e:Ch2YJZ} and $\IU$ depicted by the Adinkra\eq{e:22U} differ by the unidextrous supermultiplets\eq{e:udif}. The mapping $\rD_{\a+}\rD_{\ad-}:\IU\onto\IY_{\a\ad}$ is therefore {\em\/not\/} a strict homomorphism of off-shell supermultiplets (which may be adopted {\em\/verbatim\/} from Ref.\cite{r6-3.2}), and the two supermultiplets must be regarded as strictly inequivalent off-shell supermultiplets.

\small\raggedright
\def\rasp{\leavevmode\raise.45ex\hbox{$\rhook$}}
\providecommand{\href}[2]{#2}\begingroup\raggedright\endgroup

\begin{thebibliography}{10}

\bibitem{r1001}
S.~J. Gates, Jr., M.~T. Grisaru, M.~Ro{\v{c}}ek, and W.~Siegel, {\em
  Superspace}.
\newblock Benjamin/Cummings Pub. Co., Reading, MA, 1983.

\bibitem{rPW}
P.~West, {\em Introduction to Supersymmetry and Supergravity}.
\newblock World Scientific Publishing Co. Inc., Teaneck, NJ, 1990.

\bibitem{rWB}
J.~Wess and J.~Bagger, {\em Supersymmetry and Supergravity}.
\newblock Princeton Series in Physics. Princeton University Press, Princeton,
  NJ, 2nd~ed., 1992.

\bibitem{rBK}
I.~L. Buchbinder and S.~M. Kuzenko, {\em Ideas and Methods of Supersymmetry and
  Supergravity}.
\newblock Studies in High Energy Physics Cosmology and Gravitation. IOP
  Publishing Ltd., Bristol, 1998.

\bibitem{rES-SuGra}
A.~Salam and E.~Sezgin, eds., {\em Supergravities In Diverse Dimensions. Vol.
  1, 2.}
\newblock North-Holland, Amsterdam, 1989.

\bibitem{rGLPR}
S.~J. Gates, Jr., W.~Linch, J.~Phillips, and L.~Rana, ``The fundamental
  supersymmetry challenge remains,'' {\em Gravit. Cosmol.} {\bfseries 8}
  no.~1-2, (2002) 96--100,
  \href{http://arxiv.org/abs/hep-th/0109109}{{\ttfamily arXiv:hep-th/0109109}}.

\bibitem{rGR-1}
S.~J. Gates, Jr. and L.~Rana, ``On extended supersymmetric quantum mechanics,''
  {\em University of Maryland Report: UMDPP 93-194} (1994) unpublished.

\bibitem{rGR0}
S.~J. Gates, Jr. and L.~Rana, ``Ultramultiplets: A new representation of rigid
  2-d, ${N}$=8 supersymmetry,'' {\em Phys. Lett.} {\bfseries B342} (1995)
  132--137,
\href{http://arxiv.org/abs/hep-th/9410150}{{\ttfamily arXiv:hep-th/9410150}}.

\bibitem{r6-1}
C.~F. Doran, M.~G. Faux, S.~J. Gates, Jr., T.~H{\"u}bsch, K.~M. Iga, and G.~D.
  Landweber, ``On graph-theoretic identifications of {A}dinkras, supersymmetry
  representations and superfields,'' {\em Int. J. Mod. Phys.} {\bfseries A22}
  (2007) 869--930, \href{http://arxiv.org/abs/math-ph/0512016}{{\ttfamily
  arXiv:math-ph/0512016}}.

\bibitem{r6-3}
C.~F. Doran, M.~G. Faux, S.~J. Gates, Jr., T.~H{\"u}bsch, K.~M. Iga, G.~D.
  Landweber, and R.~L. Miller, ``Topology types of {A}dinkras and the
  corresponding representations of ${N}$-extended supersymmetry,''
  \href{http://arxiv.org/abs/0806.0050}{{\ttfamily arXiv:0806.0050}}.

\bibitem{r6-3.2}
C.~F. Doran, M.~G. Faux, S.~J. Gates, Jr., T.~H{\"u}bsch, K.~M. Iga, G.~D.
  Landweber, and R.~L. Miller, ``Adinkras for clifford algebras, and worldline
  supermultiplets,'' \href{http://arxiv.org/abs/0811.3410}{{\ttfamily
  arXiv:0811.3410}}.

\bibitem{r6-1.2}
C.~F. Doran, M.~G. Faux, S.~J. Gates, Jr., T.~H{\"u}bsch, K.~M. Iga, and G.~D.
  Landweber, ``A superfield for every dash-chromotopology,'' {\em Int. J. Mod.
  Phys.} {\bfseries A24} (2009) 5681--5695,
  \href{http://arxiv.org/abs/0901.4970}{{\ttfamily arXiv:0901.4970}}.

\bibitem{r6-3.1}
C.~F. Doran, M.~G. Faux, S.~J. Gates, Jr., T.~H{\"u}bsch, K.~M. Iga, G.~D.
  Landweber, and R.~L. Miller, ``Codes and supersymmetry in one dimension,''
  \href{http://dx.doi.org/10.4310/ATMP.2011.v15.n6.a7}{{\em Adv. Theor. Math.
  Phys.} {\bfseries 15} (2011) 1909--1970},
  \href{http://arxiv.org/abs/1108.4124}{{\ttfamily arXiv:1108.4124}}.

\bibitem{rFIL}
M.~G. Faux, K.~M. Iga, and G.~D. Landweber, ``Dimensional enhancement via
  supersymmetry,'' {\em Adv. Math. Phys.} {\bfseries 2011} (2011) 259089,
  \href{http://arxiv.org/abs/0907.3605}{{\ttfamily arXiv:0907.3605}}.

\bibitem{rFL}
M.~G. Faux and G.~D. Landweber, ``Spin holography via dimensional
  enhancement,'' {\em Phys. Lett.} {\bfseries B681} (2009) 161--165,
  \href{http://arxiv.org/abs/0907.4543}{{\ttfamily arXiv:0907.4543}}.

\bibitem{rGH-obs}
S.~J. Gates, Jr. and T.~H{\"u}bsch, ``On dimensional extension of
  supersymmetry: From worldlines to worldsheets,''
  \href{http://dx.doi.org/10.4310/ATMP.2012.v16.n6.a2}{{\em Adv. Theor. Math.
  Phys.} {\bfseries 16} no.~6, (2012) 1619--1667},
  \href{http://arxiv.org/abs/1104.0722}{{\ttfamily arXiv:1104.0722}}.

\bibitem{rA}
M.~Faux and S.~J. Gates, Jr., ``Adinkras: A graphical technology for
  supersymmetric representation theory,'' {\em Phys. Rev. D (3)} {\bfseries 71}
  (2005) 065002, \href{http://arxiv.org/abs/hep-th/0408004}{{\ttfamily
  arXiv:hep-th/0408004}}.

\bibitem{r6--1}
C.~F. Doran, M.~G. Faux, S.~J. Gates, Jr., T.~H{\"u}bsch, K.~M. Iga, and G.~D.
  Landweber, ``Off-shell supersymmetry and filtered {C}lifford supermodules,''
  \href{http://arxiv.org/abs/math-ph/0603012}{{\ttfamily
  arXiv:math-ph/0603012}}.

\bibitem{r6-3c}
C.~F. Doran, M.~G. Faux, S.~J. Gates, Jr., T.~H{\"u}bsch, K.~M. Iga, and G.~D.
  Landweber, ``Relating doubly-even error-correcting codes, graphs, and
  irreducible representations of ${N}$-extended supersymmetry,'' in {\em
  Discrete and Computational Mathematics}, F.~Liu and et~al., eds.
\newblock Nova Science Publishers, Inc., Hauppauge, NY, 2008.
\newblock \href{http://arxiv.org/abs/0806.0051}{{\ttfamily arXiv:0806.0051}}.

\bibitem{rPT}
A.~Pashnev and F.~Toppan, ``On the classification of ${N}$-extended
  supersymmetric quantum mechanical systems,'' {\em J. Math. Phys.} {\bfseries
  42} (2001) 5257--5271,
\href{http://arxiv.org/abs/hep-th/0010135}{{\ttfamily arXiv:hep-th/0010135}}.

\bibitem{rT01}
F.~Toppan, ``Classifying ${N}$-extended 1-dimensional supersymmetric systems,''
  in {\em Noncommutative Structures in Mathematics and Physics}, S.~Duplij and
  J.~Wess, eds., p.~195.
\newblock Kluwer Ac. Pub, 2001.
\newblock
\href{http://arxiv.org/abs/hep-th/0109047}{{\ttfamily arXiv:hep-th/0109047}}.
\newblock

\bibitem{rT01a}
F.~Toppan, ``Division algebras, extended supersymmetries and applications,''
  {\em Nucl. Phys. Proc. Suppl.} {\bfseries 102} (2001) 270--277,
\href{http://arxiv.org/abs/hep-th/0109073}{{\ttfamily arXiv:hep-th/0109073}}.

\bibitem{rCRT}
H.~L. Carrion, M.~Rojas, and F.~Toppan, ``Octonionic realizations of
  1-dimensional extended supersymmetries. {A} classification,'' {\em Mod. Phys.
  Lett.} {\bfseries A18} (2003) 787--798,
\href{http://arxiv.org/abs/hep-th/0212030}{{\ttfamily arXiv:hep-th/0212030}}.

\bibitem{rKRT}
Z.~Kuznetsova, M.~Rojas, and F.~Toppan, ``Classification of irreps and
  invariants of the ${N}$-extended supersymmetric quantum mechanics,'' {\em
  JHEP} {\bfseries 03} (2006) 098,
\href{http://arxiv.org/abs/hep-th/0511274}{{\ttfamily arXiv:hep-th/0511274}}.

\bibitem{rKT07}
Z.~Kuznetsova and F.~Toppan, ``Refining the classification of the irreps of the
  1{D} ${N}$-extended supersymmetry,'' {\em Mod. Phys. Lett.} {\bfseries A23}
  (2008) 37--51, \href{http://arxiv.org/abs/hep-th/0701225}{{\ttfamily
  arXiv:hep-th/0701225}}.

\bibitem{rGKT10}
M.~Gonzales, S.~Khodaee, and F.~Toppan, ``On non-minimal ${N}=4$
  supermultiplets in {1D} and their associated sigma-models,'' {\em J. Math.
  Phys.} {\bfseries 52} (2011) 013514,
  \href{http://arxiv.org/abs/1006.4678}{{\ttfamily arXiv:1006.4678}}.

\bibitem{rUMD09-1}
S.~J. Gates, Jr., J.~Gonzales, B.~MacGregor, J.~Parker, R.~Polo-Sherk, V.~G.~J.
  Rodgers, and L.~Wassink, ``4{D}, $\mathcal{N}=1$ supersymmetry genomics
  ({I}),'' {\em JHEP} {\bfseries 12} (2009) 009,
  \href{http://arxiv.org/abs/0902.3830}{{\ttfamily arXiv:0902.3830}}.

\bibitem{rUMD09-2}
S.~J. Gates, Jr., J.~Hallett, J.~Parker, V.~G.~J. Rodgers, and K.~Stiffler,
  ``4{D}, $\mathcal{N}=1$ supersymmetry genomics ({II}),'' {\em JHEP}
  {\bfseries 1206} (2012) 071, \href{http://arxiv.org/abs/1112.2147}{{\ttfamily
  arXiv:1112.2147}}.

\bibitem{rGHHS-CLS}
S.~J. Gates, Jr., J.~Hallett, T.~H{\"u}bsch, and K.~Stiffler, ``The real
  anatomy of complex linear superfields,'' {\em Int. J. Mod. Phys.} {\bfseries
  A27} (2012) 1250143, \href{http://arxiv.org/abs/1202.4418}{{\ttfamily
  arXiv:1202.4418}}.

\bibitem{rUMD12-3}
I.~Chappell, J.~Gates, S.~James, I.~Linch, William~D., J.~Parker, S.~Randall,
  {\em et~al.}, ``4{D}, ${N}=1$ supergravity genomics,''
  \href{http://arxiv.org/abs/1212.3318}{{\ttfamily arXiv:1212.3318 [hep-th]}}.

\bibitem{rGSW1}
M.~B. Green, J.~H. Schwarz, and E.~Witten, {\em Superstring Theory}, vol.~1:
  Introduction of {\em Cambridge Monographs on Mathematical Physics}.
\newblock Cambridge University Press, Cambridge, 1987.

\bibitem{rGSW2}
M.~B. Green, J.~H. Schwarz, and E.~Witten, {\em Superstring Theory}, vol.~2:
  Loop Amplitudes, Anomalies and Phenomenology of {\em Cambridge Monographs on
  Mathematical Physics}.
\newblock Cambridge University Press, Cambridge, 1987.

\bibitem{rJPS}
J.~Polchinski, {\em String theory}.
\newblock Cambridge Monographs on Mathematical Physics. Cambridge University
  Press, Cambridge, 1998.

\bibitem{rBBS}
K.~Becker, M.~Becker, and J.~H. Schwarz, {\em String Theory and {M}-Theory: A
  Modern Introduction}.
\newblock Cambridge University Press, 2007.

\bibitem{rHSS}
T.~H{\"u}bsch, ``Haploid (2,2)-superfields in 2-dimensional space-time,'' {\em
  Nucl. Phys.} {\bfseries B555} no.~3, (1999) 567--628.

\bibitem{rGSS}
R.~Q. Almukahhal and T.~H{\"u}bsch, ``Gauging {Y}ang-{M}ills symmetries in
  1+1-dimensional space-time,'' {\em Int. J. Mod. Phys. A} {\bfseries 16}
  no.~29, (2001) 4713--4768.

\bibitem{rUDSS01}
R.~Brooks, F.~Muhammad, and S.~J. Gates, Jr., ``Unidexterous ${D} = 2$
  supersymmetry in superspace,''
{\em Nucl. Phys.} {\bfseries B268} (1986) 599--620.

\bibitem{rUDSS04}
S.~J. Gates, Jr., R.~Brooks, and F.~Muhammad, ``Unidexterous superspace: The
  flax of (super)strings,''
{\em Phys. Lett.} {\bfseries B194} (1987) 35.

\bibitem{rUDSS06}
S.~J. Gates, Jr. and F.~Gieres, ``Unidexterous supergravity, beltrami
  parametrization and {BRST} quantization,''
{\em Nucl. Phys.} {\bfseries B320} (1989) 310.

\bibitem{rUDSS08}
S.~J. Gates, Jr. and T.~H{\"u}bsch, ``Unidexterous locally supersymmetric
  actions for {C}alabi-{Y}au compactifications,''
{\em Phys. Lett.} {\bfseries B226} (1989) 100.

\bibitem{rUDSS09}
S.~J. Gates, Jr. and T.~H{\"u}bsch, ``{C}alabi-{Y}au heterotic strings and
  unidexterous sigma models,''
{\em Nucl. Phys.} {\bfseries B343} (1990) 741--774.

\bibitem{rAD-Uni}
R.~Amorim and A.~K. Das, ``Unidexterous versus ambidexterous gravities,''
  \href{http://dx.doi.org/10.1103/PhysRevD.54.4177}{{\em Phys. Rev.} {\bfseries
  D54} (1996) 4177--4180}.

\bibitem{rHP1}
T.~H{\"u}bsch and I.~E. Petrov, ``Worldsheet matter superfields on
  half-shell,'' {\em J. Phys. A} {\bfseries 43} (2010) 295206,
  \href{http://arxiv.org/abs/0912.1038}{{\ttfamily arXiv:0912.1038}}.

\bibitem{rHT-UCS}
T.~H{\"u}bsch, ``Unidexterously constrained worldsheet superfields,'' {\em J.
  Phys.} {\bfseries A43} (2010) 295402,
  \href{http://arxiv.org/abs/1002.0515}{{\ttfamily arXiv:1002.0515}}.

\bibitem{rRLM-codes}
R.~L. Miller, ``Doubly-even codes.''
\newblock \url{http://www.rlmiller.org/de\_codes/}.

\bibitem{rHTSSp08}
T.~H{\"u}bsch, ``Superspace: A comfortably vast algebraic variety,'' in {\em
  Geometry and Analysis}, L.~Ji, ed., vol.~2 of {\em Advanced Lectures in
  Mathematics}, pp.~39--67.
\newblock International Press, 2010.
\newblock \href{http://arxiv.org/abs/0901.2136}{{\ttfamily arXiv:0901.2136}}.
\newblock Closing address at the {C}onference ``{G}eometric {A}nalysis:
  {P}resent and {F}uture'', Harvard University, 2008.

\bibitem{rCHVP}
W.~C. Huffman and V.~Pless, {\em Fundamentals of Error-Correcting Codes}.
\newblock Cambridge Univ. Press, 2003.

\bibitem{rWyb}
B.~G. Wybourne, {\em Classical Groups for Physicists}.
\newblock John Wiley \&\ Sons Inc., 1974.

\bibitem{rFRH}
F.~R. Harvey, {\em Spinors and Calibrations}, vol.~9 of {\em Perspectives in
  Mathematics}.
\newblock Academic Press Inc., Boston, MA, 1990.

\bibitem{rJSH}
J.-S. Huang, {\em Lectures on Representation Theory}.
\newblock World Scientific Pub. Co., 1999.

\bibitem{r6-2}
C.~F. Doran, M.~G. Faux, S.~J. Gates, Jr., T.~H{\"u}bsch, K.~M. Iga, and G.~D.
  Landweber, ``Adinkras and the dynamics of superspace prepotentials,'' {\em
  Adv. S. Th. Phys.} {\bfseries 2} no.~3, (2008) 113--164,
  \href{http://arxiv.org/abs/hep-th/0605269}{{\ttfamily arXiv:hep-th/0605269}}.

\bibitem{rTHGK12}
T.~H{\"u}bsch and G.~A. Katona, ``On the construction and the structure of
  off-shell supermultiplet quotients,''
  \href{http://dx.doi.org/10.1142/S0217751X12501734}{{\em Int. J. Mod. Phys.}
  {\bfseries A29} (2012) 1250173},
  \href{http://arxiv.org/abs/1202.4342}{{\ttfamily arXiv:1202.4342}}.

\bibitem{rGIKT12}
M.~Gonzales, K.~Iga, S.~Khodaee, and F.~Toppan, ``Pure and entangled ${N}=4$
  linear supermultiplets and their one-dimensional sigma-models,'' {\em J.
  Math. Phys.} {\bfseries 53} (2012) 103513,
  \href{http://arxiv.org/abs/1204.5506}{{\ttfamily arXiv:1204.5506}}.

\bibitem{rTHGK13}
T.~H{\"u}bsch and G.~A. Katona, ``Golden ratio controlled chaos in
  supersymmetric dynamics,'' {\em Int. J. of Mod. Phys.} {\bfseries A} (2013,
  in print) , \href{http://arxiv.org/abs/1308.0654}{{\ttfamily
  arXiv:1308.0654}}.

\bibitem{rFGH}
M.~G. Faux, S.~J. Gates, Jr., and T.~H{\"u}bsch, ``Effective symmetries of the
  minimal supermultiplet of ${N} = 8$ extended worldline supersymmetry,'' {\em
  J. Phys.} {\bfseries A42} (2009) 415206,
  \href{http://arxiv.org/abs/0904.4719}{{\ttfamily arXiv:0904.4719}}.

\bibitem{rSSSS4}
A.~Salam and J.~Strathdee, ``On superfields and {F}ermi-{B}ose symmetry,'' {\em
  Phys. Rev. D} {\bfseries 11} (1975) 1521--1535.

\bibitem{rR+T-Combi}
F.~Roberts and B.~Tesman, {\em Applied Combinatorics}.
\newblock Prentice Hall, 2nd~ed., 2003.

\bibitem{rGHR}
S.~J. Gates, Jr., C.~M. Hull, and M.~Ro{\v{c}}ek, ``Twisted multiplets and new
  supersymmetric nonlinear sigma models,''
{\em Nucl. Phys.} {\bfseries B248} (1984) 157.

\bibitem{rSChSF0}
A.~Sevrin and J.~Troost, ``Off-shell formulation of ${N}=2$ nonlinear sigma
  models,'' \href{http://dx.doi.org/10.1016/S0550-3213(97)00103-X}{{\em Nucl.
  Phys.} {\bfseries B492} (1997) 623--646},
  \href{http://arxiv.org/abs/hep-th/9610102}{{\ttfamily arXiv:hep-th/9610102}}.

\bibitem{rSChSF}
J.~Bogaerts, A.~Sevrin, S.~van~der Loo, and S.~Van~Gils, ``Properties of
  semi-chiral superfields,''
  \href{http://dx.doi.org/10.1016/S0550-3213(99)00490-3}{{\em Nucl. Phys.}
  {\bfseries B562} (1999) 277--290},
  \href{http://arxiv.org/abs/hep-th/9905141}{{\ttfamily arXiv:hep-th/9905141}}.

\bibitem{rTwSJG0}
S.~J. Gates, Jr., ``Superspace formulation of new nonlinear sigma models,''
{\em Nucl. Phys.} {\bfseries B238} (1984) 349--366.

\end{thebibliography}
\end{document}